\documentclass[aps,prl,pdflatex,twocolumn,superscriptaddress]{revtex4-1}
\usepackage[compat=1.0.0]{tikz-feynman}
\pdfoutput=1

\usepackage{amssymb}

\usepackage{multirow}
\usepackage{bm}
\usepackage{amsmath}
\usepackage{graphicx}
\usepackage{epstopdf}
\usepackage{subfigure}
\usepackage{natbib}
\usepackage{epsfig}
\usepackage{amsfonts}
\usepackage{mathrsfs}
\usepackage[toc,page,title,titletoc,header]{appendix}
\usepackage[colorlinks,linkcolor=blue,citecolor=blue,anchorcolor=blue]{hyperref}

\usepackage{braket}

\usepackage{dsfont,amsthm,amsbsy}

\def\Tr{{\rm Tr}}

\usepackage{fancyhdr}

\usepackage{ulem}

\usepackage{bbm}

\newcommand{\bea}{\begin{equation} \begin{aligned}}
\newcommand{\eea}{\end{aligned} \end{equation} }

\newcommand{\bpm}{\begin{pmatrix}}
\newcommand{\epm}{\end{pmatrix}}

\newcommand{\del}{\partial}

\begin{document}
\title{Exact models of chiral flat-band superconductors}

\author{Zhaoyu~Han}
\email{zhan@fas.harvard.edu}
\affiliation{Department of Physics, Harvard University, Cambridge, Massachusetts 02138, USA}

\author{Jonah~Herzog-Arbeitman}
\affiliation{Department of Physics, Princeton University, Princeton, NJ 08544, USA}

\author{Qiang~Gao}
\affiliation{Department of Physics, Harvard University, Cambridge, Massachusetts 02138, USA}

\author{Eslam~Khalaf}
\email{eslam_khalaf@fas.harvard.edu}
\affiliation{Department of Physics, Harvard University, Cambridge, Massachusetts 02138, USA}

\begin{abstract}

Recent experiments have reported the surprising observation of superconductivity in flavor polarized, nearly flat bands (FBs) of rhombohedral graphene. Motivated by these findings, we introduce a class of models for single-flavor FBs with inversion symmetry, where we show a local attractive interaction between orbitals with opposite parities leads to an exact superconducting ground state. We argue that this model can be relevant to realistic multi-flavor systems including short-range repulsion, since the main effect of such repulsion is to induce flavor polarization leaving possibly attractive residual interaction between different flavorless orbitals. The nature of the pairing is determined by the interplay between the FB quantum geometry and the interaction, and is often topological when the parent FB is so. Interestingly, each such model has two nearly degenerate pairing modes, whose energetic competition can be tuned by a change in the charge transfer gap between the two orbitals or electron density. These modes have the same angular momentum but different pairing amplitude structure and possibly different topology. We show that the superfluid stiffness is proportional to the attractive interaction scale using a combination of analytical variational upper bounds and numerical bootstrap lower bounds. We find empirically that the maximum superfluid stiffness is achieved when the hot spots of quantum geometry in the Brillouin zone are marginally filled. 

\end{abstract}

\maketitle

Recent experiments have reported evidence of chiral, time-reversal-breaking superconductivity (SC) in rhombohedral $L$-layer graphene (R$L$G) with $L=4,5$~\cite{han2025signatures}, $6$~\cite{nguyen2025hierarchy} and twisted MoTe2 (tMoTe2)~\cite{xu2025signatures}. While SC is, by now, rather common in moir\'e and more generally flat band (FB) systems, a remarkable feature of these superconductors is that their parent band is likely spin and valley polarized. This poses a major challenge to standard theoretical approaches to superconductivity, which typically rely on the weak-coupling instability of Fermi surfaces of time-reversal related flavors~\cite{Lian2019PhononTBG,You2019SO4TBG,Yu2023EulerObstructed,Christos2023BandOffDiagonal,Putzer2025EliashbergBOD,Chen2025tTMDpairing,Schrade2024tWSe2Theory,jiang2025quantum,chau2024originsuperconductivityrhombohedraltrilayer}. Together with evidence of other strongly correlated phenomena in these materials, this suggests that a strong coupling perspective, where the dispersion is considered small compared to interaction, is a better starting point. In such a FB limit, the physics is mostly determined by the interplay between the interaction and the FB wavefunctions, known as the quantum geometry (QG)~\footnote{Here, we use QG to refer to the full gauge-invariant information encoded in wavefunction overlaps between arbitrary momenta rather than the long wavelength quantities such as the quantum metric.}. As a result, QG is expected to control many properties of the SC phase including the pairing symmetry, single-particle gap, and phase stiffness; the latter two together limit the critical temperature of the phase.  Several recent works have aimed to study the effects of QG on pairing in single-flavor bands~\cite{geier2024chiral,guerci2025fractionalization,wang2024chiral,christos2025finite,patri2025family,PhysRevLett.134.176001,k8s3-dgfs,may2025pairing,dong2025controllable,doi:10.1126/sciadv.abh2233,guerci2025spin,zgnk-rw1p,hu2025layerpseudospinsuperconductivitytwisted} using various approaches. However, an analytically controlled theory beyond the weak coupling limit remains lacking.

On the other hand, the development of exactly solvable models, while rare in realistic setups, can play a pivotal role in understanding complex phases of matter in FB systems. A classic example is the Haldane-Trugman-Kivelson pseudopotential model~\cite{PhysRevLett.51.605,PhysRevB.31.5280} for the fractional quantum Hall effect in the lowest Landau level, which energetically justifies the Laughlin wavefunction~\cite{PhysRevLett.50.1395} in the presence of short-ranged repulsions. This has recently been extended to a wider class of FBs, called ideal or vortexable~\cite{tarnopolsky2019origin,ledwith2020fractional, Ozawa2021RelationsInsulators, meraKahlerGeometryChern2021,meraEngineeringGeometricallyFlat2021,meraRelatingTopologyDirac2021, wangExactLandauLevel2021b, ledwith2022family,ledwith2023vortexability}, allowing the development of criteria for realizing fractional Chern insulators~\cite{regnaultFractionalChernInsulator2011, neupertFractionalQuantumHall2011,shengFractionalQuantumHall2011,qi_generic_2011,parameswaranFractionalChernInsulators2012,wuBlochModelWave2013,liuRecentDevelopmentsFractional2022,parameswaranFractionalQuantumHall2013,BergholtzReview2013} in these systems. Other examples include attractive Hubbard models with spinful time-reversal symmetry~\cite{PhysRevB.94.245149,PhysRevB.106.014518,herzog2022many} for FB superconductivity~\cite{Peotta_2015, PhysRevLett.117.045303, T_rm__2022,2023arXiv230808248P, doi:10.1073/pnas.2416726122, PhysRevB.109.214518, PhysRevLett.132.026002}, where an exact $s$-wave pairing ground state is stabilized by the local attraction, as well as models for FB ferromagnetism with repulsive interactions~\cite{1993CMaPh.158..341M,PhysRevLett.122.246401,PhysRevLett.122.246402,repellinFerromagnetismNarrowBands2020, bultinckGroundStateHidden2020, lianTBGIVExact2020,PhysRevB.103.205415, ledwith2021strong,2024arXiv240207171K}. Recently, two of us have introduced a set of compatibility conditions of the FB wavefunctions, called `quantum geometric nesting' (QGN)~\cite{PhysRevX.14.041004} conditions (see also \cite{zhang2025identifying,sun2024flat}), which enable systematic constructions of infinitely many solvable locally interacting models for certain symmetry broken states that unify the previous examples. One major drawback of the QGN models, however, is that the constructed order parameters are uniform in momentum space. This rules out their applicability to SC phases in a flavor-polarized FB, where the order parameter must vanish at certain points in the Brillouin zone (BZ) by the Pauli exclusion principle.

In this work, we introduce a new class of exact models for SC in a single FB which is assumed to be isolated from other bands in the system. We find that an inversion symmetry suffices to guarantee a perfect compatibility between the wavefunctions at $\pm\bm{k}$ through a generalized nesting condition, so that any local attraction between a pair of orbitals with opposite parities gives rise to exact SC ground states. The primary effects of Coulomb repulsion are presumed to induce the flavor polarization, such that an effective attraction resulting from the coupling to collective bosons (e.g. phonons~\cite{PhysRevLett.132.226001}) can overcome the residual repulsion between distinct orbitals~\footnote{We also note that in the FB limit the retardation effects of the bosons are not significant, so the interactions can be regarded instantaneous.}. In fact, a spinful FB model with a strong on-site Hubbard repulsion plus a subleading inter-orbital density attraction of our sort would still be exactly solvable by combining the previous results on flavor ferromagnetism~\cite{1993CMaPh.158..341M,PhysRevLett.122.246401,PhysRevLett.122.246402,repellinFerromagnetismNarrowBands2020, bultinckGroundStateHidden2020, lianTBGIVExact2020,PhysRevB.103.205415, ledwith2021strong,2024arXiv240207171K} and ours. This justifies the relevance of our models to realistic systems. Our construction reveals a very simple condition for a single flavor FB to host SC in the presence of short-ranged attraction. Furthermore, it goes beyond previous works by allowing arbitrarily complex order parameters including the topological ones supporting Majorana edge modes~\cite{PhysRevB.61.10267,volovik2003universe}. Our construction enables us to obtain various properties of the SC states including analytical variational bounds on the quasi-particle gap and the superfluid stiffness based on QG. Furthermore, we numerically establish a non-zero lower bound of the superfluid stiffness, using quantum many-body bootstrap lower bound~\cite{gao2025bootstrappingSC}, which is proportional to the interaction strength. One of the most surprising implications of our construction is the existence of two almost degenerate, particle-hole conjugate pairing modes which may have different Majorana Chern numbers in two dimensions. This suggests that the existence of two distinct SC domes switchable by tuning electron density and/or displacement field is possibly a general feature of QG-stabilized single-flavor FB SC and may explain the appearance of two distinct SC domes in R$L$G with $L=4,5,6$~\cite{han2025signatures,nguyen2025hierarchy}. We illustrate our results with a minimal two-band model that reproduces many features of the FBs of R$L$G with $L=5$, suggesting our formalism may be relevant to the observed chiral SC in this system. 

{\bf The model. }  We begin by discussing the general construction before presenting a concrete model and discussing its possible relevance to physical systems. We consider a single isolated FB with inversion symmetry. The interaction scale is assumed to be much smaller than the band gap to the remote bands, allowing us to project the interaction onto the FB subspace. We express the electronic structure with an inversion-symmetric set of local orbitals $\{\phi_{\bm{R}\alpha}(\bm{r})\}$ in unit cell $\bm{R}$, which are eigenstates of the inversion operation $\mathcal{I}$ in the sense that $\mathcal{I} \phi_{\bm{R}\alpha}(\bm{r}) = p_\alpha \phi_{-\bm{R}\alpha}(\bm{r})$ where $p_\alpha =\pm1$ is the orbital parity. We emphasize that these are {\it not} Wannier orbitals describing the FB, which could be non-local due to the topology of the FB. Instead, they are the microscopic, local basis states on which the FB wavefunctions have support. The corresponding electron annihilation operator on orbital $\alpha$ and momentum $\bm{k}$ satisfies $\mathcal{I} \hat{c}_{\bm{k}\alpha } \mathcal{I} = p_\alpha \hat{c}_{-\bm{k}\alpha }$~\footnote{Here we implicitly embed all the orbital positions to the inversion center.}. 
The FB electron operators  $\hat{\gamma}_{\bm{k}} \equiv \sum_{\alpha} u^*_\alpha(\bm{k}) \hat{c}_{\bm{k}\alpha}$ are defined by the Bloch wavefunction of the FB, $\vec{u}(\bm{k})$. Inversion assures they obey a `nesting' relation
\begin{align}\label{eq: inversion}
    p_\alpha u_\alpha(\bm{k}) =  u_\alpha(-\bm{k}) \mathrm{e}^{\mathrm{i}\xi(\bm{k})}
\end{align}
where $\xi(\bm{k})$ is the gauge-dependent phase (sewing matrix~\cite{PhysRevB.89.155114,PhysRevB.86.115112}) that is odd in $\bm{k}$  (SM Sec.~III~A). This is a generalized QGN condition differing from the original one~\cite{PhysRevX.14.041004}. A more general discussion for other FB systems can be found in SM Sec.~II.

The central result of this work is the following: for {\it any} pair of orbitals $A,B$ with opposite parities $p_A p_B =-1$, an attractive interaction between them is exactly solvable (after projecting onto the FB subspace): 
\begin{align}\label{eq: Hint simple}
    \hat{H}_{AB}= & -V \sum_{\bm{R}} \hat{n}_{\bm{R}A}\hat{n}_{\bm{R}B} + \hat{H}_{AB,2} 
\end{align}
where $V>0$, $\hat{n}_{\bm{R}\alpha} \equiv \hat{c}^\dagger_{\bm{R} \alpha}\hat{c}_{\bm{R} \alpha}$, and $\hat{H}_{AB,2}= V \sum_{\bm{R}} \left[ \rho^\text{R}_B (\hat{n}_{\bm{R}A}+\hat{n}_{\bm{R}B}) +\rho^\text{FB}_A\hat{n}_{\bm{R}B} \right]$ is a single-particle counter term ensuring that the projected Hamiltonian has no bare dispersion, i.e. it only contains fermion quartic terms $\sim \hat{\gamma}^\dagger \hat{\gamma}\hat{\gamma}^\dagger \hat{\gamma}$. $\rho^\text{R}_\alpha$ and $\rho_\alpha^\text{FB}$ are, respectively, the average electron density on $\alpha$ orbital in the occupied (lower) remote bands and the FB,  e.g. $ \rho_\alpha^\text{FB}\equiv \int\frac{\mathrm{d}\bm{k} }{\Omega_\text{BZ}}|u_\alpha(\bm{k})|^2$ where the integration is over the first BZ with volume $\Omega_\text{BZ}$. This counter term is a simple modification to the charge transfer gaps of the single-particle Hamiltonian and admits an apparent simplification when the FB is the lowest or the highest in the spectrum.

\begin{figure}[t]
    \centering
    \includegraphics[width=0.8\linewidth]{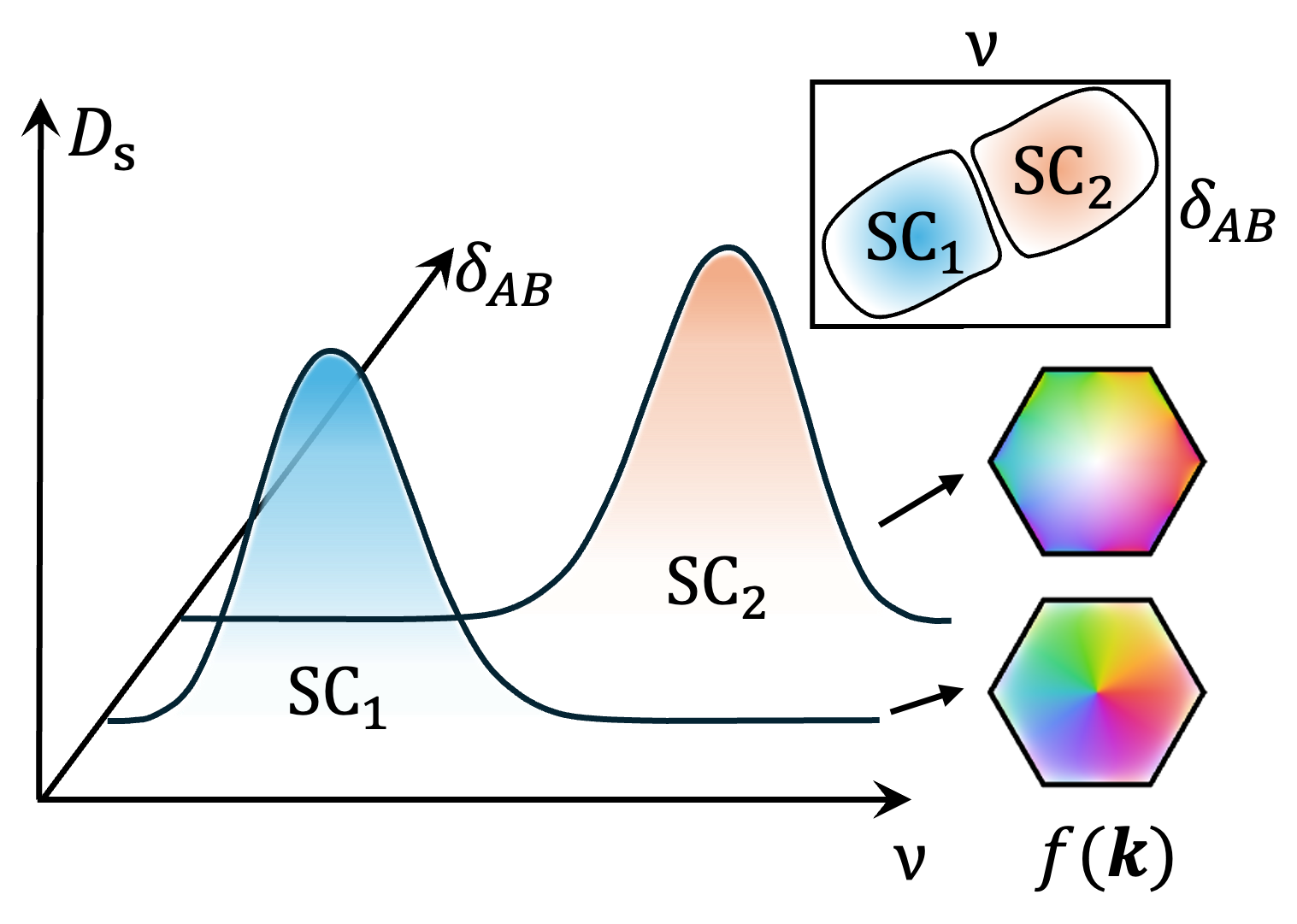}
    \caption{An illustration of the particle-hole duality. Two exact SC states are stabilized at two values of charge transfer gap between two orbitals $A,B$, which we call $\delta_{AB}$. The stiffness of the two states are related by $D_\text{s,1} (\nu) = D_\text{s,2} (1-\nu)$ (similarly for pairing gap). Two symmetric SC domes on the $\delta_{AB}-\nu$ plane thus can be expected, which is reminiscent of the observations in R$L$G with $L=4,5,6$~\cite{han2025signatures,nguyen2025hierarchy}. The form factors $f(\bm{k})$ are plotted for the two band model defined in Eq.~\ref{eq: two band} with $M=1$ (regularized onto a lattice by Eq.~\ref{eq: compactification}) in order to demonstrate the phase (color) winding and amplitude (shading) structures of the two conjugate orders. }
    \label{fig: duality}
\end{figure}

{\bf A particle-hole duality. } The counter term in the construction is asymmetric for orbitals $A,B$ although the interaction is symmetric. Thus, exchanging $A \leftrightarrow B$ will lead to a {\it different} exactly solvable Hamiltonian, $\hat{H}_{BA}$, that has the same interaction but differs in values of the relative potentials (charge transfer gaps):
\begin{align}
    \hat{H}_{BA} - \hat{H}_{AB}  = V \sum_{\bm{R}} \left[  \rho^\text{FB}_B\hat{n}_{\bm{R}A}  - \rho^\text{FB}_A\hat{n}_{\bm{R}B} \right] \ .
\end{align}
The ground states of $\hat{H}_{BA}$ and $\hat{H}_{AB}$ also have different SC orders with possibly different topology as we will discuss below. The presence of two competing SC phases is thus a fundamental feature of this problem. 

In the FB subspace, the two Hamiltonians are related by the combination of a FB particle-hole transformation $\mathcal{C}: \hat{\gamma}\rightarrow \hat{\gamma}^\dagger$ (implying $\nu \rightarrow 1-\nu$) and a time-reversal transformation $\mathcal{T}:\bm{k}\rightarrow -\bm{k}, \mathrm{i}\rightarrow -\mathrm{i}$ (SM Sec.~III~C):
\begin{align}
    \mathcal{T}\mathcal{C} \hat{P}\hat{H}_\text{AB} \hat{P}(\mathcal{T}\mathcal{C} )^{-1}  = \hat{P}\hat{H}_\text{BA} \hat{P}
\end{align}
and thus all the properties of the two models, including the excitation spectra and the stiffness, are related accordingly; see Fig.~\ref{fig: duality} for an illustration. In the below discussions we will focus on $\hat{H}_{AB}$ without loss of generality.

{\bf Exact solvability. } The pairing operator that generate the {\it exact} many-body ground states (GS) of the FB-subspace projected (implemented by $\hat{P}$) Hamiltonian, $\hat{P}\hat{H}_\text{AB}\hat{P}$, is:
\begin{align}
   \hat{\eta}^\dagger \equiv  \int\frac{\mathrm{d}\bm{k} }{\Omega_\text{BZ}}\frac{u_{A}^*(\bm{k})}{u_{B}(-\bm{k})} \hat{\gamma}_{\bm{k}}^\dagger \hat{\gamma}_{-\bm{k}}^\dagger.
\end{align}
Following Eq.~\ref{eq: inversion} it is apparent that the form factor $f(\bm{k})=u_{A}^*(\bm{k})/u_{B}(-\bm{k})$ is antisymmetric in $\bm{k}$. Note that $f(\bm{k})$ may have 
singularities in the BZ where $u_{B}(-\bm{k}) = 0$, but we will show that this does not lead to any divergence in any physical observables. Exchanging $A\leftrightarrow B$ leads to the particle-hole conjugate form factor, which has the same phase winding about the origin (and thus the same angular momentum) but inverse amplitude. The GS in the $2N$-particle sector is $(\hat{\eta}^\dagger)^N|\text{vac}\rangle$~\footnote{The vacuum state here is defined as the lower remote bands being occupied but the FB and the higher remote bands being empty.} which has zero energy. A detailed proof of this claim is presented in SM Sec.~II. We sketch the strategy here: upon projection onto the FB subspace, the interactions will become (up to unimportant constant and chemical potential shifts)
$
 \hat{P}\hat{H}_{AB} \hat{P} =  V \sum_{\bm{R}} \hat{P} \hat{c}^\dagger_{\bm{R}B}\hat{c}_{\bm{R}A}\hat{P} \hat{c}^\dagger_{\bm{R}A}\hat{c}_{\bm{R}B}\hat{P}
$
which is positive semi-definite (PSD). Then one can check that $
    [\hat{P} \hat{c}^\dagger_{\bm{R}A}\hat{c}_{\bm{R}B}\hat{P},\hat{\eta}^\dagger] = 0 $
therefore $ \hat{P}\hat{H}_{AB} \hat{P} (\hat{\eta}^\dagger)^N|\text{vac}\rangle =0$ since the combination $\hat{P} \hat{c}^\dagger_{\bm{R}A}\hat{c}_{\bm{R}B}\hat{P}$ can move to the right of $(\hat{\eta}^\dagger)^N$ and annihilate the vacuum. Using the PSD property of the projected Hamiltonian, we conclude that these states are GSs. 

Several technical remarks are in order: (1) the terms at different $\bm{R}$s generally do not commute or form a closed algebra, therefore this model is not a commuting projector Hamiltonian or a `generalized mean-field Hamiltonian~\cite{PhysRevLett.97.190501,PhysRevB.79.214440}' (which consists of terms that form a polynomially small algebra) but only belong to the frustration-free (FF) family; (2) Different from the QGN models~\cite{PhysRevX.14.041004}, here $\hat{\eta}^\dagger$ does not necessarily commute with the projected Hamiltonian, so that more degrees of freedom for the possible GSs are liberated at the expense of no pseudospin ${\rm SU}(2)$ symmetry being exact in this model; (3) Crucially, the (unprojected) interactions here are strictly local in real space, distinct from those with `local' structures only in momentum space~\cite{phillips2020exact,PhysRevB.79.180501,PhysRevB.84.100503,PhysRevLett.87.066403,RevModPhys.76.643,ORTIZ2005421}.

{\bf The ground states. } Since all the GSs in different number sectors are degenerate, any linear combination of them is still a GS. In particular it is convenient to define a coherent state with an arbitrary complex number $z$:
\begin{align}
    |z\rangle \equiv & \sum_{N} \mathrm{e}^{-z N} \left(\hat{\eta}^\dagger/2\right)^N |\text{vac}\rangle 
\end{align}
which can be verified to be precisely a BCS wavefunction, given by an occupied Bogoliubov-de-Gennes (BdG) band with wavefunction : $\vec{w}^T(\bm{k}) = \left(e^{-z} u^*_{A}(\bm{k}) \vec{u}^T(\bm{k}) \ , \  u_{B}(\bm{k}) \vec{u}^\dagger(-\bm{k}) \right)^T $ (SM Sec.~III~B). This simple result suggests that certain BCS MF theories of the current model could yield exact result~\cite{PhysRevB.96.115110} despite the strong coupling nature of the problem. 

With this mapping, in two dimensional cases, one can also define the majorana Chern number of the SC GS:
\begin{align}
    C^\text{maj}= C^\text{BdG} +C^\text{FB} +2 C^<
\end{align}
where $C^\text{FB}$ and $C^<$ are the Chern numbers of the FB and the occupied remote bands, respectively, and $C^\text{BdG}$ is the Chern number of the wavefunction $\vec{w}(\bm{k})$, which is independent of $z$. In two-orbital systems, we prove in SM Sec.~III~D that $C^{\text{BdG}} $ is always $0$. If the FB is the lower band, $C^<=0$, implying that $C^\text{maj} =  C^\text{FB}$ is directly determined by the topology of the parent FB ($C^\text{maj} = - C^\text{FB}$ for the upper-band case). Exchanging $A\leftrightarrow B$, the conjugate SC order may have a different $C^\text{maj}$, as we show with an example in App.~\ref{app: L band}.

These states have simple structures that allow for analytical computations of various important properties. For later convenience, we define the occupation function
\begin{align}
    \nu(\bm{k};\Re z)\equiv \frac{\langle z|\hat{\gamma}^\dagger_{\bm{k}}\hat{\gamma}_{\bm{k}}|z\rangle}{ \langle z|z\rangle }  = \frac{e^{-2\Re z}|u_A(\bm{k})|^2}{|u_B(\bm{k})|^2+e^{-2\Re z} |u_A(\bm{k})|^2}
\end{align}
where $\Re$ means the real part. In terms of this function we define the FB electron-averaged expectation for any function $O(\bm{k})$:
\begin{align}
    \langle O\rangle_z \equiv \int\frac{\mathrm{d}\bm{k} }{\Omega_\text{BZ}}O(\bm{k}) \nu(\bm{k};\Re z).
\end{align}
For example, $\nu(\Re z) \equiv \langle 1 \rangle_z $ is the FB filling fraction of the state $|z\rangle$, which is monotonically tuned by $\Re z$. The off-diagonal-long-range-order $\langle z|\hat{\eta}^\dagger|z\rangle =\nu(\Re z)\mathrm{e}^{z}$ is nonzero and has an arbitrary phase, confirming that this state breaks the charge $U(1)$ symmetry in the ground states.

\begin{figure*}[t]
    \centering
\includegraphics[width=0.48\linewidth]{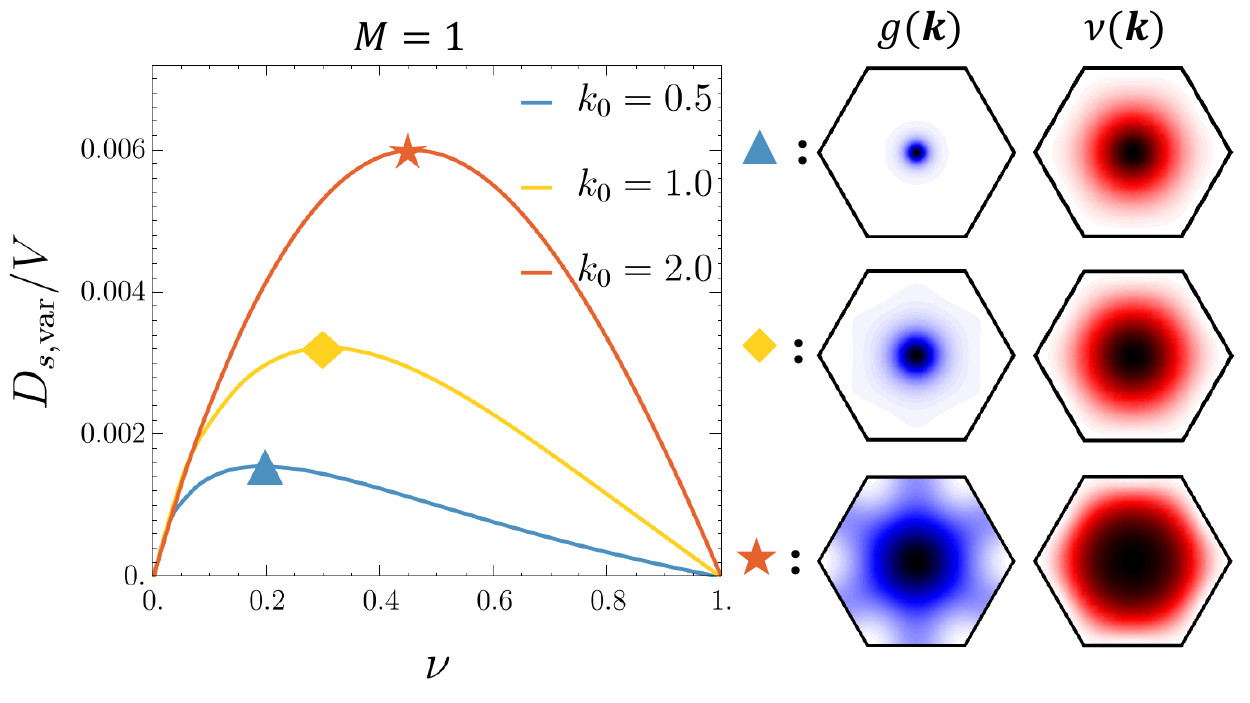}
\includegraphics[width=0.48\linewidth]{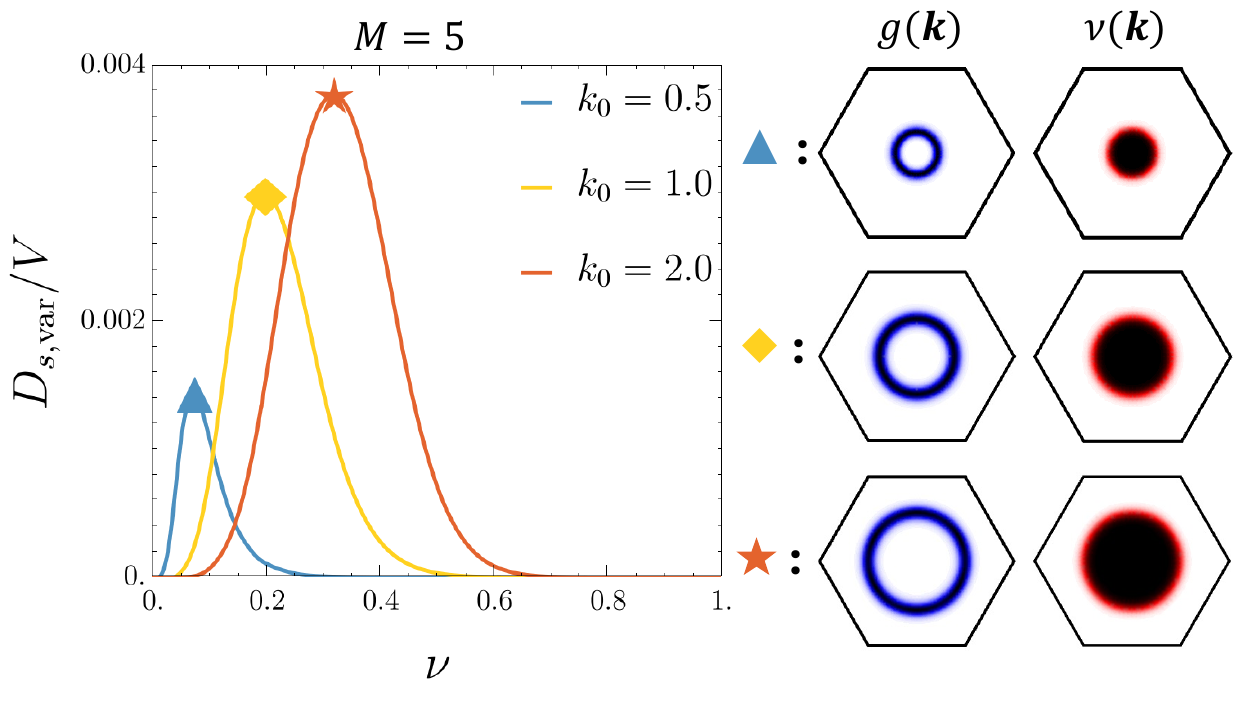}
    \caption{Variational upper bound on superfluid stiffness, $D_\text{s,var}$, of model in Eq.~\ref{eq: two band} choosing $A=1$, $B=2$ (see Eq.~\ref{eq: compactification} for its compactification on lattice) for the cases with (left) $M=1$ and (right) $M=5$. The $A=2$, $B=1$ case is obtained by particle-hole conjugation. For each $k_0$, the optimal doping is marked with solid symbols and the corresponding distributions of the Fubini-Study metric $g(\bm{k})\equiv  [\partial_i \vec{u}^\dagger \cdot (\mathbbm{1}-\vec{u}  \vec{u}^\dagger) \cdot \partial_i \vec{u} ](\bm{k})$ and mode specific filling fraction $\nu(\bm{k})$ are plotted within the first BZ.}
    \label{fig: upper bounds}
\end{figure*}

{\bf Analytical upper bounds on excitation gap and superfluid stiffness. } With the exact GSs one can infer several important properties. First, we use the single mode approximation to upper bound and estimate the excitation spectrum in the charge-$1$ sector. The strategy is to construct simple trial state $\hat{\gamma}^\dagger_{\bm{k}}|z\rangle$ (or equivalently $\hat{\gamma}_{-\bm{k}}|z\rangle$) and evaluate its energy (SM Sec.~V):
\begin{align}
    \epsilon_{1,\text{var}}(\bm{k};\Re z) =&V \langle |u_B|^2\rangle_z \left[e^{2\Re z}|u_B(\bm{k})|^2+ |u_A(\bm{k})|^2\right]
\end{align}
We note that taking $\Re z \rightarrow \pm \infty$ produces the exact result for the spectrum in the single-paritcle/hole sector of the problem: $\epsilon_{1\text{p}}(\bm{k}) = V \rho^\text{FB}_A |u_B(\bm{k})|^2$ and $\epsilon_{1\text{h}}(\bm{k}) =V \rho^\text{FB}_B |u_A(\bm{k})|^2$. This bound can serve as an estimate of the pairing gap. It can be systematically improved with more complicated trial states (e.g. trions) or through numerical calculations. It should be noted that even if $\epsilon_{1,\text{var}}$ has a gapped spectrum, this does not necessarily mean the real charge-$1$ spectrum is gapped; however, there can {\it at most} be coincidental gap closing points, since the BdG wavefunction $\vec{w}(\bm{k})$ is patch-wise continuously well-defined so that no nodes necessarily arise. 

Next we turn to upper bounding the superfluid stiffness. The approach is to consider a flat gauge connection $\bm{A}$ and then optimize the energy response over all possible BCS states. Surprisingly, we prove in SM Sec.~IV that the GS state at $\bm{A}=0$ is already the lowest-energy state within this ansatz space for an infinitesimal $\bm{A}$ due to the inversion symmetry, yielding a simple, gauge-invariant diamagnetic bound~\cite{PhysRevB.47.7995,doi:10.1073/pnas.2217816120} on the stiffness (in the sense that $D_\text{s,var} -D_\text{s} \succeq 0$ is PSD):
\begin{align}
    D^{ij}_\text{s,var}(\Re z)=& \frac{V}{2\Omega_\text{UC}}\Re\Bigg{[}\langle \partial_{i} (u^*_A u_B)\rangle^*_z  \langle \partial_{j} (u^*_A u_B)\rangle_z  \nonumber\\
    &\ \ \ + 
    e^{2\Re z} \left\langle\frac{\partial_{i} (u^*_A u_B)^*  \partial_{j} (u^*_A u_B)}{|u_A|^2}\right\rangle_z\langle|u_B|^2\rangle_z
\Bigg{]}.
\end{align}
where $\Omega_\text{UC}$ is the unit cell volume. This remarkable tightness of the diamagnetic bound within this broad variational space should be contrasted with the case of QGN models where the form factor must be varied with $\bm{A}$ in order to produce the tightest bound (see SM Sec.~IV~F and Refs.~\cite{gao2025bootstrappingSC, PhysRevB.106.014518}). This expression contains a filling-weighted average of the orbital-selective quantum metric~\cite{2024NatPh..20.1262Y} and can be further bounded by conventional geometric quantities (App.~\ref{app: Ds bound}). Intuitively, it measures on average how much mismatch between the wavefunctions at opposite occupied momenta is introduced by $\bm{A}$. We note that in the $\Re z \rightarrow \pm \infty$ limit this result relates to the bound-state mass in the two-particle/hole sector, $m_{2\text{p}/2\text{h}}$, e.g. $D^{ij}_\text{s,var}(\Re z\rightarrow+\infty) = \frac{\nu \rho^\text{FB}_A}{2\Omega_\text{UC}}[m_{2\text{p}}^{-1}]_{ij}$. See App.~\ref{app: bound state} for discussions on the exact spectrum of the bound state.

{\bf Examples and numerical investigations. } Here we demonstrate our construction with simple FBs in two dimensions. We start from an inversion symmetric continuum two-band model with Hamiltonian
\begin{align}\label{eq: two band}
    t(\bm{k})  &=  \Delta \begin{bmatrix}
        |\bm{k}|^{2M} & \bar{k}^M \\
       k^M & \delta
    \end{bmatrix} 
\end{align}
where $M$ is a positive integer, and $k,\bar{k}\equiv k_x \pm \mathrm{i} k_y$. When $\delta=1$, this model has an isolated FB at zero energy separated from a higher band with gap $\Delta$, which has wavefunction $\vec{u}^T(\bm{k}) = (1,- k^M)/\sqrt{1+|\bm{k}|^{2M}}$ and $C^\text{FB}=-M$. When $M$ is odd, we can apply the general construction to construct exact models for chiral SC that contain simply the local attraction between the two orbitals. Taking $A=1,B=2$, we find that the exact pairing order parameter is $\hat{\eta}^\dagger =  \int_{\bm{k} }\frac{1}{k^M} \hat{\gamma}_{\bm{k}}^\dagger \hat{\gamma}_{-\bm{k}}^\dagger$ (which gives rise to the Moore-Read $p-\mathrm{i} p$ `Pfaffian' state  when $M=1$~\cite{PhysRevB.96.115110,PhysRevB.79.180501,MOORE1991362}); whereas $A=2,B=1$ amounts to the particle-hole dual $\hat{\eta}^\dagger =  \int_{\bm{k} }\bar{k}^M \hat{\gamma}_{\bm{k}}^\dagger \hat{\gamma}_{-\bm{k}}^\dagger$. In both cases $C^\text{maj} = C^\text{FB}= -M$ is odd and thus the SC is topological and supports majorana edge mode. 

The above model is defined in open momentum space with an arbitrary length unit. To regularize its QG, we compactify the model onto a hexagonal BZ by simply replacing~\cite{PhysRevX.15.021087}
\begin{align}\label{eq: compactification}
    k \rightarrow [k_0 \hat{\zeta}(\bm{k}) ]^{-1} 
\end{align}
in Eq.~\ref{eq: two band} and the wavefunction $\vec{u}(\bm{k})$, where $\hat{\zeta}(\bm{k})  \equiv \zeta(k) -\pi \bar{k} /\Omega_\text{BZ} $ is a non-holomorphic, periodic modification of the Weierstrass $\zeta$ function~\cite{10.1063/1.5042618}, whose period is fixed to $2\pi(1\pm \mathrm{i}/\sqrt{3})$ in the numerical study of this work (i.e. the real-space primitive vectors are taken to be $\bm{a}_{1,2} =\frac{1}{2} (1, \pm \sqrt{3})$). $k_0$ is a tunable momentum scale within which the wavefunction is approximately holomorphic. The QG of the FB is mainly concentrated at $|\bm{k}|\lesssim k_0$, whose total strength positively depends on $k_0$. The corresponding hopping amplitudes exponentially decays with a rate $\sim k_0$ in real space (we note that a strictly local lattice realization of Chern band is impossible~\cite{chen2014impossibility}).

The Hamiltonian $t(\bm{k})$ is a flattened version of the celebrated BHZ model~\cite{doi:10.1126/science.1133734} that is broadly relevant to topological phenomena in two dimensions (see e.g.~\cite{PhysRevLett.133.206601,soejima2025jellium}). One perspective is that this wavefunction $\vec{u}(\bm{k})$ can be regarded as that of a topological band built from  $s$ and $p$ orbitals in the $M=1$ case with concentrated charge and Berry curvature distribution~\cite{PhysRevX.15.021087,PhysRevLett.129.047601,herzog2024topological}, which describes the QG of twisted bilayer graphene. On the other hand, the $M=5$ model is related to a widely used effective two-band model of the flat band bottom of R$5$G~\cite{PhysRevB.109.205122,PhysRevB.110.205130,PhysRevX.14.041040,patri2025family,PhysRevB.110.205124,2025arXiv250309692B}, capturing the essential feature of its QG, such as concentrated Berry curvature in a ring-shaped region. Interpreting the two orbitals as located on the bottom and top layers, their charge transfer gap can be directly tuned by a perpendicular displacement field. We mention that there exist a simple $L$-orbital generalization of the two-orbital $M=1$ model in Eq.~\ref{eq: two band}, which may more faithfully describe the physics of R$L$G for arbitrary $L$ (App.~\ref{app: L band}). In this model, choosing $A,B=1,2$ or $2,1$ will yield different $C^\text{maj} = 1$ or $2L-3$.

\begin{figure}
    \centering
    \includegraphics[width=0.95\linewidth]{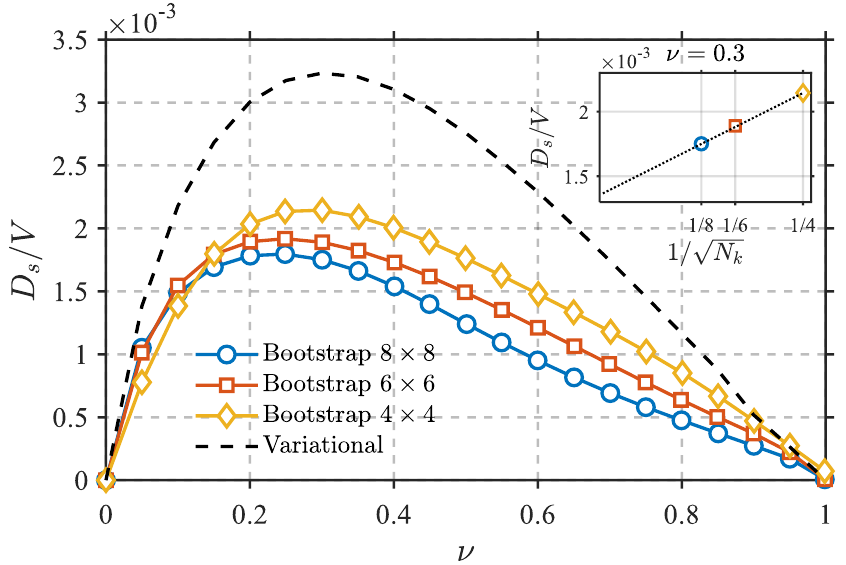}
    \caption{The lower bound on superfluid stiffness obtained by quantum many-body bootstrap, for the example model with defined in Eq.~\ref{eq: two band} choosing $A=1$, $B=2$ (see Eq.~\ref{eq: compactification} for its compactification on lattice), for the case with $M=1$ and $k_0=1$. The inset shows the finite-size scaling of the lower bound at $\nu=0.3$ with the number of unit cell number $N_k$. We note that our reduced-density-matrix bootstrap method works with fixed number sectors, but here we manually superposed the results in different charge sectors in order to make a fair comparison with the BCS state.}
    \label{fig:upper_and_lower_bounds}
\end{figure}

Within the well-defined lattice model, we plot $D_\text{s,var}$ for various choices of $k_0$ for the cases of $M=1$ and $M=5$ in Fig.~\ref{fig: upper bounds}. We find that, interestingly, the maximum stiffness is approximately achieved when the hot spot of QG is marginally filled. This reveals a more interesting relation between QG and the stiffness than the cases of QGN models. We note that for all cases studied, the single-particle spectrum $\epsilon_{1,\text{var}}$ always has a much larger gap than the superfluid stiffness, suggesting they are strongly coupled SC. 

Although the variational upper bound demonstrates a connection to QG akin to the case of $s$-wave FB superconductivity~\cite{Peotta_2015,PhysRevB.106.014518,herzog2022many,PhysRevB.109.024507,hazra2018upper}, it cannot prove the finiteness of the stiffness. We thus performed quantum many-body bootstrap~\cite{PhysRevLett.108.200404,2012NJPh...14b3027B,2019arXiv191000560L,PhysRevA.102.052819,2020arXiv200106510H,2020arXiv200106510H,2019arXiv191000560L,PhysRevLett.130.153001,PhysRevX.14.031006,gao2025bootstrappingQH,scheer2024hamiltonian,khoo2024scalable} calculations to rigorously lower bound this quantity with results shown in Fig.~\ref{fig:upper_and_lower_bounds} using the method introduced in Ref.~\cite{gao2025bootstrappingSC}. We find that the lower bound extrapolates to a finite value in the thermodynamic limit and exhibits qualitatively similar behavior as the upper bounds.

{\bf Discussions. } Our work establishes that a single polarized FB with attractive short-range interactions can host SC order with properties crucially determined by its QG. In particular, the resulting SC can be topological and its stiffness is set by the interaction scale, leading to a parametrically large transition temperature. We stress that only an effective inversion symmetry structure in the FB QG, i.e. the property Eq.~\ref{eq: inversion}, is needed in our construction. This enables the usage of our approach in systems with emergent or approximate inversion symmetry, in e.g. charge concentrated bands, tMoTe$_2$, or the $\bm{k}\cdot \bm{p}$ low-energy model of R$L$G which has an approximate $SO(2)$ rotation symmetry (the symmetry breaking terms in the realistic Hamiltonian primarily affects the dispersion~\cite{PhysRevB.109.205122}, which is small in the FB region, but not the QG). While in this work we have mainly focused on presenting the general approach, we find it quite remarkable that the general property and the simple two-band model we study already mimic certain observations in R$L$G~\cite{han2025signatures,nguyen2025hierarchy}, such as the existence of two different SC domes tunable by displacement field (charge transfer gap) and electron density (Fig.~\ref{fig: duality}). We will defer a more realistic modeling of the R$L$G systems using the solvable limit revealed in this work to future works.


{\bf Acknowledgement. } We thank Patrick Ledwith, Ashvin Vishwanath, Qingchen Li, Tixuan Tan, Tonghang Han, Maine Christos, Zhihuan Dong and Steven A. Kivelson for the helpful discussions. We thank Felix Desrochers for pointing out a typo in the first version. Z.~H. is supported by a Simons Investigator award, the
Simons Collaboration on Ultra-Quantum Matter, which is a grant from the Simons Foundation (Ashvin Vishwanath, 651440). J.H.-A. is supported by a Hertz Fellowship. E.~K. is supported by NSF MRSEC DMR-2308817 through the Center for Dynamics and Control of Materials. The authors thank the Harvard FAS Reaserch Computing (FASRC) for computational support.

\bibliographystyle{apsrev4-1} 
\bibliography{ref}
\onecolumngrid 
\newpage
\appendix
\vspace{4ex} 
{\centering\bfseries\large End Matter\par}
\vspace{4ex} 
\twocolumngrid 
\setcounter{secnumdepth}{1}

\section{A bound on $D_{\text{s,var}}$}

\label{app: Ds bound}

We can obtain an elementary upper bound for the $D_{\text{s,var}}$ as follows. For convenience we define the FB projection matrix $P_{\alpha\beta}(\bm{k}) \equiv u_\alpha u^*_\beta (\bm{k}) $ and $x\equiv \mathrm{e}^{-2\Re z}$. We also define the BZ average $\langle O\rangle = \int \frac{\mathrm{d}\bm{k}}{\Omega_\text{BZ}}  O(\bm{k})$.

First we use Cauchy-Schwartz inequality to write
\bea
&\langle \partial_{i} (u^*_A u_B)\rangle^*_z  \langle \partial_{i} (u^*_A u_B)\rangle_z \\
= & |\braket{\frac{x P_{AA}}{P_{BB}+x P_{AA}} \del_i P_{AB}
}|^2 \\
\leq &\braket{\frac{x^2 P_{AA}^2}{P_{BB}+x P_{AA}} }\braket{\frac{|\del_i P_{AB}|^2}{P_{BB}+x P_{AA}}
} \\
\eea
and thus
\bea
D^{ii}_\text{s,var} &\leq  \braket{\frac{x^2 P_{AA}^2}{P_{BB}+x P_{AA}} }\braket{\frac{|\del_i P_{AB}|^2}{P_{BB}+x P_{AA}}
} \\
&+ \frac{1}{x}\braket{\frac{x P_{AA}}{P_{BB}+x P_{AA}} \frac{|\del_i P_{AB}|^2}{P_{AA}}} \braket{\frac{x P_{AA}}{P_{BB}+x P_{AA}} P_{BB}} \\
&= \braket{\frac{x |\del_i P_{AB}|^2}{P_{BB}+x P_{AA}}
}\braket{ P_{AA}} \ .
\eea
Note that $|\del_i P_{AB}|^2 \leq g_{ii} = \frac{1}{2}\Tr[\partial_i P \partial_i P] $ leads to a further bound that relates to the Fubini-Study metric. For a two-orbital model where $P_{AA} + P_{BB}=1$, $x P_{AA} + P_{BB} \geq \min(1,x)$, which leads to a simple bound $D^{ii}_{\text{s,var}} \leq \langle g_{ii}\rangle \langle P_{AA}\rangle \max(1,x)$. It would also be interesting to obtain topological lower bounds for $D_\text{s,var}$, as have been obtain for the integrated Fubini-Study metric \cite{PhysRevB.90.165139,PhysRevLett.124.167002,PhysRevLett.128.087002,2025arXiv250100100Y,PhysRevB.109.L161111,2024arXiv240615343K,2025arXiv250116428J,2025arXiv250716909P,2025arXiv250618081Z}.

\section{The bound-state spectrum in the two-particle/hole sector}

\label{app: bound state}
The bound-state energy in the two-paritcle, momentum-$\bm{p}$ sector of $\hat{P}\hat{H}_{AB}\hat{P}$, which we call $\epsilon_{2\text{p}}(\bm{p})$, can be exactly solved by the implicit equation:
\begin{widetext}
\begin{align}\label{eq: implicit}
 \int \frac{\mathrm{d}\bm{k}}{\Omega_\text{BZ}}\frac{\left|u^*_Au_B(\bm{k}) -u^*_Au_B(\bm{k}+\bm{p}) \right|^2 - \left[|u_A(\bm{k})|^2+|u_A(\bm{k}+\bm{p})|^2\right]\epsilon_{2\text{p}}(\bm{p})/V}{|u_B(\bm{k})|^2+ |u_{B}(\bm{k}+\bm{p})|^2 - \epsilon_{2\text{p}}(\bm{p})/V } = 0  
\end{align}
\end{widetext}
which is manifestly gauge invariant. A small-$\bm{p}$ expansion of this equation yields the exact result for the two-particle mass (dependence on $\bm{k}$ omitted in the integrand):
\begin{align}\label{eq: two particle mass}
    [m_{2\text{p}}^{-1}]_{ij}  \equiv & \left. \partial_{\bm{p}_i} \partial_{\bm{p}_j} \epsilon_{2\text{p}}\right|_{\bm{p}=\bm{0}} \\
    =& V \Re \int \frac{\mathrm{d}\bm{k}}{\Omega_\text{BZ}} \frac{ \partial_{i} \left[  u_A u_B^*\right] \partial_{j} \left[  u_B u_A^* \right]  }{|u_B|^2}  / \int \frac{\mathrm{d}\bm{k}}{\Omega_\text{BZ}}  \frac{|u_A|^2}{|u_B|^2}
\end{align}

Applying the particle-hole duality, we find that the two-hole sector result can be obtained by simply substituting $z\rightarrow -z$, $A\leftrightarrow B$ and $\nu\leftrightarrow 1-\nu $ in all the expressions in this section.

\section{A $L$-orbital FB model for R$L$G. }\label{app: L band}
 Here we expand the discussion on a better characterization of the QG of R$L$G than the two-orbital one adopted in Eq.~\ref{eq: two band}. This model can be viewed as a $L$-band generalization of Eq.~\ref{eq: two band} with $M=1$. 

We start with a $\bm{k}\cdot \bm{p}$ minimal model of the band structure of R$L$G around the $K$ point. It is given by a $2L \times 2L$ hopping matrix:
\begin{align}
    t(\bm{k}) =\begin{bmatrix}
        u_d & v_F \bar{k} \\
        v_F k & u_d  & t_1 \\
         & t_1 & 2 u_d & v_F \bar{k} \\
         & & v_F k & 2 u_d  & t_1 \\
        & &  & \ddots & \ddots & \ddots  \\
         & & & & t_1 & L u_d & v_F \bar{k} \\
        & & & & & v_F k & L u_d 
    \end{bmatrix}
\end{align}
where the basis is chosen such that
\begin{align}
    \psi \equiv & (\psi_{\mathcal{A}1}, \psi_{\mathcal{B}1},\psi_{\mathcal{A}2}, \psi_{\mathcal{B}2},\dots ,\psi_{\mathcal{A}L}, \psi_{\mathcal{B}L})^T.
\end{align}
where $\mathcal{A},\mathcal{B}$ label the graphene sub-lattice index and the number represent the layer. Note that by writing this model we have only kept the nearest-neighbor hopping elements of R$L$G, which gives rise to an emergent $SO(2)$ symmetry at low-energy that contains the inversion symmetry ($\pi$ rotation). The relative parities can then be determined: $\mathcal{A}1,\mathcal{B}2,\mathcal{A}3,\mathcal{B}4,\dots$ have opposite parity to $\mathcal{B}1,\mathcal{A}2,\mathcal{B}3,\mathcal{A}4,\dots$.

We are mainly concerned with the parameter regime $ u_d \lesssim 30 \text{meV} \ll t_1 \approx 355 \text{meV}$ which is relevant to the experiments. In this limit, the dispersion of the two lowest-energy bands are remarkably flat inside the regime $|\bm{k}| \lesssim k_0 \equiv t_1/v_F$, forming a well defined flat band-bottom energetically separated from the dispersive regime in the BZ. On the other hand, we note the non-trivial QG of the band is also concentrated within the scale $|\bm{k}| \lesssim k_0$. Based on these observations, it is reasonable to assume that the physics of the flat band-bottom can be effectively modeled by an isolated FB whose QG is similarly confined within  $|\bm{k}| \lesssim k_0$. Guided by this thought, we note a simpler effective $L$-orbital model can be motivated as follows.

For concreteness we will herein after consider the lowest-energy hole band. 
We note that this band has concentrated wavefunction weight on only the $\mathcal{A}$ sublattice. Specifically, when $|\bm{k}|\ll k_0$ the wavefunction is given by
\begin{align}
    \psi_{\mathcal{A}\alpha} \approx & (-k/k_0)^{\alpha-1} \\
    \psi_{\mathcal{B}\alpha} \approx & -l \frac{u_d}{t_1}(-k/k_0)^{\alpha}.
\end{align}
This motivates us to neglect the $\mathcal{B}$ sublattice and consider a simpler $L$-band model for only the $\mathcal{A}$ orbitals. Remarkably, such a model can be obtained in a self-consistent way: to `integrate out' the $\mathcal{B}$ sublattice, we can perform the Schur complement of the $\mathcal{B}$-sublattice hopping matrix in the full hopping matrix:
\begin{align}
    t_\text{Schur}(\bm{k}) \equiv& t_{\mathcal{A}\mathcal{A}} - t_{\mathcal{A}\mathcal{B}} t_{\mathcal{B}\mathcal{B}}^{-1} t_{\mathcal{B}\mathcal{A}}
\end{align}
to yield a model purely for the $\mathcal{A}$ sublattice. This is typically a uncontrolled approximation, but here it is justifiable since the flat band bottoms of this band has almost zero-energy $\epsilon\sim  u_d$ and the Schur complement would yield the exact result if the energy were $0$. This generates a complicated $L$-orbital model, but it is remarkably close to an exactly solvable one (set $k_0=1$ as the momentum unit):
\begin{align}\label{eq: multi band}
   \tilde{t}(\bm{k}) & \equiv - \frac{t^2_1}{u_d}  \begin{bmatrix}
        \frac{\bm{k}^2}{1} &  \bar{k} \\
        k & \frac{\bm{k}^2}{2}  + 1 & \frac{\bar{k}}{2} \\
         & \frac{k}{2} & \frac{\bm{k}^2}{3}  + \frac{1}{2}  & \frac{\bar{k}}{3} \\
        & & \ddots & \ddots & \ddots  \\
         &  & & \frac{k}{L-2} & \frac{\bm{k}^2}{L-1}  + \frac{1}{L-2}  & \frac{\bar{k} }{L-1}\\
        & & & & \frac{k}{L-1} &  \frac{1}{L-1}  
    \end{bmatrix}\nonumber\\
    & \approx   t_\text{Schur}(\bm{k})
\end{align}
where the difference is merely
\begin{align}
    [t_\text{Schur}(\bm{k})]_{\alpha\alpha} - [\tilde{t}(\bm{k})]_{\alpha\alpha} = \begin{cases}
        \alpha u_d  & \alpha \neq L \\
        L u_d - \frac{t^2_1}{u_d} \frac{(\bm{k}/k_0)^2}{L} & \alpha =L
    \end{cases}
\end{align}
which is negligible in the limit we considered, and is especially so when $L$ is large. Physically, this amounts to muting the displacement field on $\mathcal{B}$ sublattice and turning off the hopping between $\mathcal{A}L$ and $\mathcal{B}L$. One can easily verify that $\tilde{t}(\bm{k})$ features
\begin{align}
    u_{\alpha=1,\dots,L} \propto  (-k/k_0)^{\alpha-1}
\end{align}
as an exact zero-energy state, and when $L=2$ it reduces to the two-band model Eq.~\ref{eq: two band} with $M=1$. This is the highest band in the spectrum with $C^\text{FB}= -L+1$. One can also verify that the QG of this band is also concentrated within $|\bm{k}|\lesssim k_0$, and gives a more accurate characterization of the Fubini-Study metric distribution of R$L$G than the two-orbital model Eq.~\ref{eq: two band}.

Now we apply our systematic construction for chiral SC to this effective model. Since in the small $\bm{k}$ regime most of the wavefunction weight is located on the first two layers, we may assume the interaction between the first two layers is the most important. Then, applying the general construction approach by choosing $A = 1$, $B=2$, we find the suggested pairing operator is $\hat{\eta}^{\dagger} =\int_{\bm{k}} \frac{1}{\bar{k}} \hat{\gamma}^\dagger_{1,\bm{k}} \hat{\gamma}_{1,-\bm{k}}$ which has $C^\text{BdG}=-L+2$ and thus the total Chern number of the superconductor $C^\text{maj} = 1$. Instead, if we choose choose $A = 2$, $B=1$, we find the suggested pairing operator is $\hat{\eta}^{\dagger} = \int_{\bm{k}} k \hat{\gamma}^\dagger_{1,\bm{k}} \hat{\gamma}_{1,-\bm{k}}$ which has $C^\text{BdG}=L-2$ and thus the total Chern number of the superconductor $C^\text{maj} = 2L-3$, which is different from the previous one! We thus have a crude but concrete prediction for the Chern numbers of the two superconducting domes observed in R$L$G with $L=4,5,6$. We note that the form factors are certainly not to be trusted for $|\bm{k}|>k_0$, and an ultra-violet regularization will be needed for the more quantitative properties by e.g. a lattice compactification similar to Eq.~\ref{eq: compactification}.

\end{document}


\title{Exact models of chiral flat-band superconductors: Supplemental Materials}

\author{Zhaoyu~Han}
\email{zhan@fas.harvard.edu}
\affiliation{Department of Physics, Harvard University, Cambridge, Massachusetts 02138, USA}

\author{Jonah~Herzog-Arbeitman}
\affiliation{Department of Physics, Princeton University, Princeton, NJ 08544, USA}

\author{Qiang~Gao}
\affiliation{Department of Physics, Harvard University, Cambridge, Massachusetts 02138, USA}

\author{Eslam~Khalaf}
\email{eslam_khalaf@fas.harvard.edu}
\affiliation{Department of Physics, Harvard University, Cambridge, Massachusetts 02138, USA}

\maketitle

\onecolumngrid

\tableofcontents

\section{Preliminaries}

We consider the electronic structure encoded in a tight-binding model with translation symmetry:
\begin{align}
\hat{H}_0 = \sum_{\bm{R}\bm{R}', \alpha\beta} t_{\alpha\beta, \bm{R}-\bm{R}'} \hat{c}^\dagger_{\bm{R},\alpha} \hat{c}_{\bm{R}',\beta}
\end{align}
where $\bm{R}$ indicate the unit cell center position, Greek letters $\alpha,\beta,\dots$ denote the orbital index in each unit cell. For simplicity, we combine the orbital and possible flavor indices unless otherwise specified. We note that by taking appropriate large orbital number limit, one can always reduce the discrete space within each unit cell to a continuum. In momentum space, we define basis $\hat{c}_{\bm{k},\alpha}\equiv\sum_{\bm{R}}\mathrm{e}^{\mathrm{i} \bm{k}\cdot (\bm{R}+\bm{r}_\alpha)} \hat{c}_{\bm{R},\alpha}/\sqrt{\mathsf{V}}$ and Fourier transform $t_{\alpha\beta} (\bm{k}) \equiv\sum_{\bm{R}}\mathrm{e}^{\mathrm{i} \bm{k}\cdot (\bm{R}+\bm{r}_\alpha-\bm{r}_\beta)} t_{\alpha\beta,\bm{R}}$ ($\mathsf{V}$ is the number of unit cells in the system, $\bm{r}_\alpha$ is the intra-unit-cell position of orbital $\alpha$) to rewrite
\begin{align}
\hat{H}_0 = \sum_{\bm{k}, \alpha\beta} t_{\alpha\beta}(\bm{k}) \hat{c}^\dagger_{\bm{k},\alpha} \hat{c}_{\bm{k},\beta}
\end{align}
Then we can diagonalize the Hamiltonian into the form
\begin{align}
\hat{H}_0 &= \sum_{\bm{k}, n} \epsilon_n(\bm{k}) \hat{\gamma}^\dagger_{\bm{k},n} \hat{\gamma}_{\bm{k},n}, \\
\hat{\gamma}_{\bm{k},n} &= U_{n\alpha}^\dagger(\bm{k}) \hat{c}_{\bm{k},\alpha},
\end{align}
where we use Latin letters ($n,m,\dots$) to label the bands. We assume there are $\mathsf{N}$ orbitals, and thus $\mathsf{N}$ bands in total.

We consider an ideal setup where there are $\mathsf{N}_\text{flat}$ nearly degenerate and flat bands isolated from all the other bands by a large gap $\Delta$, and the relevant interaction strength $V$ is small compared to $\Delta$ but large compared to the flat bands' bare bandwidth $W$:
\begin{align}
    \Delta \gg V \gg W
\end{align}
Without loss of generality, we can relabel all the bands and assume the $n=1\dots \mathsf{N}_\text{flat}$ bands are the degenerate flat bands, and $n=\mathsf{N}_\text{flat}+1\dots \mathsf{N}$ bands are the others. 

For the single-particle basis, we can thus define a projector matrix:
\begin{align}\label{eq: projector}
    P_{\alpha \beta} (\bm{k})\equiv  \sum_{n\le \mathsf{N}_\text{flat}} U_{\alpha n }(\bm{k}) U_{n\beta}^\dagger(\bm{k}),
\end{align}
Similarly, we also define a projector matrix onto the remaining bands, $Q_{\alpha \beta} (\bm{k})\equiv \delta_{\alpha\beta} - P_{\alpha \beta} (\bm{k})$.

For the ideal scenario we are assuming, we define the ``flat band subspace'', $\mathcal{H}_\text{flat}$, as the Fock space of the electron modes $\gamma_{\bm{k},n=1\dots \mathsf{N}_\text{flat}}$ tensoring with the ``vacuum'' state of the other modes $\gamma_{\bm{k},n=\mathsf{N}_\text{flat}+1\dots \mathsf{N}}$ (i.e. all the modes with energy higher/lower than the flat bands are empty/occupied). Especially we will define $|\text{vac}\rangle$ of $\mathcal{H}_\text{flat}$ as the ``vacuum'' state for the remote bands tensoring the empty state for the flat bands.

We formally define the projector onto $\mathcal{H}_\text{flat}$ as $\hat{P}$ and the projector onto the complement $\hat{Q}\equiv 1-\hat{P}$. To the leading order of a perturbation series treating both $V/\Delta$ and $W/\Delta$ as small parameters, we can set $\epsilon_{n=1\dots \mathsf{N}_\text{flat}}(\bm{k}) =0$ and project the interacting Hamiltonian onto $\mathcal{H}_\text{flat}$. The solvable model we construct should be viewed as exact up to $\mathcal{O}[(V/\Delta)^2,(W/V)]$ errors.

For later convenience, we define a $(\mathsf{N}-\mathsf{N}_\text{flat})\times (\mathsf{N}-\mathsf{N}_\text{flat}) $ diagonal matrix $\Lambda$ denoting the occupation of the remote bands ($\bm{k}$ arbitrary in the below definition):
\begin{align}\label{eq: Lambda}
    \Lambda = \text{diag}\left(\langle \gamma^\dagger_{\bm{k},\mathsf{N}_\text{flat}+1} \gamma_{\bm{k},\mathsf{N}_\text{flat}+1}\rangle, \langle \gamma^\dagger_{\bm{k},\mathsf{N}_\text{flat}+2} \gamma_{\bm{k},\mathsf{N}_\text{flat}+2}\rangle, \dots\right)
\end{align}

We emphasize that most results of discussed below do not rely on symmetries other than translation symmetry and charge conservation, and they apply to flat-band systems in arbitrary dimensions.

\section{Solvable models with superconducting ground states: a general scheme}
\label{sec: interactions}

\subsection{Overview on the construction strategy}

The goal of this work is to systematically construct solvable models whose ground state in the charge-$2N$ sector is of the form:
\begin{align}\label{eq: gs}
    |N\rangle = (\eta^\dagger)^{N} |\text{vac}\rangle.
\end{align}
which is defined by a pairing operator
\begin{align}\label{eq: pairing operator}
    \hat{\eta}^\dagger &\equiv \frac{1}{\mathsf{V}} \sum_{\bm{k}; n,m\le \mathsf{N}_\text{flat}} F^{\bm{Q}}_{nm} (\bm{k})\hat{\gamma}^\dagger_{\bm{Q}/2+\bm{k}, n} \hat{\gamma}^\dagger_{\bm{Q}/2-\bm{k},m}   
\end{align}
where $\bm{Q}$ is an arbitrary momentum and $F^{\bm{Q}}_{nm} (\bm{k})$ is the form factor defined in the flat band subspace satisfying 
\begin{align}\label{eq: F antisymmetric}
 F^{\bm{Q}} (\bm{k}) = - F^{\bm{Q},T} (-\bm{k})   
\end{align}

For each possible pairing operator (with a definite $\bm{Q}$ and a form factor $F^{\bm{Q}}(\bm{k})$), there is an infinite set of interacting Hamiltonians of which Eq.~\ref{eq: gs} are the ground states. In general, those ``interactions'' take the form
\begin{align}\label{eq: interaction}
    \hat{H}_\text{int} \equiv  \sum_{\bm{R},\bm{R}'} V_{\bm{R}-\bm{R}'}  \hat{S}^{\dagger}_{\bm{R}}\hat{S}_{\bm{R}'}  + \hat{H}_{\text{int},2}
\end{align} 
where $\hat{S}_{\bm{R}}$ is a local, neutral, fermion bilinear operator ($ \hat{c}^\dagger\hat{c}$) centered at $\bm{R}$. For the hermicity of the Hamiltonian, we need $V_{\bm{R}-\bm{R}'} = [V_{\bm{R}'-\bm{R}}]^*$ and thus $V(\bm{q})$ is real on all $\bm{q}$; besides this, we further require $V(\bm{q})$ to be {\it non-negative} on all $\bm{q}$. $ \hat{H}_{\text{int},2}$, which will be explicitly constructed below, contain some suitably designed fermion bilinear terms such that the projected Hamiltonian takes the form
\begin{align}\label{eq: projected H}
    \hat{P}\hat{H} \hat{P} & =   \sum_{\bm{R},\bm{R}'} V_{\bm{R}-\bm{R}'}  \hat{P}\hat{S}^{\dagger}_{\bm{R}}\hat{P}\hat{S}_{\bm{R}'}\hat{P} 
\end{align}

The crucial condition that underlies the solvability of the model is that $\hat{S}_{\bm{R}}$ needs to commute with the pair creation operator within the $\mathcal{H}_\text{flat}$:
\begin{align}\label{eq: commute}
    [\hat{\eta}^\dagger, \hat{P}\hat{S}_{\bm{R}}\hat{P}] = 0 
\end{align}
It should be noted that these operators are defined in the entire Hilbert space, whereas the above commutation relation only needs to hold after projection.  Different $\hat{S}_{\bm{R}}$ at different $\bm{R}$ are not necessarily mutually commuting. We will show that there are infinitely many such constructions for each pair operator, so that any combination of the corresponding models is also a valid solvable model. 

It is important to note that, in this work we do not restrict $\hat{S}_\text{R}$ to be hermitian. Therefore, $\hat{P}\hat{S}^\dagger_{\bm{R}}\hat{P}$ and thus the Hamiltonian does not commute with $\hat{\eta}^\dagger$ within $\mathcal{H}_\text{flat}$. In this regard, these models encompass more possibilities than the previously proposed `quantum geometric nesting' models~\cite{PhysRevX.14.041004} for flat band systems.

To see why this commutation property Eq.~\ref{eq: commute} implies solvability, we recall that the terms in the projected Hamiltonian Eq.~\ref{eq: projected H} always ends with $\hat{P}\hat{S}_{\bm{R}}\hat{P}$. So when the Hamiltonian acts on the desired groundstate, the operator combination $\hat{P}\hat{S}_{\bm{R}}\hat{P}$ can always be moved to the right and directly annihilate $|\text{vac}\rangle$
\begin{align}
 & \hat{P} \hat{S}_{\bm{R}}\hat{P} (\hat{\eta}^\dagger)^N |\text{vac}\rangle = (\hat{\eta}^\dagger)^N \hat{P} \hat{S}_{\bm{R}}\hat{P} |\text{vac}\rangle =0
    \implies \hat{P} \hat{H}\hat{P}(\hat{\eta}^\dagger)^N|\text{vac}\rangle = 0 
\end{align}
Since we have assumed that $V(\bm{q})$ is non-negative, $\hat{H}$ is positive semi-definite, the superconducting states are thus the ground states.

Before closing this section let's explicitly construct the general expression of:
\begin{align}
    \hat{H}_{\text{int},2} \equiv -\sum_{\bm{R},\bm{R}'} V_{\bm{R}-\bm{R}'}   \hat{P}\hat{S}^{\dagger}_{\bm{R}} \left( 1-\hat{P}\right) \hat{S}_{\bm{R}'}\hat{P}  + \hat{H}_{\text{arbitrary},2} \hat{Q} + \hat{Q}\hat{H}'_{\text{arbitrary},2}
\end{align}
where $\hat{H}_{\text{arbitrary},2}$ and $\hat{H}_{\text{arbitrary},2}'$ are two arbitrary quadratic Hamiltonians. This form immediately implies the projected Hamiltonian is given by Eq.~\ref{eq: projected H}. To see why this is a fermion bilinear term, we recognize that any fermion operator $\hat{c}$ can be notationally decomposed as $\hat{c} \equiv \Bar{c} +\tilde{c}$ where $\Bar{c}$ is the part that involves $\gamma_{\bm{k} n\le \mathsf{N}_\text{flat}}$ and $\Tilde{c}$ is the remaining part. Then, all four fermion terms in $\hat{S}^\dagger\hat{S}$ admits the following decomposition:
\begin{align}\label{eq: projection details}
    \hat{P}\hat{c}_1^\dagger \hat{c}_2(1-\hat{P}) \hat{c}_3^\dagger \hat{c}_4\hat{P} =  \langle \tilde{c}_2 \tilde{c}_3^\dagger\rangle \bar{c}_1^\dagger \bar{c}_4 + \bar{c}_2 \bar{c}_3^\dagger\langle \tilde{c}_1^\dagger \tilde{c}_4\rangle 
\end{align}
and thus only contain quadratic terms in $\hat{c}$.

\subsection{The construction of the $\hat{S}$ operators}

Now, we construct the desired $\hat{S}_{\bm{R}}$ operators:
\begin{align}
    \hat{S}_{\bm{R}} = \sum_{\mu\nu,\bm{R}_1,\bm{R}_2} S_{\mu\nu}(\bm{R}_1,\bm{R}_2) \hat{c}^{\dagger}_{\bm{R}+\bm{R}_1,\mu} \hat{c}_{\bm{R}+\bm{R}_2,\nu}
\end{align}
which can be written in band basis as
\begin{align}
    \hat{S}_{\bm{R}} &= \frac{1}{\mathsf{V}}\sum_{\mu\nu,\bm{p}\bm{q}}  \mathrm{e}^{\mathrm{i}\bm{R}\cdot (\bm{p}-\bm{q})}S_{\mu\nu}(\bm{p},\bm{q}) \hat{c}^{\dagger}_{\bm{p},\mu} \hat{c}_{\bm{q},\nu} \\
    &= \frac{1}{\mathsf{V}} \sum_{nm,\bm{p}\bm{q}}  \mathrm{e}^{\mathrm{i}\bm{R}\cdot (\bm{p}-\bm{q})} S_{nm}(\bm{p},\bm{q}) \hat{\gamma}^{\dagger}_{\bm{p},n} \hat{\gamma}_{\bm{q},m} \label{eq: S eigenbasis}
\end{align}
where 
\begin{align}
    S_{\mu\nu}(\bm{p},\bm{q}) &\equiv  \sum_{\bm{R}_1,\bm{R}_2} S_{\mu\nu}(\bm{R}_1,\bm{R}_2) \mathrm{e}^{\mathrm{i}\left[(\bm{R}_1+\bm{r}_\mu) \cdot \bm{p} -(\bm{R}_2+\bm{r}_\nu) \cdot \bm{q} \right]} \\
    S_{nm}(\bm{p},\bm{q}) &\equiv U^\dagger_{ n \mu}(\bm{p}) S_{\mu\nu}(\bm{p},\bm{q})  U_{\nu m}(\bm{q}).
\end{align}

Decomposing $ S_{nm}(\bm{p},\bm{q})$ into blocks: 
\begin{align}\label{eq: blocks}
    S_{nm}(\bm{p},\bm{q}) = \left[\begin{array}{ c | c }
    B(\bm{p},\bm{q}) & W(\bm{p},\bm{q})  \\
    \hline
     \tilde{W}(\bm{p},\bm{q}) & \tilde{B}(\bm{p},\bm{q})
  \end{array}\right] 
\end{align}
we find the commutativity condition Eq.~\ref{eq: commute} can be satisfied by a large class of construction
\begin{align}\label{eq: commute construction}
    B(\bm{p},\bm{q}) =F^{\bm{Q}}(\bm{p}-\frac{\bm{Q}}{2}) A(\bm{p}-\frac{\bm{Q}}{2},\frac{\bm{Q}}{2}-\bm{q}) 
\end{align}
where $A(\bm{k},\bm{k}') = A^T(\bm{k}',\bm{k})$. To see this indeed implies Eq.~\ref{eq: commute}, we compute
\begin{align}
    &[\hat{\eta}^\dagger, \hat{P}\hat{S}_{\bm{R}}\hat{P}] \\
    =& \frac{1}{\mathsf{V}^2} \sum_{\substack{ nmkl \le \mathsf{N}_\text{flat} \\ \bm{k}\bm{p}\bm{q}}} \mathrm{e}^{\mathrm{i} \bm{R}\cdot (\bm{q}-\bm{p})} F^{\bm{Q}}_{nm}(\bm{k}) S_{kl}(\bm{p},\bm{q}) \left[\hat{\gamma}^\dagger_{\bm{Q}/2+\bm{k}, n} \hat{\gamma}^\dagger_{\bm{Q}/2-\bm{k},m}, \hat{\gamma}^{\dagger}_{\bm{p},k} \hat{\gamma}_{\bm{q},l} \right] \\
    =& \frac{1}{\mathsf{V}^2} \sum_{\substack{ nmkl \le \mathsf{N}_\text{flat} \\ \bm{k}\bm{p}\bm{q}}} \mathrm{e}^{\mathrm{i} \bm{R}\cdot (\bm{q}-\bm{p})} F^{\bm{Q}}_{nm}(\bm{k}) S_{kl}(\bm{p},\bm{q}) \hat{\gamma}^\dagger_{\bm{p},k} \left[\delta_{\bm{q},\frac{\bm{Q}}{2}-\bm{k}} \delta_{ml} \hat{\gamma}^\dagger_{\frac{\bm{Q}}{2}+\bm{k},n}  - \delta_{\bm{q},\frac{\bm{Q}}{2}+\bm{k}}  \delta_{nl} \hat{\gamma}^\dagger_{\frac{\bm{Q}}{2}-\bm{k},m}  \right] \\
    =& \frac{1}{\mathsf{V}^2} \sum_{\substack{ nmk \le \mathsf{N}_\text{flat} \\ \bm{k}\bm{q}}} \mathrm{e}^{-\mathrm{i} \bm{R}\cdot (\bm{k}+\bm{p} )} 
 \hat{\gamma}^\dagger_{\frac{\bm{Q}}{2}+\bm{p},k} \left[  S_{km}(\frac{\bm{Q}}{2}+\bm{p},\frac{\bm{Q}}{2}-\bm{k}) F^{\bm{Q}}_{nm}(\bm{k}) - S_{km}(\frac{\bm{Q}}{2}+\bm{p},\frac{\bm{Q}}{2}-\bm{k}) F^{\bm{Q}}_{mn}(-\bm{k}) \right]  \hat{\gamma}^\dagger_{\frac{\bm{Q}}{2}+\bm{k},n} \\
=& \frac{1}{\mathsf{V}^2} \sum_{\substack{ nmk \le \mathsf{N}_\text{flat} \\ \bm{k}\bm{q}}} \mathrm{e}^{-\mathrm{i} \bm{R}\cdot (\bm{k}+\bm{p} )} 
 \hat{\gamma}^\dagger_{\frac{\bm{Q}}{2}+\bm{p},k} \left[  -F^{\bm{Q}}_{km}(\bm{p}) S_{nm}(\frac{\bm{Q}}{2}+\bm{k},\frac{\bm{Q}}{2}-\bm{p})  - S_{km}(\frac{\bm{Q}}{2}+\bm{p},\frac{\bm{Q}}{2}-\bm{k}) F^{\bm{Q}}_{mn}(-\bm{k}) \right]  \hat{\gamma}^\dagger_{\frac{\bm{Q}}{2}+\bm{k},n} \\
 =& \frac{1}{\mathsf{V}^2} \sum_{\substack{nmk \le \mathsf{N}_\text{flat} \\ \bm{k}\bm{q}}} \mathrm{e}^{-\mathrm{i} \bm{R}\cdot (\bm{k}+\bm{p} )} 
 \hat{\gamma}^\dagger_{\frac{\bm{Q}}{2}+\bm{p},k} \left[  -F^{\bm{Q}}(\bm{p})A^T(\bm{k},\bm{p}) F^{\bm{Q},T}(\bm{k})  - F^{\bm{Q}}(\bm{p}) A(\bm{p},\bm{k})  F^{\bm{Q}}(-\bm{k}) \right]_{kn} \hat{\gamma}^\dagger_{\frac{\bm{Q}}{2}+\bm{k},n} \\
 =&0
\end{align}
In the third equality, we changed the dummy variable $\bm{k}\rightarrow -\bm{k}$ in the second term, and  $\bm{p}\rightarrow \frac{\bm{Q}}{2}+\bm{p}$ in all terms. In the fourth equality, we interchanged dummy variables $\bm{p}\leftrightarrow\bm{k}$, $n\leftrightarrow k$ in the first term. In the fifth equality, we substitute in the definition of $S_{nm}$. In the final equality, we used the property of $F(\bm{k})=-F^{T}(-\bm{k})$.

The essential structure for this construction, which also proves its necessity, is that for valid pair operator defined by a full-rank, antisymmetric matrix $F$ and a fermion bilinear operator operator defined by a matrix $S$, their commutator equals to (here $ijkl$ represent abstract indices)
\begin{align}
    [F_{ij} \hat{\gamma}^\dagger_i\hat{\gamma}^\dagger_j,S_{kl}\hat{\gamma}^\dagger_k\hat{\gamma}_l] = 2 (SF)_{ij} \hat{\gamma}^\dagger_i\hat{\gamma}^\dagger_j ,
\end{align}
which vanishes if and only if $S=FM$ where $M=M^T$ is a symmetric matrix.

We now compute the projected part of $\hat{H}_{\text{int} ,2}$ using Eq.~\ref{eq: projection details}:
\begin{align}
   \hat{P} \hat{H}_{\text{int} ,2} \hat{P}
& \equiv \sum_{\bm{R},\bm{R}'} V_{\bm{R}-\bm{R}'}  \hat{P}\left(\hat{S}^{\dagger}_{\bm{R}}\hat{S}_{\bm{R}'} -\hat{S}^{\dagger}_{\bm{R}}\hat{P}\hat{S}_{\bm{R}'} \right)\hat{P} \\
& = - \frac{1}{\mathsf{V}}  \sum_{\substack{\bm{p}\bm{q},\bm{p}'\bm{q}' \\ nmkl}} \delta_{\bm{p}'-\bm{q}',\bm{q}-\bm{p}} V(\bm{q}-\bm{p})  S^*_{mn}(\bm{q},\bm{p}) S_{kl}(\bm{p}',\bm{q}') \hat{P}\hat{\gamma}^{\dagger}_{\bm{p},n} \hat{\gamma}_{\bm{q},m}   (1-\hat{P})\hat{\gamma}^{\dagger}_{\bm{p}',k} \hat{\gamma}_{\bm{q}',l} \hat{P} \\
&=  - \frac{1}{\mathsf{V}}  \sum_{\substack{\bm{p}  \bm{q} \\ nl\leq \mathsf{N}_\text{flat}, m>\mathsf{N}_\text{flat}}} V(\bm{q}-\bm{p})  S^\dagger_{nm}(\bm{q},\bm{p})  (1-\Lambda_{mm}) S_{ml}(\bm{q},\bm{p}) \hat{\gamma}^{\dagger}_{\bm{p},n}  \hat{\gamma}_{\bm{p},l} \nonumber\\
& \ \ \ \ \ - \frac{1}{\mathsf{V}}  \sum_{\substack{\bm{p} \bm{q} \\  mk\leq \mathsf{N}_\text{flat},n >\mathsf{N}_\text{flat}}} V(\bm{q}-\bm{p})   S_{kn}(\bm{q},\bm{p}) \Lambda_{nn} S^\dagger_{nm}(\bm{q},\bm{p})
\hat{\gamma}_{\bm{q},m}  \hat{\gamma}^{\dagger}_{\bm{q},k}  \\
&\equiv \sum_{\bm{p} ,  nm\leq \mathsf{N}_\text{flat}} [h_{\text{int},2}(\bm{p})]_{nm} \hat{\gamma}^{\dagger}_{\bm{p},n}
\hat{\gamma}_{\bm{p},m}   + \delta E
\end{align} 
where 
\begin{align}
\label{eq: h2}
    h_{\text{int},2}(\bm{p}) & =  -\frac{1}{\mathsf{V}}\sum_{\bm{q}} V(\bm{q}-\bm{p})\tilde{W}^\dagger(\bm{q},\bm{p}) (\mathbbm{1}-\Lambda) \tilde{W}(\bm{q},\bm{p}) +  \frac{1}{\mathsf{V}}\sum_{\bm{q}} V(\bm{p}-\bm{q}) W(\bm{p},\bm{q}) \Lambda W^\dagger(\bm{p},\bm{q})  \\
    \delta E & = - \frac{1}{\mathsf{V}}\sum_{\bm{p}\bm{q}} V(\bm{p}-\bm{q})\Tr[ W(\bm{p},\bm{q}) \Lambda W^\dagger(\bm{p},\bm{q})]
\end{align}
where $\tilde{W}^\dagger(\bm{p},\bm{q}), W(\bm{p},\bm{q})$ are $\mathsf{N}_{\text{flat}} \times (\mathsf{N}- \mathsf{N}_{\text{flat}})$ matrices, and $\Lambda$ is defined in Eq.~\ref{eq: Lambda}.

\subsection{The resulting solvable Hamiltonian in the flat-band subspace}

For completeness, here we give explicit expression for the ideal solvable Hamiltonian in momentum space:
\begin{align}
    \hat{P}\hat{H}_\text{int}\hat{P} =\mathsf{V} \sum_{\bm{K}} V(\bm{K}) \hat{P} \hat{S}^{\dagger}(\bm{K}) \hat{P}\hat{S}(\bm{K})\hat{P}
\end{align}
where
\begin{align}
    V(\bm{K}) \equiv &  \sum_{\bm{R}} \mathrm{e}^{-\mathrm{i} \bm{K}\cdot \bm{R}} \ V_{\bm{R}} \\
    \hat{S}(\bm{K}) \equiv &  \frac{1}{\mathsf{V}} \sum_{\bm{R}} \mathrm{e}^{- \mathrm{i} \bm{K}\cdot \bm{R}} \hat{S}_{\bm{R}}\\
    = & \frac{1}{\mathsf{V}}  \sum_{nm,\bm{p}\bm{q}}  \delta_{\bm{K},\bm{p}-\bm{q}} S_{nm}(\bm{p},\bm{q}) \hat{\gamma}^{\dagger}_{\bm{p},n} \hat{\gamma}_{\bm{q},m}
\end{align}
Substituting in the constructed $S$ yields the following expression of the projected Hamiltonian:
\begin{align}
    \hat{H}_\text{solvable} =& \frac{1}{\mathsf{V}}  \sum_{\substack{\bm{p}\bm{q},\bm{p}'\bm{q}' \\ nmkl\leq \mathsf{N}_\text{flat}}} \delta_{\bm{p}'-\bm{q}',\bm{q}-\bm{p}} V(\bm{q}-\bm{p})  S^*_{mn}(\bm{q},\bm{p}) S_{kl}(\bm{p}',\bm{q}') \hat{\gamma}^{\dagger}_{\bm{p},n} \hat{\gamma}_{\bm{q},m}   \hat{\gamma}^{\dagger}_{\bm{p}',k} \hat{\gamma}_{\bm{q}',l} \\
    =&  \frac{1}{\mathsf{V}}  \sum_{\substack{\bm{p}\bm{q},\bm{p}'\bm{q}' \\ nmkl\leq \mathsf{N}_\text{flat} }}  V_{nmkl}(\bm{p},\bm{q},\bm{p}',\bm{q}') \hat{\gamma}^{\dagger}_{\bm{p},n}\hat{\gamma}_{\bm{q},m}   \hat{\gamma}^{\dagger}_{\bm{p}',k} \hat{\gamma}_{\bm{q}',l} \label{eq: solvable H} \\
   V_{nmkl}(\bm{p},\bm{q},\bm{p}',\bm{q}') \equiv &  \delta_{\bm{p}'-\bm{q}',\bm{q}-\bm{p}} V(\bm{q}-\bm{p})  [F^{\bm{Q}}(\bm{q}-\frac{\bm{Q}}{2}) A(\bm{q}-\frac{\bm{Q}}{2},\frac{\bm{Q}}{2}-\bm{p}) ]^\dagger_{nm}   [F^{\bm{Q}}(\bm{p}'-\frac{\bm{Q}}{2}) A(\bm{p}'-\frac{\bm{Q}}{2},\frac{\bm{Q}}{2}-\bm{q}')  ]_{kl} \label{eq: V}
\end{align}

\section{Simple constructions in inversion symmetric systems}

There is a class of simple constructions guaranteed by inversion symmetry $\mathcal{I}: \bm{r} \rightarrow -\bm{r}$, which we discuss here.

\subsection{Electronic structure of inversion symmetric systems}

In inversion symmetric systems we can always choose a set of local basis states $\{\hat{c}_{\bm{R}\alpha }\}$ such that they are eigenstates of the inversion symmetry operation (note that we have chosen a periodic embedding by defining all the orbital to be at the inversion center)
\begin{align}
    \mathcal{I} \hat{c}_{\bm{k}\alpha } \mathcal{I} = p_\alpha \hat{c}_{-\bm{k}\alpha }
\end{align}
where $p_\alpha =\pm $ is the parity of the orbital. In this basis, the hopping matrix admits a special structure. To see this, we recall:
\begin{align}
   \hat{H}_0 =  &\sum_{\bm{k}, \alpha\beta} t_{\alpha\beta}(\bm{k}) \hat{c}^\dagger_{\bm{k},\alpha} \hat{c}_{\bm{k},\beta} = 
    \mathcal{I} \hat{H}_0\mathcal{I}  = \sum_{\bm{k}, \alpha\beta} p_\alpha p_\beta t_{\alpha\beta}(\bm{k}) \hat{c}^\dagger_{-\bm{k},\alpha} \hat{c}_{-\bm{k},\beta} 
\end{align}
which implies that
\begin{align}
    t_{\alpha\beta}(\bm{k}) = p_\alpha p_\beta t_{\alpha\beta}(-\bm{k}).
\end{align}
Reordering all the orbitals such that the even-parity ones are before the odd-parity ones, we can rewrite the hopping matrix into a block-diagonal form:
\begin{align}
    t(\bm{k}) = \begin{bmatrix}
        t_+(\bm{k}) & t'(\bm{k})\\
        t'^\dagger(\bm{k}) & t_-(\bm{k})
    \end{bmatrix}
\end{align}
with $t_{\pm}(\bm{k}) = t_{\pm}(-\bm{k})$ and $t'(\bm{k}) = -t'(-\bm{k})$. The equations to solve the eigenstate $u = (u_{+}, u_{-})$ and eigenvalue $\epsilon$ reads
\begin{align}
    (t_+ -\epsilon) u_+ +t' u_- &=0 \\
    t'^\dagger u_+ + (t_- - \epsilon) u_- & = 0
\end{align}
which implies
\begin{align}
    (t_+ -\epsilon) u_+ &= t' (t_--\epsilon)^{-1}  t'^\dagger u_+ \label{eq: u+ eq} \\
    u_- &= -t'^{-1} (t_+ -\epsilon) u_+ \label{eq: u- eq}
\end{align}
The parities of $t_\pm, t'$ means that the solution for $\epsilon$ in Eq.~\ref{eq: u+ eq} must be even in $\bm{k}$
\begin{align}
    \epsilon(\bm{k})= \epsilon(-\bm{k})
\end{align}
and there always exist a gauge in which $u_+$ is also even
\begin{align}\label{eq: p}
    u_+(\bm{k}) = u_+(-\bm{k}) 
\end{align}
which further implies from Eq.~\ref{eq: u- eq} that $u_-$ must be odd in this gauge
\begin{align}\label{eq: m}
    u_-(\bm{k}) = -u_-(-\bm{k}) .
\end{align}

We now prove the existence of such a gauge. For convenience let's define $D$ to be the inversion representation on these orbitals, which satisfies $D^\dagger = D$ and $D^2 =\mathbbm{1}$ (in our basis it is nothing but the parity matrix $\text{diag}(p_1, \dots, p_{\mathsf{N}})$). It implements the symmetry in the way that $D t(\bm{k}) D = t(-\bm{k})$. This implies that for any eigenvector $u(\bm{k})$, $D u(\bm{k}) $ must be the corresponding eigenvector at $-\bm{k}$ up to a phase:
\begin{align}\label{eq: inversion}
    D u(\bm{k}) = e^{\mathrm{i}\varphi(\bm{k})} u(-\bm{k}).
\end{align}
Now using $D^2= \mathbbm{1}$, we deduce that
\begin{align} 
    u(\bm{k})=D^2 u(\bm{k}) = e^{\mathrm{i}\varphi(\bm{k})}  D u(-\bm{k}) = e^{\mathrm{i}[\varphi(\bm{k})+\varphi(-\bm{k})]} u(\bm{k})
\end{align}
implying $\varphi(\bm{k})$ is odd in $\bm{k}$ in any gauge. In particular we can perform a gauge transformation $ u(\bm{k})\rightarrow  u(\bm{k}) \mathrm{e}^{\mathrm{i} \varphi(\bm{k})/2}$ which absorb this phase and simplify the relation Eq.~\ref{eq: inversion} to 
\begin{align}
    D u(\bm{k}) =  u(-\bm{k})
\end{align}
which exactly gives rise to the above relation Eq.~\ref{eq: p}\&\ref{eq: m}.

From here on we will stick to this gauge choice unless otherwise specified. It should be noted that no conclusion below will depend on this gauge choice; it is just that many structures will become transparent in it.

\subsection{A construction for the case of a single flat band}

\label{sec: construction: inversion symmetric systems}

Here we construct a class of simple solvable models for inversion symmetric systems with a single (flavor-less) flat band. Since there is only one band involved, we define hereinafter
\begin{align}
    u_{\alpha} (\bm{k}) \equiv U_{\alpha 1}(\bm{k})
\end{align}
for simplicity.

Now let's pick two orbitals $A,B$ with opposite parities $p_Ap_B=-1$, and consider the local fermion bilinear operator:
\begin{align}\label{eq: one flatband S}
    \hat{S}_{\bm{R}} = \hat{c}^\dagger_{\bm{R}A} \hat{c}_{\bm{R}B}
\end{align}
which can be used to construct a simple  interaction
\begin{align}
    \hat{H}_\text{int} = V \sum_{\bm{R}}  \hat{S}^\dagger_{\bm{R}} \hat{S}_{\bm{R}} + \hat{H}_{\text{int},2} \sim  - V \sum_{\bm{R}} \hat{n}_{\bm{R}B}\hat{n}_{\bm{R}A}
\end{align}
which describes a local attraction. The expression of $\hat{H}_{\text{int},2}$ will be given below.

The $\hat{S}$ operator in the band basis reads:
\begin{align}
    S_{nm}(\bm{p},\bm{q}) = u_{A}^*(\bm{p}) u_{B}(\bm{q})
\end{align}
We find the commutativity condition in Eq.~\ref{eq: commute construction} discussed above is satisfied at $\bm{Q}=0$, since
\begin{align}
   B(\bm{p},\bm{q}) &=  S_{11}(\bm{p},\bm{q}) =  F(\bm{p}) A(\bm{p},-\bm{q}) \\
   F(\bm{p}) & = u_{A}^*(\bm{p})/u_{B}(-\bm{p}) = p_A/p_B u_{A}^*(-\bm{p})/u_{B}(\bm{p}) = -F(-\bm{p})\\
   A(\bm{p},-\bm{q}) & = u_{B}(-\bm{p}) u_{B}(\bm{q}) = p_A p_B u_{B}(\bm{p}) u_{B}(-\bm{q}) = A(-\bm{q},\bm{p})
\end{align}
which suggests that the system has exact SC GS with order parameter
\begin{align}
\hat{\eta}^{\dagger} = \frac{1}{\mathsf{V}}\sum_{\bm{k}} \frac{u_{A}^*(\bm{k})}{u_{B}(-\bm{k}) } \hat{\gamma}^\dagger_{1,\bm{k}} \hat{\gamma}_{1,-\bm{k}}.
\end{align}

The requisite $\hat{H}_{\text{int},2}$ has to satisfy Eq.~\ref{eq: h2}:
\begin{align}
    [h_{\text{int},2}(\bm{p})]_{11} = - \frac{1}{\mathsf{V}}\sum_{\bm{q}} \left[ |U_{B1}(\bm{p})|^2 \sum_{n=2\dots \mathsf{N} } (1-\Lambda_{nn})|U_{An} (\bm{q})|^2  - |U_{A1} (\bm{p})|^2\sum_{n=2\dots \mathsf{N} } |U_{Bn}(\bm{q})|^2  \Lambda_{nn} \right] 
\end{align}
This is always quite simple to satisfy. For example, one can take:
\begin{align}
     \hat{H}_{\text{int},2} = \frac{V}{\mathsf{V}}\sum_{\bm{q}} \left[\rho^\text{R,e}_B \hat{n}_A - \rho^\text{R,h}_{A} \hat{n}_B \right]
\end{align}
where we defined:
\begin{align}
    \rho^\text{R,e}_\alpha \equiv & \frac{1}{\mathsf{V}} \sum_{\bm{q}; n=2,\dots,\mathsf{N}}\Lambda_{nn} |U_{\alpha n} (\bm{q})|^2 \\
   \rho^\text{R,h}_\alpha \equiv & \frac{1}{\mathsf{V}}\sum_{\bm{q}; n=2,\dots,\mathsf{N}}(1-\Lambda_{nn}) |U_{\alpha n} (\bm{q})|^2 \\
    \rho^\text{FB}_\alpha \equiv & \frac{1}{\mathsf{V}}\sum_{\bm{q}} |U_{\alpha 1} (\bm{q})|^2
\end{align}
to be the average numbers of the remote-band electron/hole and the average FB wavefunction weight on the $\alpha$ orbital. Note that $\rho^\text{FB}_\alpha + \rho^\text{R,e}_\alpha+\rho^\text{R,h}_\alpha=1$. Including this term, we reach a simple construction of $\hat{H}_\text{int}$:
\begin{align}\label{eq: Hint simple}
    \hat{H}_\text{int} =   V \sum_{\bm{R}} \left[ -\hat{n}_{\bm{R}A}\hat{n}_{\bm{R}B} + \rho^\text{R,e}_B \hat{n}_{\bm{R}A}+ (1-\rho^\text{R,h}_A)\hat{n}_{\bm{R}B} \right]
\end{align}

If the flat band is the lowest, the construction is particularly simple:
\begin{align} \label{eq: one flatband Hint2 lowest}
    \hat{H}_{\text{int}} = V \sum_{\bm{R}} \left[ -\hat{n}_{\bm{R}A}\hat{n}_{\bm{R}B} + \rho^\text{FB}_A \hat{n}_{\bm{R}B} \right] 
\end{align}
Similarly, if the flat band is the highest, the construction is also particularly simple:
\begin{align} \label{eq: one flatband Hint2 highest}
    \hat{H}_{\text{int}}=  V \sum_{\bm{R}} \left[ - (1-\hat{n}_{\bm{R}A})(1-\hat{n}_{\bm{R}B}) -\rho^\text{FB}_B \hat{n}_{\bm{R}A} \right]  
\end{align}

In any case, the projected, exactly solvable Hamiltonian is simply:
\begin{align}\label{eq: inversion solvable H}
\hat{H}_\text{solvable} = \frac{V}{\mathsf{V}}\sum_{\bm{k}_1+\bm{k}_3 = \bm{k}_2+\bm{k}_4} u^*_B(\bm{k}_1) u_A(\bm{k}_2) u^*_A(\bm{k}_3) u_B(\bm{k}_4) \hat{\gamma}^\dagger_{\bm{k}_1} \hat{\gamma}_{\bm{k}_2}\hat{\gamma}^\dagger_{\bm{k}_3}\hat{\gamma}_{\bm{k}_4}
\end{align}

It is possible that this model describes a topological superconductor in two dimensions. To see this, we superpose the SC states within different number sectors to construct a BCS state (defined by a complex number $z$):
\begin{align}
    |z\rangle \equiv & \sum_{N} \mathrm{e}^{-z N} (\hat{\eta}^\dagger/2)^N|\text{vac}\rangle  \\
    =& \prod'_{\bm{k}; i = 1 \dots  \mathsf{N}_\text{flat}} \left[1 +\mathrm{e}^{-z } f(\bm{k})\hat{\gamma}^\dagger_{\bm{k}} \hat{\gamma}^\dagger_{-\bm{k}}  \right]|\text{vac}\rangle
\end{align}
where $\prod'$ means the product is over half of the momentum space to avoid double counting. This it precisely a BCS state with occupied Bogoliubov-de-Gennes bands defined by 
\begin{align}
    \hat{\Xi}_{\bm{k},i}^\dagger = & \left[ e^{-z} u^*_{A}(\bm{k}) \hat{\gamma}^\dagger_{\bm{k}} + u_{B}(-\bm{k})\hat{\gamma}_{-\bm{k}}\right] / \sqrt{|U_{B1}(-\bm{k})|^2+e^{-2\Re z} |U_{A1}(\bm{k})|^2}\\
    = & \begin{pmatrix}
        \hat{c}^\dagger_{\bm{k}\mu} & 
        \hat{c}_{-\bm{k}\nu}
    \end{pmatrix}  \begin{pmatrix}
       e^{-z} u^*_{A}(\bm{k}) u_{\mu }(\bm{k}) \\ u_{B}(-\bm{k})u^*_{\nu}(-\bm{k}) 
    \end{pmatrix}  \equiv  \begin{pmatrix}
        \hat{c}^\dagger_{\bm{k}} &
        \hat{c}_{-\bm{k}}
    \end{pmatrix} w (\bm{k})
\end{align}
where $\Re z$ means the real part of $z$. The Chern number of the superconductor is given by
\begin{align}
    C =& C^\text{BdG} + C_< - C_>    =  2 C_< + C_\text{FB} + C^\text{BdG} 
\end{align}
where
\begin{align}
    C^\text{BdG} =& \frac{1}{2\pi} \int_\text{BZ} \mathrm{d}^2 \bm{k} \ \mathrm{i} \left( \partial_x w^\dagger \cdot \partial_y w - \partial_y w^\dagger \cdot \partial_x w\right)
\end{align}
is the contribution from the flat band pairing, and $C_{<,>}$ are the total Chern number of the remote bands below or above the flat band, which are fully occupied by electrons or holes and thus contribute oppositely to the BdG band Chern number. $C_\text{FB}$ is the Chern number of the flat band. 

\subsection{The relation between the two almost degenerate SC orders: particle-hole duality}

As one can see from the above general construction scheme in Eq.~\ref{eq: Hint simple}, exchanging $A\leftrightarrow B$ can lead to another $\hat{H}_\text{int}$ that is almost identical:
\begin{align}
        \hat{H}'_\text{int} =   V \sum_{\bm{R}} \left[ -\hat{n}_{\bm{R}B}\hat{n}_{\bm{R}A} + \rho^\text{R,e}_A \hat{n}_{\bm{R}B}+ (1-\rho^\text{R,h}_B)\hat{n}_{\bm{R}A} \right]
\end{align}
since they only differ by a two-particle term:
\begin{align}
     \hat{H}'_\text{int} -\hat{H}_\text{int} = V \sum_{\bm{R}} \left[  \rho^\text{FB}_B\hat{n}_{\bm{R}A}  - \rho^\text{FB}_A\hat{n}_{\bm{R}B} \right].
\end{align}
This model is also exactly solvable and correspond to the order parameter
\begin{align}
    \hat{\eta}'^{\dagger} = \frac{1}{\mathsf{V}}\sum_{\bm{k}} \frac{u_{B}^*(\bm{k})}{u_{A}(-\bm{k}) } \hat{\gamma}^\dagger_{1,\bm{k}} \hat{\gamma}_{1,-\bm{k}}.
\end{align}
Thus the form factors of the two order parameters are simply related through:
\begin{align}
    [f(\bm{k})]^* = 1/f'(\bm{k}).
\end{align}

Upon projection onto the flat band subspace, the two Hamiltonians take very similar forms:
\begin{align}
    \hat{H}_\text{solvable}  =& \hat{P} \hat{H}_\text{int} \hat{P}  = \frac{V}{\mathsf{V}}\sum_{\bm{k}_1+\bm{k}_3 = \bm{k}_2+\bm{k}_4} u^*_B(\bm{k}_1) u_A(\bm{k}_2) u^*_A(\bm{k}_3) u_B(\bm{k}_4) \hat{\gamma}^\dagger_{\bm{k}_1} \hat{\gamma}_{\bm{k}_2}\hat{\gamma}^\dagger_{\bm{k}_3}\hat{\gamma}_{\bm{k}_4} \\
    \hat{H}'_\text{solvable} =& \hat{P} \hat{H}'_\text{int} \hat{P} = \frac{V}{\mathsf{V}}\sum_{\bm{k}_1+\bm{k}_3 = \bm{k}_2+\bm{k}_4} u^*_A(\bm{k}_1) u_B(\bm{k}_2) u^*_B(\bm{k}_3) u_A(\bm{k}_4) \hat{\gamma}^\dagger_{\bm{k}_1} \hat{\gamma}_{\bm{k}_2}\hat{\gamma}^\dagger_{\bm{k}_3}\hat{\gamma}_{\bm{k}_4}
\end{align}
and they are exactly related by a particle-hole transformation $\mathcal{C}: \hat{\gamma}\rightarrow \hat{\gamma}^\dagger$ and a time-reversal transformation $\mathcal{T}$:
\begin{align}
    \mathcal{T}\mathcal{C} \hat{H}_\text{solvable} (\mathcal{T}\mathcal{C} )^{-1} =&  \frac{V}{\mathsf{V}}\sum_{\bm{k}_1+\bm{k}_3 = \bm{k}_2+\bm{k}_4} u_B(\bm{k}_1) u^*_A(\bm{k}_2) u_A(\bm{k}_3) u^*_B(\bm{k}_4) \hat{\gamma}_{-\bm{k}_1} \hat{\gamma}^\dagger_{-\bm{k}_2}\hat{\gamma}_{-\bm{k}_3}\hat{\gamma}^\dagger_{-\bm{k}_4} \\
    =&  \frac{V}{\mathsf{V}}\sum_{\bm{k}_1+\bm{k}_3 = \bm{k}_2+\bm{k}_4} u_B(\bm{k}_1) u^*_A(\bm{k}_2) u_A(\bm{k}_3) u^*_B(\bm{k}_4) \left[\delta_{\bm{k}_1 \bm{k}_2}-\hat{\gamma}^\dagger_{\bm{k}_2}\hat{\gamma}_{\bm{k}_1} \right]\left[\delta_{\bm{k}_1 \bm{k}_2}-\hat{\gamma}^\dagger_{\bm{k}_4}\hat{\gamma}_{\bm{k}_3} \right]  \\   
    =& \hat{H}'_\text{solvable}
\end{align}
where the last equality used the oddity of $u_B(\bm{k}) u^*_A(\bm{k})$ to ensure the additional terms to vanish. This implies that all the properties of the two models, such as the stiffness and the single particle excitation spectrum, are related by these two transformations.

\subsection{Peculiarities of the two-orbital systems}

The  two-orbital cases have a peculiar feature: $C^\text{BdG}=0$. This is now to show that the wavefunction $w(\bm{k}) = \left(e^{-z} u^*_{A}(\bm{k}) u(\bm{k}) \ , \  u_{B}(\bm{k}) u^*(-\bm{k}) \right)^T $ is always Chern $0$. We note that the Chern number is independent of the embedding (although the Berry curvature is not), so taking the periodic embedding doesn't affect this lemma. 

To prove our claim, we expand
\bea
w(\bm{k}) &= \bpm 
e^{-z} u^*_{A}(\bm{k}) u_A(\bm{k}) & e^{-z} u^*_{A}(\bm{k}) u_B(\bm{k}) &  u_{B}(\bm{k}) u^*_A(-\bm{k}) &  u_{B}(\bm{k})  
\epm^T \\
&= \bpm 
e^{-z} u^*_{A}(\bm{k}) u_A(\bm{k}) & e^{-z} u^*_{A}(\bm{k}) u_B(\bm{k}) &  p_A e^{\mathrm{i}\varphi(\bm{k})}  u_{B}(\bm{k}) u^*_A(\bm{k})  & p_B e^{\mathrm{i}\varphi(\bm{k})} u_{B}(\bm{k}) u^*_B(\bm{k})  
\epm^T  \ .\\
\eea
We now make the following observation. The components $u_\al(\bm{k}) u^*_{\be}(\bm{k})$ are exactly the elements of the flat band projector $P(\bm{k})$, which are smooth on the BZ and gauge-independent. This gives us the freedom to pick a periodic gauge choice for $u_\al(\bm{k})$: if the flat band has nonzero chern number, then $u_\al(\bm{k})$ cannot be smooth in a periodic gauge. If we can now show that $ e^{\mathrm{i}\varphi(\bm{k})}$ is smooth in the periodic gauge, then all entries of $w(\bm{k})$ are smooth and periodic. Hence the chern number of $w(\bm{k}) $ is zero. 

We can show this explicitly. Recall that the lowest eigenvector of the hopping matrix $t(\bm{k})=d_i(\bm{k}) \sigma_i$ ($\sigma_i$ are Pauli matrices) is
\bea
u(\bm{k}) = \frac{1}{\sqrt{\mathcal{N}(\bm{k})}} (d_3 - |d|, d_1 + i d_2)^T \ .
\eea
Since $h$ is periodic in $k$, $u(\bm{k})$ is periodic. If $u(\bm{k})$ has a Chern number, there is a point where an entry of $u(\bm{k}) \to 0/0$ and a vortex appears, making the wavefunction not smooth. We now note that by assumption $t(\bm{k})$ has an inversion symmetry implemented by $D=\sigma_3$ so that $
d_3(\bm{k}) = d_3(\bm{k}), |d|(\bm{k}) = |d|(\bm{k})$ and $d_i(\bm{k}) = - d_i(\bm{k})$ for $i=1,2$. This means that the gauge choice of $u(\bm{k})$ is explicitly inversion symmetric, so
\bea
B(\bm{k}) = u^\dag(-\bm{k}) \sigma_3 u(\bm{k})  = 1 \ .
\eea
Hence, we have exhibited a gauge where $e^{\mathrm{i}\varphi(\bm{k})} = 1$ is smooth (constant), so that all entries of $w(\bm{k})$ are smooth and periodic, implying $C^\text{BdG}=0$.

\subsection{Exact solution to the two-particle sector}

We start with the Hamiltonian (define $\bar{O} = \hat{P}\hat{O}\hat{P}$, $S_{\bm{p},\bm{q}}\equiv B(\bm{p},\bm{q})$ in this section)
\bea
\bar{H} &= \frac{V}{\mathsf{V}} \sum_{\bm{q}} \bar{S}_{\bm{q}}^\dag \bar{S}_{\bm{q}}  \\
\bar{S}_{\bm{q}} &= \sum_{\bm{k}} u^*_{A}(\bm{k}+\bm{q}) u_{B}(\bm{k})  \gamma^\dag_{\bm{k}+\bm{q}} \gamma_{\bm{k}} \\
&= \sum_{\bm{k}} S_{\bm{k},\bm{q}}  \gamma^\dag_{\bm{k}+\bm{q}} \gamma_{\bm{k}} \\
\eea
and compute the two body spectrum. We organize the calculation as
\bea
\bar{S}_{\bm{q}} \gamma^\dag_{\bm{k}'+\bm{p}} \gamma^\dag_{-\bm{k}'}\ket{0} &=\sum_{\bm{k}} S_{\bm{k},\bm{q}}  \gamma^\dag_{\bm{k}+\bm{q}} \gamma_{\bm{k}} \gamma^\dag_{\bm{k}'+\bm{p}} \gamma^\dag_{-\bm{k}'}\ket{0} \\
&=\sum_{\bm{k}} S_{\bm{k},\bm{q}}  \gamma^\dag_{\bm{k}+\bm{q}}  ( \delta_{\bm{k},\bm{k}'+\bm{p}} \gamma^\dag_{-\bm{k}'} -  \gamma^\dag_{\bm{k}'+\bm{p}} \delta_{\bm{k},-\bm{k}'})\ket{0} \\
&=  (S_{\bm{k}'+\bm{p},\bm{q}} \gamma^\dag_{\bm{k}'+\bm{p}+\bm{q}}  \gamma^\dag_{-\bm{k}'} - S_{-\bm{k}',\bm{q}} \gamma^\dag_{-\bm{k}'+\bm{q}}  \gamma^\dag_{\bm{k}'+\bm{p}})\ket{0} \\
\eea
and thus
\bea
\bra{0} \gamma_{-\bm{k}}  \gamma_{\bm{k}+\bm{p}} \bar{H} \gamma^\dag_{\bm{k}'+\bm{p}} \gamma^\dag_{-\bm{k}'}\ket{0} &= \frac{V}{\mathsf{V}} \sum_{\bm{q} }\braket{
 (S^*_{\bm{k}+\bm{p},\bm{q}} \gamma^\dag_{-\bm{k}}  \gamma_{\bm{k}+\bm{p}+\bm{q}} - S^*_{-\bm{k},\bm{q}}  \gamma_{\bm{k}+\bm{p}} \gamma_{-\bm{k}+\bm{q}} ) 
 (S_{\bm{k}'+\bm{p},\bm{q}} \gamma^\dag_{\bm{k}'+\bm{p}+\bm{q}}  \gamma^\dag_{-\bm{k}'} - S_{-\bm{k}',\bm{q}} \gamma^\dag_{-\bm{k}'+\bm{q}}  \gamma^\dag_{\bm{k}'+\bm{p}}) } \\
 &= \frac{V}{\mathsf{V}} \sum_{\bm{q}} \Big(S^*_{\bm{k}+\bm{p},\bm{q}} S_{\bm{k}'+\bm{p},\bm{q}} (\delta_{\bm{k},\bm{k}'} - \delta_{\bm{k}+\bm{p}+\bm{q},-\bm{k}'}) + S^*_{-\bm{k},\bm{q}}  S_{-\bm{k}',\bm{q}}  (\delta_{\bm{k},\bm{k}'} - \delta_{-\bm{k}+\bm{q},\bm{k}'+\bm{p}}) \\
 &\quad - S^*_{-\bm{k},\bm{q}} S_{\bm{k}'+\bm{p},\bm{q}} (\delta_{-\bm{k}+\bm{q},\bm{k}'+\bm{p}+\bm{q}} - \delta_{-\bm{k}+\bm{q},-\bm{k}'}) - S^*_{\bm{k}+\bm{p},\bm{q}} S_{-\bm{k}',\bm{q}}  (\delta_{\bm{k}+\bm{p}+\bm{q},-\bm{k}'+\bm{q}}-\delta_{\bm{k}+\bm{p}+\bm{q},\bm{k}'+\bm{p}} ))
  \Big)
\eea
We now simplify the delta functions
\bea
 &= \frac{V}{\mathsf{V}} \sum_{\bm{q}} \Big(S^*_{\bm{k}+\bm{p},\bm{q}} S_{\bm{k}'+\bm{p},\bm{q}} (\delta_{\bm{k},\bm{k}'} - \delta_{\bm{k}+\bm{p}+\bm{q},-\bm{k}'}) + S^*_{-\bm{k},\bm{q}}  S_{-\bm{k}',\bm{q}}  (\delta_{\bm{k},\bm{k}'} - \delta_{-\bm{k}+\bm{q},\bm{k}'+\bm{p}}) \\
 &\quad - S^*_{-\bm{k},\bm{q}} S_{\bm{k}'+\bm{p},\bm{q}} (\delta_{-\bm{k},\bm{k}'+\bm{p}} - \delta_{-\bm{k}+\bm{q},-\bm{k}'}) - S^*_{\bm{k}+\bm{p},\bm{q}} S_{-\bm{k}',\bm{q}}  (\delta_{\bm{k}+\bm{p},-\bm{k}'}-\delta_{\bm{k}+\bm{q},\bm{k}'} ))
  \Big) \\
&= \frac{V}{\mathsf{V}} \sum_{\bm{q}} (S^*_{\bm{k}+\bm{p},\bm{q}} S_{\bm{k}'+\bm{p},\bm{q}}  + S^*_{-\bm{k},\bm{q}}  S_{-\bm{k}',\bm{q}} ) \delta_{\bm{k},\bm{k}'} - (S^*_{-\bm{k},\bm{q}} S_{\bm{k}'+\bm{p},\bm{q}} + S^*_{\bm{k}+\bm{p},\bm{q}} S_{-\bm{k}',\bm{q}})  \delta_{\bm{k}+\bm{p},-\bm{k}'}  \\
&\qquad -  \frac{V}{\mathsf{V}} \sum_{\bm{q}  }\Big(S^*_{\bm{k}+\bm{p},\bm{q}} S_{\bm{k}'+\bm{p},\bm{q}} \delta_{\bm{k}+\bm{p}+\bm{q},-\bm{k}'} + S^*_{-\bm{k},\bm{q}}  S_{-\bm{k}',\bm{q}} \delta_{-\bm{k}+\bm{q},\bm{k}'+\bm{p}} - S^*_{-\bm{k},\bm{q}} S_{\bm{k}'+\bm{p},\bm{q}} \delta_{-\bm{k}+\bm{q},-\bm{k}'} - S^*_{\bm{k}+\bm{p},\bm{q}} S_{-\bm{k}',\bm{q}}  \delta_{\bm{k}+\bm{q},\bm{k}'} )
  \Big)
\eea
The first line is a ``diagonal" term
\bea
\frac{V}{\mathsf{V}} \sum_{\bm{q}} (|S_{\bm{k}+\bm{p},\bm{q}}|^2 + |S_{-\bm{k},\bm{q}}|^2) (\delta_{\bm{k},\bm{k}'} -\delta_{\bm{k}+\bm{p},-\bm{k}'})
\eea
and the second line is a scattering matrix
\bea
-  \frac{V}{\mathsf{V}} \Big(S^*_{\bm{k}+\bm{p},-\bm{k}-\bm{k}'-\bm{p}} S_{\bm{k}'+\bm{p},-\bm{k}-\bm{k}'-\bm{p}}  + S^*_{-\bm{k},\bm{k}'+\bm{k}+\bm{p}}  S_{-\bm{k}',\bm{k}'+\bm{k}+\bm{p}}- S^*_{-\bm{k},\bm{k}-\bm{k}'} S_{\bm{k}'+\bm{p},\bm{k}-\bm{k}'} - S^*_{\bm{k}+\bm{p},\bm{k}'-\bm{k}} S_{-\bm{k}',\bm{k}'-\bm{k}})
  \Big) \ .
\eea
We have derived a form of the 2-body matrix $H_{\bm{k},\bm{k}'}$. It appears complicated because the basis of states $\gamma^\dag_{\bm{k}'+\bm{p}} \gamma^\dag_{-\bm{k}'}\ket{0}$ is over-complete due to fermion anti-symmetry. First we shift $\bm{k} \to \bm{k}-\bm{p}/2$ and similarly for $\bm{k}'$ so that the basis is $\gamma^\dag_{\bm{k}'+\bm{p}/2} \gamma^\dag_{-\bm{k}'+\bm{p}/2}\ket{0}$. Now we see that at each momentum $\bm{p}$, the basis has $\sim L^2/2$ states since we can take $\bm{k}$ to be in half the BZ. Then we have
\bea
H_{\bm{k}-\bm{p}/2,\bm{k}'-\bm{p}/2} &= \frac{V}{\mathsf{V}} \sum_{\bm{q}} (|S_{\bm{k}+\bm{p}/2,\bm{q}}|^2 + |S_{-\bm{k}+\bm{p}/2,\bm{q}}|^2) (\delta_{\bm{k},\bm{k}'} -\delta_{\bm{k},-\bm{k}'}) \\
&\qquad -  \frac{1}{\mathsf{V}} \Big( S^*_{\bm{k}+\bm{p}/2,-\bm{k}-\bm{k}'} S_{\bm{k}'+\bm{p}/2,-\bm{k}-\bm{k}'}  + S^*_{-\bm{k}+\bm{p}/2,\bm{k}'+\bm{k}}  S_{-\bm{k}'+\bm{p}/2,\bm{k}'+\bm{k}} \\
&\qquad- S^*_{-\bm{k}+\bm{p}/2,\bm{k}-\bm{k}'} S_{\bm{k}'+\bm{p}/2,\bm{k}-\bm{k}'} - S^*_{\bm{k}+\bm{p}/2,\bm{k}'-\bm{k}} S_{-\bm{k}'+\bm{p}/2,\bm{k}'-\bm{k}}\Big) \ .
\eea
Next we turn to the scattering matrix. We see
\bea
S^*_{\bm{k}+\bm{p}/2,-\bm{k}-\bm{k}'} S_{\bm{k}'+\bm{p}/2,-\bm{k}-\bm{k}'}  &= u^*_B(\bm{p}/2+\bm{k}) u_A^*(\bm{p}/2-\bm{k}) \, u_A(\bm{p}/2-\bm{k}')  u_B(\bm{p}/2+\bm{k}') \\
S^*_{-\bm{k}+\bm{p}/2,\bm{k}'+\bm{k}}  S_{-\bm{k}'+\bm{p}/2,\bm{k}'+\bm{k}} &=   u^*_B(\bm{p}/2-\bm{k}) u_A^*(\bm{p}/2+\bm{k}) \, u_A(\bm{p}/2+\bm{k}') u_B(\bm{p}/2-\bm{k}') \\
- S^*_{-\bm{k}+\bm{p}/2,\bm{k}-\bm{k}'} S_{\bm{k}'+\bm{p}/2,\bm{k}-\bm{k}'} &= - u_B^*(\bm{p}/2-\bm{k})  u_A^*(\bm{p}/2+\bm{k}) \, u_A(\bm{p}/2-\bm{k}') u_B(\bm{p}/2+\bm{k}') \\
- S^*_{\bm{k}+\bm{p}/2,\bm{k}'-\bm{k}} S_{-\bm{k}'+\bm{p}/2,\bm{k}'-\bm{k}} &= -u_A^*(\bm{p}/2-\bm{k})  u_B^*(\bm{p}/2+\bm{k}) u_A(\bm{p}/2-\bm{k}') u_B(\bm{p}/2-\bm{k}') \ .
\eea
We define 
\bea
\mathcal{U}^*_{\bm{k}'}(\bm{p})  &= u_A(\bm{p}/2-\bm{k}')  u_B(\bm{p}/2+\bm{k}') \ .
\eea
so that we can write the matrix as
\bea
H_{\bm{k}-\bm{p}/2,\bm{k}'-\bm{p}/2} &= \mathcal{D}_{\bm{k}}(\bm{p}) (\delta_{\bm{k},\bm{k}'} -\delta_{\bm{k},-\bm{k}'}) - \frac{V}{\mathsf{V}} \Big( \mathcal{U}_{\bm{k}}(\bm{p})  \mathcal{U}^*_{\bm{k}'}(\bm{p}) +\mathcal{U}_{-\bm{k}}(\bm{p})  \mathcal{U}^*_{-\bm{k}'}(\bm{p})  - \mathcal{U}_{-\bm{k}}(\bm{p}) \mathcal{U}^*_{\bm{k}'}(\bm{p}) - \mathcal{U}_{\bm{k}}(\bm{p})  \mathcal{U}^*_{-\bm{k}'}(\bm{p}) \Big) \\
 &= \mathcal{D}_{\bm{k}}(\bm{p}) (\delta_{\bm{k},\bm{k}'} -\delta_{\bm{k},-\bm{k}'}) - \frac{V}{\mathsf{V}} \big( (\mathcal{U}_{\bm{k}}(\bm{p})   - \mathcal{U}_{-\bm{k}}(\bm{p}) ) \, ( \mathcal{U}^*_{\bm{k}'}(\bm{p}) -\mathcal{U}^*_{-\bm{k}'}(\bm{p})) \,  \big) \ . \\
\eea
If we now restrict to a non-redundant basis where the index $\bm{k}$ is taken only over half the BZ (inversion-symmetric points are excluded), then
\bea
H^{\text{half BZ}}_{\bm{k}-\bm{p}/2,\bm{k}'-\bm{p}/2} &= \mathcal{D}_{\bm{k}}(\bm{p}) \delta_{\bm{k},\bm{k}'} - \frac{V}{\mathsf{V}} ( \mathcal{U}_{\bm{k}}(\bm{p})   - \mathcal{U}_{-\bm{k}}(\bm{p}) ) ( \mathcal{U}^*_{\bm{k}'}(\bm{p}) -\mathcal{U}^*_{-\bm{k}'}(\bm{p})  ) \\
&\equiv (D - \frac{V}{\mathsf{V}} \mathcal{V}\mathcal{V}^\dag)_{\bm{k},\bm{k}'} \
\eea
where $\mathcal{V}$ is rank 1. A convenient expression for $\mathcal{V}$ is 
\bea
\mathcal{V}_{\bm{k}}(\bm{p}) &= \eps_{\al \be} u^*_\al(\bm{p}/2-\bm{k})u^*_\be(\bm{p}/2+\bm{k})
\eea
with summation over the orbitals implied, and $\epsilon$ is the unit anti-symmetric tensor in the $A,B$ sub-block.

To solve this matrix analytically, we use the matrix determinant lemma $\det (\mathbbm{1} + \mathcal{A} \mathcal{B}) = \det (\mathbbm{1} +\mathcal{B} \mathcal{A})$ where $\mathcal{A}, \mathcal{B}$ are rectangular matrices such that the determinants are over different spaces on the left and right side. We know write the eigenvalue problem as
\bea
\det (\mathcal{D} - \la \mathbbm{1} - \frac{1}{\mathsf{V}} \mathcal{V}\mathcal{V}^\dag) &= 0 \\
\det (\mathcal{D} - \la \mathbbm{1}) \det \lp \mathbbm{1} - \frac{V}{\mathsf{V}} \frac{1}{ (\mathcal{D}- \la \mathbbm{1}) }  \mathcal{V}\mathcal{V}^\dag \rp &= 0 \ . \\
\eea
If $\lambda$ is not in the continuous spectrum of $\mathcal{D}$, then the second factor must be zero. This means
\bea
0 &= \det \lp \mathbbm{1} - \frac{1}{\mathsf{V}} \frac{1}{ (\mathcal{D} - \la \mathbbm{1}) }  \mathcal{V}\mathcal{V}^\dag \rp = \det \lp \mathbbm{1} - \frac{V}{\mathsf{V}} \mathcal{V}^\dag \frac{1}{ (\mathcal{D} - \la \mathbbm{1}) }  \mathcal{V}\rp = \lp \mathbbm{1} - \frac{V}{\mathsf{V}} \mathcal{V}^\dag \frac{1}{ (\mathcal{D} - \la \mathbbm{1}) }  \mathcal{V}\rp \ .
\eea
Thus we obtain a simple implicit expression for the bound state energies $\la$. They obey
\bea
1 &= \frac{V}{\mathsf{V}} \mathcal{V}^\dag \frac{1}{ (\mathcal{D}- \la \mathbbm{1}) }\mathcal{V} \\
 &=  \frac{V}{\mathsf{V}} \sum_{\bm{k} \in \frac{1}{2} BZ} \eps_{\al' \be'} u_{\al'}(\bm{p}/2-\bm{k})u_{\be'}(\bm{p}/2+\bm{k}) \frac{1}{\mathcal{D}_{\bm{k}}(\bm{p}) - \la(\bm{p}) } \eps_{\al \be} u^*_\al(\bm{p}/2-\bm{k})u^*_\be(\bm{p}/2+\bm{k}) \\
&=  \frac{V}{2\mathsf{V}} \sum_{\bm{k} \in  BZ}  \frac{\left|u_A(\bm{k}-\bm{p}/2)u_B(\bm{k}+\bm{p}/2) +u_B(\bm{k}-\bm{p}/2)u_A(\bm{k}+\bm{p}/2) \right|^2}{\mathcal{D}_{\bm{k}}(\bm{p}) - \la(\bm{p}) } 
\eea
Lastly let's recall that
\bea
\mathcal{D}_{\bm{k}+\bm{p}/2}(\bm{p}) &=  \frac{V}{\mathsf{V}} \sum_{\bm{q}} (|S_{\bm{k}+\bm{p}/2,\bm{q}}|^2 + |S_{-\bm{k}+\bm{p}/2,\bm{q}}|^2) \\
 &=  \frac{V}{\mathsf{V}} \sum_{\bm{q}} ( |u_A(\bm{k}+\bm{p}/2+\bm{q})|^2 |u_B(\bm{k}+\bm{p}/2)|^2 + |u_A(\bm{q}-\bm{k}+\bm{p}/2)|^2 |u_B(-\bm{k}+\bm{p}/2)|^2 ) \\
  &= V  \overline{|u_A|^2} \left[|u_B(\bm{k}+\bm{p}/2)|^2 +   |u_B(-\bm{k}+\bm{p}/2)|^2\right] \\
\eea
Thus we reach the final implicit equation:
\bea\label{eq: implicit}
1 &=  \frac{1}{2 \overline{|u_A|^2}  \mathsf{V}} \sum_{\bm{k} \in BZ}  \frac{\left|u_A(\bm{k}-\bm{p}/2)u_B(\bm{k}+\bm{p}/2) +u_B(\bm{k}-\bm{p}/2)u_A(\bm{k}+\bm{p}/2) \right|^2}{|u_B(\bm{k}-\bm{p}/2)|^2+ |u_{B}(\bm{k}+\bm{p}/2)|^2 - \la(\bm{p})/V } 
\eea
which could be re-expressed in terms of $P$:
\bea
0= \sum_{\bm{k} \in BZ}  \frac{ |P_{AB}(\bm{k})-P_{AB}(\bm{k}+\bm{p})|^2 - [P_{AA}(\bm{k})+P_{AA}(\bm{k}+\bm{p})]\la(\bm{p})/V }{P_{BB}(\bm{k}-\bm{p}/2)+ P_{BB}(\bm{k}+\bm{p}/2) - \la(\bm{p})/V }\\
\eea
It is easy to check that $\lambda(\bm{p}=0)=0$ satisfies the above equation.

Finally we consider the small $\bm{p}$ expansion to Eq.~\ref{eq: implicit}:
\begin{align}
 \partial_i \partial_j \lambda(\bm{q}=0) \cdot \sum_{\bm{k} \in BZ}  \left.\frac{|u_A|^2}{2|u_B|^2}\right|_{\bm{k}}  &= V \sum_{\bm{k} \in BZ}  \left\{ \frac{1}{2|u_B|^2}  \Re\left[  \partial_i u_B^* \partial_j u_B  |u_A|^2 + \partial_i u_B^* \partial_j u_A^* u_A u_B + \partial_i u_A \partial_j u_B u_A^* u_B^*
 \right]\right. \nonumber \\
 &\ \ \ \ \ \ \ \ \ \ \ \ \ \ \ \ \ \ \ \ \ \ \ \ \ \  \ \ \ \ \ \ \ \ \ \ \ \ \  \ \ \ \ \ \ \ \ \ \ \ \ \  \ \ \ \ \ \ \ \ \ \ \ \ \  \ \ \ \ \ \ \ \ \ \ \ \ \   - \left.\left.\frac{1}{4}\left(\partial_i \partial_j u_A^* u_A+u_A^* \partial_i \partial_j u_A  \right) \right\}\right|_{\bm{k}}
\end{align}
By integration by part, we can lower the second derivatives on $u_A$ in the last two terms:
\begin{align}
 \partial_i \partial_j \lambda(\bm{q}=0) \cdot \sum_{\bm{k} \in BZ}  \left.\frac{|u_A|^2}{|u_B|^2}\right|_{\bm{k}}  &= V\sum_{\bm{k} \in BZ}\left. \left\{ \frac{1}{|u_B|^2}  \Re\left[  \partial_i u_B^* \partial_j u_B  |u_A|^2 + \partial_i u_B^* \partial_j u_A^* u_A u_B + \partial_i u_A \partial_j u_B u_A^* u_B^* + \partial_i u_A  \partial_j u_A^*  |u_B|^2 \right] \right\}\right|_{\bm{k}} \\
 & = V \sum_{\bm{k} \in BZ}\left.  \frac{\partial_i \left(  u_A u_B^*\right)\partial_j \left( u_A^* u_B\right) }{2|u_B|^2}  \right|_{\bm{k}}
\end{align}
We thus reach an exact expression for the inverse mass of the two-particle bound state:
\begin{align}\label{eq: two particle mass}
    [m_{2\text{p}}^{-1}]_{ij}  = \partial_i \partial_j \lambda(\bm{q}=0) =V  \sum_{\bm{k} \in BZ}\left.  \frac{\Re \partial_i \left(  u_A u_B^*\right)\partial_j \left( u_A^* u_B\right) }{|u_B|^2}  \right|_{\bm{k}} / \sum_{\bm{k} \in BZ}  \left.\frac{|u_A|^2}{|u_B|^2}\right|_{\bm{k}}
\end{align}

Applying the particle-hole duality, we find that the two-hole sector result can be obtained by simply exchanging $A\leftrightarrow B$ in all the expressions in this section. For example, the two-hole bound state mass is given by:
\begin{align}\label{eq: two hole mass}
    [m_{2\text{h}}^{-1}]_{ij}   =V  \sum_{\bm{k} \in BZ}\left.  \frac{\Re\partial_i \left(  u_A u_B^*\right)\partial_j \left( u_A^* u_B\right) }{|u_A|^2}  \right|_{\bm{k}} / \sum_{\bm{k} \in BZ}  \left.\frac{|u_B|^2}{|u_A|^2}\right|_{\bm{k}}
\end{align}

These bound state masses will be related to the stiffness bound in Sec.~\ref{sec: stiffness inversion}

\section{Upper bound on phase stiffness}

\subsection{The strategy}

The superfluid stiffness is defined by the energy response to a flat gauge connection $\bm{A}$:
\begin{align}
    \kappa_{ij} = \lim_{\mathsf{V}\rightarrow \infty}\frac{1}{4\mathsf{V}} \left. \frac{\partial^2 E(\bm{A})}{\partial {A_i}\partial {A_j}} \right|_{\bm{A}=0}
\end{align}
where $E(\bm{A})$ is the ground state energy in the presence of constant vector potential $\bm{A}$. [The above expression is in the convention that the volume of a unit cell $\Omega_\text{UC}=1$. Otherwise a $1/\Omega_\text{UC}$ factor should be included.] It is important to note that the order of taking the thermodynamic limit and the derivatives matters, since when a component $A_i$ is a multiple of $2\pi/L_i$, its effect can be completely absorbed by a rigid shift of momentum grid ($L_i$ is the linear system size in that direction). For a finite system size and a small but nonzero $|\bm{A}|\ll \frac{2\pi}{L}$, we cannot exactly solve the model but can use variational method to upper bound $E(\bm{A})$. Specifically, any variational state yields an upper bound 
\begin{align}
    E_\text{var}(\bm{A}) = \min_{F}\frac{\langle N| \hat{H}(\bm{A})|N\rangle_\text{var} }{\langle N|N\rangle_\text{var} } \geq E(\bm{A})
\end{align}

In this work we will use BCS states or projected BCS states (also known as Antisymmetric Geminal Product - AGP -states) as our variational ansatz. When $\bm{A}=0$, we know our model is solvable and the true GS is indeed a BCS/AGP state, so the variational ansatz is exact at $\bm{A}=0$, where the energy is simply $0$. On the other hand, by gauge invariance, we know that $\partial_{A_i} E = 0$ at $\bm{A}=0$ for both the true energy and the variational energy, so that the quadratic order in the Taylor's expansion is the leading non-zero order. Combining these two facts, we reach the conclusion:
\begin{align}
    \kappa_\text{var} \succeq \kappa \succeq 0
\end{align}
meaning that the stiffness tensor obtained by variational energy is greater or equal to the real one, thus providing a rigorous upper bound ($H(\bm{A})$ is always positive, so $\kappa$ is guaranteed to be non-negative). In particular, this implies the geometric mean of the stiffness $\bar{\kappa}\equiv \sqrt[d]{\det{\kappa}} $, which is a relevant quantity for the coherence energy scale of superconductor, is rigorously bounded from above
\begin{align}
    \sqrt[d]{\det{\kappa_\text{var}}} \geq \sqrt[d]{\det{\kappa}} .
\end{align}

\subsection{The variational ansatzes and the observables}

\subsubsection{The projected BCS states}
Here we discuss the properties of Antisymmetric Geminal Product (AGP) states, or projected BCS states. For simplicity, we only consider the case where the cooper pair momentum $\bm{Q}=0$:
\begin{align}
    |N\rangle & = \left(\hat{\eta}^\dagger\right)^N |\text{vac}\rangle /\mathcal{N}_N \\
    \hat{\eta}^\dagger &\equiv  \sum_{\bm{k}; n,m\le \mathsf{N}_\text{flat}} F_{nm} (\bm{k})\hat{\gamma}^\dagger_{\bm{k}, n} \hat{\gamma}^\dagger_{-\bm{k},m}   
\end{align}
where $F$ is variational form factor satisfying $F(\bm{k}) = - F^T(-\bm{k})$. The advantage of this ansatz is that it is an eigenstate of number operator, i.e. has a definite number, while we will see below that it is cumbersome to evaluate expectation values of observables for this form of states. 
To make the structure of these states clear, we singular-value decompose (SVD) the form factor $F(\bm{k})$ as 
\begin{align}
    F_{nm}(\bm{k}) =   \sum_{i} F^{(i)}_{nm}(\bm{k}) = \sum_{i} f^{(i)}(\bm{k}) v^{(i)}_n(\bm{k}) v^{(i)}_m(-\bm{k})
\end{align}
where $f^{(i)}(\bm{k}) = -f^{(i)}(-\bm{k})$ and on each $\bm{k}$, the set of vectors $\{v^{(i)}\}$ are orthonormal. [Technically speaking, this is not a SVD since $f^{(i)}$ can be negative here; however, one can easily achieve such a decomposition through SVD followed by absorbing phases in the unitary matrices into the definition of $f^{(i)}$ spectrum.] Defining a new basis for the flat band orbitals
\begin{align}
    \hat{\xi}_{\bm{k},i}^\dagger =  v^{(i)}_n(\bm{k}) \hat{\gamma}^\dagger_{\bm{k},n}
\end{align}
we can recast the $\hat{\eta}$ operator into a compact form:
\begin{align}
    \hat{\eta}_\text{var}^\dagger &\equiv 2\sum'_{\bm{k}; i = 1 \dots  \mathsf{N}_\text{flat}} f^{(i)} (\bm{k})\hat{\xi}^\dagger_{\bm{k}, i} \hat{\xi}^\dagger_{-\bm{k}, i}   
\end{align}
The $\sum'$ is not over the whole momentum space but only half of it due to the redundancy of the representation; the prefactor $2$ also arise for the same reason. 

Then each pair $\hat{\xi}^\dagger_{\bm{k}, i} \hat{\xi}^\dagger_{-\bm{k}, i} $ can be viewed as the creation operator onto a hard-core boson mode~\cite{10.1063/1.5127850}. With this simple interpretation, we find the norm of the state
\begin{align}
    \mathcal{N}_N \equiv \langle vac| \left(\hat{\eta}\right)^N \left(\hat{\eta}^\dagger\right)^N |vac\rangle_\text{var} = 4^N S^{\mathsf{N}_0}_N(\{|f^{(i)}(\bm{k})|^2\})
\end{align}
where $S^{\mathsf{N}_0}_N$ is the Elementary Symmetric Polynomial (ESP) of degree $N$ with $\mathsf{N}_0$ variables $\{|f^{(i)}(\bm{k})|^2\}$ ($\bm{k}$ is again restricted in half of the momentum space), and $\mathsf{N}_0 \equiv \mathsf{N}_\text{flat} \mathsf{V}/2$ is the total number of hard-core boson (cooper pair) modes. 

Now we are ready to evaluate the two-body reduced density matrix (2RDM) for this state, which is particularly convenient in the $\xi$ basis:
\begin{align}
    \Gamma_{IJKL} \equiv \langle \hat{\xi}_I^\dagger \hat{\xi}_J \hat{\xi}_K^\dagger \hat{\xi}_L\rangle / \mathcal{N}
\end{align}
where $I =(\bm{k},i)$ is a compact index for electron modes. Due to the special structure of the AGP state, most of the entries of this tensor is zero. The only non-zero ones take the form ($\bar{I}$ means the paired mode of mode $I$):
\begin{align}
    \Gamma_{IIJJ} =&  \frac{1}{S^{\mathsf{N}_0}_N(\{|f^{(I)}|^2\})}\begin{cases}
        |f^{(I)}|^2|f^{(J)}|^2  S^{\mathsf{N}_0-2}_{N-2}\left(\{|f|^2\} \backslash |f^{(I)}|^2, |f^{(J)}|^2 \right) 
        &  I\neq J , \bar{J} \\
        |f^{(I)}|^2  S^{\mathsf{N}_0-1}_{N-1}\left(\{|f|^2\} \backslash |f^{(I)}|^2 \right)   & \text{otherwise}
    \end{cases}  \\
    \Gamma_{IJJI} =& \frac{1}{S^{\mathsf{N}_0}_N(\{|f^{(I)}|^2\})} \begin{cases}
         |f^{(I)}|^2  S^{\mathsf{N}_0-2}_{N-1}\left(\{|f|^2\} \backslash |f^{(I)}|^2, |f^{(J)}|^2 \right)  &  I\neq J , \bar{J} \\
        0  & \text{otherwise}
    \end{cases}   \\
      \Gamma_{IJ\bar{I} \bar{J}} =& \frac{1}{S^{\mathsf{N}_0}_N(\{|f^{(I)}|^2\})} \begin{cases}
        f^{(I),*} f^{(J)}  S^{\mathsf{N}_0-2}_{N-1}\left(\{|f|^2\} \backslash |f^{(I)}|^2, |f^{(J)}|^2 \right)   &  I\neq J , \bar{J}  \\
        |f^{(I)}|^2 S^{\mathsf{N}_0-1}_{N-1}\left(\{|f|^2\} \backslash |f^{(I)}|^2 \right)
        & I = J \\
         0 & I = \bar{J}
    \end{cases}
\end{align}
where $\backslash$ means excluding certain elements from a set.

For $N=1$ (two-particle sector), the results can be simply evaluated as 
\begin{align}\label{eq: two particle sector}
    \Gamma_{IIJJ} =&  \frac{1 }{\mathsf{N}_0 }\begin{cases}
         0 &  I\neq J , \bar{J} \\
         |f^{(I)}|^2 / \bar{f}_\text{sq}^2  & \text{otherwise}
    \end{cases}  \\
    \Gamma_{IJJI} =& \frac{1 }{\mathsf{N}_0 } \begin{cases}
        |f^{(I)}|^2/ \bar{f}_\text{sq}^2 &  I\neq J , \bar{J} \\
        0  & \text{otherwise}
    \end{cases}   \\
    \Gamma_{IJ\bar{I} \bar{J}} =& \frac{1 }{\mathsf{N}_0 } \begin{cases}
         f^{(I),*} f^{(J)}/ \bar{f}_\text{sq}^2 &  I\neq J , \bar{J}  \\
        |f^{(I)}|^2 / \bar{f}_\text{sq}^2 & I = J \\
         0 & I = \bar{J}
    \end{cases}
\end{align}
where 
\begin{align}
    \bar{f}_\text{sq}^2 \equiv \frac{1}{2\mathsf{N}_0} \sum_{I} |f^{(I)}|^2.
\end{align}

\subsubsection{The BCS states}

The AGP states are better suited for the study of finite systems with numerical methods. However, we are eventually interested in the thermodynamic limit $N,\mathsf{N}_0\rightarrow \infty$ with a fixed filling fraction $\nu=N/\mathsf{N}_0$. The AGP states are no longer convenient in this limit. Instead, we find it convenient to work with the grand-canonical version of this state defined by a complex number $z$:
\begin{align}
    |z\rangle \equiv & \sum_{N} \mathrm{e}^{-z N} \left(\hat{\eta}^\dagger/2\right)^N |\text{vac}\rangle  \\
    =& \prod'_{\bm{k}; i = 1 \dots  \mathsf{N}_\text{flat}} \left[1 +\mathrm{e}^{-z } f^{(i)} (\bm{k})\hat{\xi}^\dagger_{\bm{k}, i} \hat{\xi}^\dagger_{-\bm{k}, i}  \right]|\text{vac}\rangle
\end{align}
this is precisely a BCS state with occupied Bogoliubov-de-Gennes bands defined by 
\begin{align}
    \hat{\Xi}_{\bm{k},i}^\dagger = & \left[ e^{-z} f^{(i)}(\bm{k}) \hat{\xi}^\dagger_{\bm{k},i} + \hat{\xi}_{-\bm{k},i}\right] / \sqrt{1+e^{-2\Re z} |f^{(i)}(\bm{k})|^2}
\end{align}
where $\Re z$ means the real part of $z$. 

The normalization factor of this state is 
\begin{align}
    \mathcal{N} = \langle z|z\rangle = \prod'_{\bm{k};i} \left( 1+e^{-2\Re z} |f^{(I)}(\bm{k})|^2\right)
\end{align}
which clearly admits the interpretation of the grand partition function of a classical hard-core boson gas. Now that the particle number and thus the filling fraction is only fixed on average but we expect the fluctuation to vanish in the thermodynamic limit:
\begin{align}
    \nu = \frac{1}{2\mathsf{N}_0  \mathcal{N} }\frac{\partial \mathcal{N} }{\partial (\Re z)} = \frac{1}{\mathsf{N}_0} \sum'_{\bm{k};i} \frac{e^{-2\Re z} |f^{(I)}(\bm{k})|^2}{1+e^{-2\Re z} |f^{(I)}(\bm{k})|^2} \equiv \frac{1}{\mathsf{N}_0} \sum'_{\bm{k};i} \nu^{(i)}(\bm{k})
\end{align}
where we also introduced the filling fraction on each mode $\nu^{(i)}(\bm{k})$ for later convenience.

The off-diagonal-long-range order is easy to compute in this state:
\begin{align}
    \langle z |\hat{\eta}^\dagger |z \rangle = \frac{1}{\mathsf{N}_0} \sum'_{\bm{k};i} \frac{e^{-z^*} |f^{(I)}(\bm{k})|^2}{1+e^{-2\Re z} |f^{(I)}(\bm{k})|^2} =\nu \mathrm{e}^{z}
\end{align}

The 2RDM for this state is now simple to evaluate:
\begin{align}
  \text{(Hartree)} \ \ \ \ \ \  \Gamma_{IIJJ} =&  \begin{cases}
        \nu^{(I)} \nu^{(J)} &  I\neq J , \bar{J} \\
        \nu^{(I)}  & \text{otherwise}
    \end{cases}  \\
  \text{(Fock)} \ \ \ \ \ \ \ \ \    \Gamma_{IJJI} =& \begin{cases}
        \nu^{(I)} (1- \nu^{(J)}) & I\neq J , \bar{J} \\
        0  & \text{otherwise}
    \end{cases}   \\
  \text{(Cooper)} \ \ \ \ \ \    \Gamma_{IJ\bar{I} \bar{J}} =& \begin{cases}
        (1-\nu^{(I)})(1-\nu^{(J)})  [f^{(I),*} f^{(J)} e^{-2\Re z}]  &  I\neq J , \bar{J}  \\
         \nu^{(I)} & I = J \\
         0 & I = \bar{J}
    \end{cases}
\end{align}
We give the three types of terms different names according to the well-known convention for Wick's contraction of four fermion terms. These BCS states are indeed Slator determinant states of BdG fermions so they indeed satisfy the Wick's theorem.

\subsection{The variational energy}

To derive the projectected Hamiltonian in the presence of $\bm{A}$, let's go back to the definition of the electronic structure:
\begin{align}
\hat{H}_0 & = \sum_{\bm{R}\bm{R}',\alpha\beta} t_{\alpha\beta,\bm{R}-\bm{R}'} \mathrm{e}^{\mathrm{i}\bm{A}\cdot (\bm{R}+\bm{r}_\alpha - \bm{R}'-\bm{r}_\beta)}\hat{c}^\dagger_{\bm{R}\alpha}  \hat{c}_{\bm{R}'\beta} 
\end{align}
where the flat connection is introduced by Periels substitution. The momentum space eigenbasis is still defined by $\hat{c}_{\bm{k}\alpha}\equiv\sum_{\bm{R}}\mathrm{e}^{\mathrm{i} \bm{k}\cdot (\bm{R}+\bm{r}_\alpha)} \hat{c}_{\bm{R},\alpha}/\sqrt{\mathsf{V}}$, so the hopping matrix is modified to
\begin{align}
    t^\sigma_{\alpha\beta} (\bm{k};\bm{A}) &\equiv\sum_{\bm{R}}\mathrm{e}^{\mathrm{i} (\bm{k}+\bm{A})\cdot (\bm{R}+\bm{r}_\alpha-\bm{r}_\beta)} t_{\alpha\beta,\bm{R}} \\
    &= t^\sigma_{\alpha\beta} (\bm{k}+\bm{A};\bm{A}=0)
\end{align}
and thus the diagonalized Hamiltonian reads
\begin{align}
\hat{H}_0 &= \sum_{\bm{k}, n} \epsilon_n(\bm{k}+\bm{A}) \hat{\gamma}^\dagger_{\bm{k}n} \hat{\gamma}_{\bm{k}n}, \\
\hat{\gamma}_{\bm{k}n} &=   U_{n\alpha}^{\dagger}(\bm{k}+\bm{A}) \hat{c}_{\bm{k}\alpha}.
\end{align}
Similarly we have the substituted results for $\hat{S}_{\bm{R}}$
\begin{align}
    \hat{S}_{\bm{R}} &= \sum_{\mu\nu,\bm{R}_1,\bm{R}_2} S_{\mu\nu}(\bm{R}_1,\bm{R}_2) \mathrm{e}^{\mathrm{i}\bm{A}\cdot (\bm{R}_1+\bm{r}_\mu - \bm{R}_2-\bm{r}_\nu)} \hat{c}^{\dagger}_{\bm{R}+\bm{R}_1,\mu} \hat{c}_{\bm{R}+\bm{R}_2,\nu} \\
    &= \frac{1}{\mathsf{V}}\sum_{\mu\nu,\bm{p}\bm{q}}  \mathrm{e}^{\mathrm{i}\bm{R}\cdot (\bm{p}-\bm{q})}S_{\mu\nu}(\bm{p}+\bm{A},\bm{q}+\bm{A}) \hat{c}^{\dagger}_{\bm{p},\mu} \hat{c}_{\bm{q},\nu} 
\end{align}

This means that the model Hamiltonian in the presence of $\bm{A}$ can be simply obtained as
\begin{align}
    \hat{H} = \frac{1}{\mathsf{V}}  \sum_{\substack{\bm{p}\bm{q},\bm{p}'\bm{q}' \\ nmkl\leq \mathsf{N}_\text{flat} }} V_{nmkl}(\bm{p}+\bm{A},\bm{q}+\bm{A},\bm{p}'+\bm{A},\bm{q}'+\bm{A}) \hat{\gamma}^{\dagger}_{\bm{p},n}\hat{\gamma}_{\bm{q},m}   \hat{\gamma}^{\dagger}_{\bm{p}',k} \hat{\gamma}_{\bm{q}',l}
\end{align}
with the $V$ defined in Eq.~\ref{eq: V}. 

Now using the expression for 2RDM, we evaluate the expectation value of the energy on the BCS ansatz state (only keeping the extensive part of the energy):
\begin{align}
    E_\text{var}(\bm{A}) =&  \frac{1}{\mathsf{V}}  \sum_{\bm{p},\bm{q} } \left[E_{\text{Hartree}}(\bm{p},\bm{q})+ E_{\text{Fock}}(\bm{p},\bm{q})+ E_{\text{Cooper}}(\bm{p},\bm{q}) \right] \\
    E_{\text{Hartree}}(\bm{p},\bm{q}) =  &   
    V(0) \sum_{n=1,m=1}^\infty  \Tr\left\{\left[ - e^{-2\Re z} FF^{\dagger}(\bm{p}) \right]^n B^\dagger(\bm{p}+\bm{A},\bm{p}+\bm{A}) \right\}  \cdot \Tr \left\{B(\bm{q}+\bm{A},\bm{q}+\bm{A}) \left[ - e^{-2\Re z} FF^{\dagger}(\bm{q}) \right]^m \right\}    \\
    E_{\text{Fock}}(\bm{p},\bm{q}) = & -  V(\bm{q}-\bm{p}) \sum_{n=1,m=0}^\infty  \Tr \left\{\left[ - e^{-2\Re z} FF^{\dagger}(\bm{p}) \right]^n    B^\dagger(\bm{q}+\bm{A},\bm{p}+\bm{A})  \left[ - e^{-2\Re z} FF^{\dagger}(\bm{q}) \right]^m  B(\bm{q}+\bm{A},\bm{p}+\bm{A})  \right\} \\
    E_{\text{Cooper}}(\bm{p},\bm{q}) = & V(\bm{q}-\bm{p}) \sum_{n=0,m=0}^\infty  e^{-2\Re z} \Tr\left\{  F^{\dagger}(\bm{p})  \left[ - e^{-2\Re z} FF^{\dagger}(\bm{p}) \right]^n  B^\dagger(\bm{q}+\bm{A},\bm{p}+\bm{A}) \right. \nonumber\\
     & \ \ \ \ \ \ \ \ \ \  \ \ \ \ \ \ \ \ \ \ \ \ \ \ \  \ \ \ \ \ \ \ \ \ \ \ \ \ \ \left. \left[ - e^{-2\Re z} FF^{\dagger}(\bm{q}) \right]^m  F(\bm{q})  B^T(-\bm{p}+\bm{A}, -\bm{q}+\bm{A}) \right\}
\end{align}
where $B$ is the flat-band block of $S$ (see Eq.~\ref{eq: blocks}).

These terms can be expressed in a succinct form:
\begin{align}
    E_{\text{Hartree}}(\bm{p},\bm{q}) =  &   
    V(0) \Tr\left[\Xi^\dagger(\bm{p})\right] \Tr\left[
    \Xi(\bm{q})\right] \\
    \Xi(\bm{q})\equiv &B(\bm{q}+\bm{A},\bm{q}+\bm{A})\nu_R(\bm{q})
\end{align}
where we have made matrix generalization to the filling factor:
\begin{align}
    \nu_R(\bm{q}) \equiv e^{-2\Re z} FF^{\dagger}(\bm{q}) \left[\mathbbm{1}+ e^{-2\Re z} FF^{\dagger}(\bm{q}) \right]^{-1} \\ 
    \nu_L(\bm{q}) \equiv e^{-2\Re z} F^\dagger F(\bm{q}) \left[\mathbbm{1}+ e^{-2\Re z} F^\dagger F(\bm{q}) \right]^{-1}
\end{align}
and the Fock and Cooper term can be combined as:
\begin{align}
    E_{\text{Fock}+\text{Cooper}}(\bm{p},\bm{q}) = &   V(\bm{q}-\bm{p}) \sum_{n=0,m=0}^\infty   e^{-2\Re z} \Tr \left\{F^{\dagger}(\bm{p})\left[ - e^{-2\Re z} FF^{\dagger}(\bm{p}) \right]^n    B^\dagger(\bm{q}+\bm{A},\bm{p}+\bm{A})  \left[ - e^{-2\Re z} FF^{\dagger}(\bm{q}) \right]^m  \right. \nonumber\\
&\ \ \ \ \ \ \ \ \ \ \ \ \ \ \ \ \ \ \ \ \ \ \ \ \ \ \ \ \ \ \ \ \ \ \ \ \ \ \ \ \ \ \ \ \ \ \ \ \ \ \ \ \ \ \ \left. \left[B(\bm{q}+\bm{A},\bm{p}+\bm{A}) F(\bm{p}) +F(\bm{q})  B^T(-\bm{p}+\bm{A}, -\bm{q}+\bm{A})\right] \right\}  \\
= &  \frac{1}{2} V(\bm{q}-\bm{p}) \sum_{n=0,m=0}^\infty   e^{-2\Re z} \Tr \left\{\left[ - e^{-2\Re z} F^{\dagger}F(\bm{p}) \right]^n \left[B(\bm{q}+\bm{A},\bm{p}+\bm{A}) F(\bm{p}) +F(\bm{q})  B^T(-\bm{p}+\bm{A}, -\bm{q}+\bm{A})\right]^\dagger  \right. \nonumber\\
&\ \ \ \ \ \ \ \ \ \ \ \ \ \ \ \ \ \ \ \ \ \ \ \ \ \ \ \ \ \ \ \ \ \ \ \ \ \left. \left[ - e^{-2\Re z} FF^{\dagger}(\bm{q}) \right]^m  
\left[B(\bm{q}+\bm{A},\bm{p}+\bm{A}) F(\bm{p}) +F(\bm{q})  B^T(-\bm{p}+\bm{A}, -\bm{q}+\bm{A})\right] \right\}  \\
=  &\frac{1}{2} V(\bm{q}-\bm{p})  \Tr \left[\Upsilon(\bm{p},\bm{q})^\dagger\Upsilon(\bm{p},\bm{q}) \right] \\
\Upsilon(\bm{p},\bm{q}) \equiv &e^{-\Re z}\left(\mathbbm{1}-\nu_R(\bm{q})\right)^{1/2}\left[B(\bm{q}+\bm{A},\bm{p}+\bm{A}) F(\bm{p}) +F(\bm{q})  B^T(-\bm{p}+\bm{A}, -\bm{q}+\bm{A})\right] \left(\mathbbm{1}-\nu_L(\bm{p})\right)^{1/2} 
\end{align}
where in the second line we recombined the values at $(\bm{p},\bm{q})$ and $(-\bm{q},-\bm{p})$.

As a sanity check, let's plug in $F=F_0$ for the $\bm{A} =0$ case. In this case the expressions can be greatly simplified 
\begin{align}
    \Tr[\Xi_{0}(\bm{q})] =& \sum_{m=1}^\infty \Tr \left\{A(\bm{q},-\bm{q})\left[ - e^{-2\Re z} F_0F_0^{\dagger}(\bm{q}) \right]^m F_0(\bm{q}) \right\} \nonumber\\
    =& -\sum_{m=1}^\infty \Tr \left\{ A(-\bm{q},\bm{q})  F_0(-\bm{q}) \left[ - e^{-2\Re z} F_0^{\dagger}F_0(-\bm{q}) \right]^m  \right\} = -\Tr[\Xi_{0}(-\bm{q})]\\
    \Upsilon_0(\bm{p},\bm{q}) \equiv &e^{-\Re z}\left(\mathbbm{1}-\nu_{0,R}(\bm{q})\right)^{1/2}\left[A(\bm{q})A(\bm{q},-\bm{p}) F(\bm{p}) +F_0(\bm{q})  A^T(-\bm{p}, \bm{q}) F_0^T(-\bm{p})\right] \left(\mathbbm{1}-\nu_{0,L}(\bm{p})\right)^{1/2} =0\label{eq: Upsilon0}
\end{align}
Note that we have used $F_0$ for the exact groundstate's form factor at $\bm{A}=0$ (which is a part of the Hamiltonian). These equalities imply that the total energy of this state indeed sums to zero.

\subsection{Perturbative treatment}

We now evaluate and optimize the variation energy to the second order in an expansion in small $\bm{A}$. We parametrize
\begin{align}
    F(\bm{k}) =& F_0(\bm{k}) + \delta F(\bm{k}) \cdot  \bm{A} + \dots \\
    B(\bm{p}+\bm{A},\bm{q}+\bm{A}) =& B(\bm{p},\bm{q}) + \partial B(\bm{p},\bm{q}) \cdot \bm{A}  + \dots\\
    E_\text{var}=&  \bm{A}\cdot \delta^2 E_\text{var} \cdot \bm{A} +\dots
\end{align}
such that
\begin{align}
    \kappa_\text{var} = \lim_{\mathsf{V}\rightarrow\infty}\frac{1}{2\mathsf{V}} \delta^2 E_\text{var}.
\end{align}
Note that $E$ should not have linear $\bm{A}$ dependence due to gauge invariance. The antisymmetric nature of form factor ensures that
\begin{align}
\delta F(\bm{k}) = -\delta F^T(-\bm{k}).
\end{align}

The goal is to evaluate 
\begin{align}
    \delta^2 E_\text{var}=&  \frac{1}{\mathsf{V}}  \sum_{\bm{p},\bm{q} } \left[\delta^2 E_{\text{Hartree}}(\bm{p},\bm{q})+ \delta^2 E_{\text{Fock}}(\bm{p},\bm{q})+ \delta^2 E_{\text{Cooper}}(\bm{p},\bm{q}) \right]
\end{align}
since it will directly upper bound the stiffness.

We first evaluate the Hartree energy. We note two useful facts: 1. there should be no terms that have $\delta$ in only one of the traces, since all these terms always trivially vanish after momentum summation. 2. There are two identities:
\begin{align}
    \Tr\left\{[F_0F_0^\dagger(\bm{q}) ]^n \delta F (\bm{q}) F_0^\dagger (\bm{q})[F_0F_0^\dagger(\bm{q}) ]^mB(\bm{q},\bm{q})\right\} &= - \Tr\left\{[F_0F_0^\dagger(-\bm{q})]^{m+1} \delta F (-\bm{q}) F_0^\dagger (-\bm{q})[F_0F_0^\dagger(-\bm{q})]^{n-1} B(-\bm{q},-\bm{q})\right\} \\
     \Tr\left\{[F_0F_0^\dagger(\bm{q}) ]^n  F_0 (\bm{q}) \delta F^\dagger (\bm{q})[F_0F_0^\dagger(\bm{q}) ]^mB(\bm{q},\bm{q})\right\} &= - \Tr\left\{[F_0F_0^\dagger(-\bm{q})]^{m} F_0 (-\bm{q}) \delta  F^\dagger (-\bm{q})[F_0F_0^\dagger(-\bm{q})]^{n} B(-\bm{q},-\bm{q})\right\} 
\end{align}
which guarantee that most terms that contain $\delta F$ dependence will be canceled by their time-reversal partner terms. Then we reach the result:
\begin{align}
     \delta^2E_{\text{Hartree}}(\bm{p},\bm{q}) =  &   
    V(0)  \Tr\left[ \delta \Xi(\bm{p})^\dagger\right] \Tr\left[\delta \Xi(\bm{q})\right] \\
    \delta \Xi(\bm{q}) \equiv & -\sum_{m=1}^\infty \partial B(\bm{q},\bm{q}) \nu_{0,R}(\bm{q}) - e^{-2\Re z}  B(\bm{q},\bm{q})  \delta F(\bm{q})F_0^{\dagger}(\bm{q})\left[\mathbbm{1}-\nu_{0,R}(\bm{q})\right]
\end{align}

We next treat the Fock and Cooper parts together. Note that $\Upsilon(\bm{p},\bm{q})$ start with linear in $\bm{A}$ term (cf. Eq.~\ref{eq: Upsilon0}); therefore, we only need to keep the leading term therein:
\begin{align}
\delta^2 E_{\text{Fock}+\text{Cooper}}  (\bm{p},\bm{q}) = &\frac{1}{2} V(\bm{q}-\bm{p})  \Tr \left[\delta \Upsilon(\bm{p},\bm{q})^\dagger \delta\Upsilon(\bm{p},\bm{q}) \right] \\
\delta\Upsilon(\bm{p},\bm{q}) \equiv &e^{-\Re z}\left(\mathbbm{1}-\nu_{0,R}(\bm{q})\right)^{1/2}\left[\partial B(\bm{q},\bm{p}) F_0(\bm{p}) +\delta F(\bm{q})  B^T(-\bm{p}, -\bm{q})+B(\bm{q},\bm{p})\delta F(\bm{p})\right.\nonumber\\
& \ \ \ \ \ \ \ \ \ \ \ \ \ \ \ \ \ \ \ \ \ \  \ \ \ \ \ \ \ \ \ \ \ \ \ \ \ \ \ \ \ \ \ \ \ \ \ \ \ \  \ \ \ \ \ \ \ \ \ \ \ \left.+F_0(\bm{q})  \partial B^T(-\bm{p}, -\bm{q})\right] \left(\mathbbm{1}-\nu_{0,L}(\bm{p})\right)^{1/2} 
\end{align}

\subsection{Simplifications for the simple constructions in inversion symmetric system}\label{sec: stiffness inversion}

Here we discuss the case we constructed in Sec.~\ref{sec: construction: inversion symmetric systems}. The expression for the variational energy response reads:
\begin{align}
        \delta^2 E_\text{var}=&  \frac{1}{\mathsf{V}}  \sum_{\bm{p},\bm{q} } \left[\delta^2 E_{\text{Hartree}}(\bm{p},\bm{q})+ \delta^2 E_{\text{Fock}}(\bm{p},\bm{q})+ \delta^2 E_{\text{Cooper}}(\bm{p},\bm{q}) \right] \\
         \delta^2 E_{\text{Hartree}}(\bm{p},\bm{q}) = & V \nu_0(\bm{p}) \nu_0(\bm{q}) [u^*_Au_B(\bm{p})]^*[u^*_Au_B(\bm{q})] \nonumber\\
        & \ \ \ \ \ \ \ \ \ \ \ \ \ \left[\partial \ln u^*_Au_B(\bm{p}) + \delta \ln f(\bm{p})\right]^* \left[\partial \ln u^*_Au_B(\bm{q}) + \delta \ln f(\bm{q})\right] \\
        \delta^2 E_{\text{Fock}+\text{Cooper}}  (\bm{p},\bm{q}) = & \frac{V}{2} (1-\nu_0(\bm{q}))(1-\nu_0(\bm{p})) |u_A(\bm{p})|^2 |u_A(\bm{q})|^2 e^{-2 \Re z} \nonumber\\
        & \ \ \ \ \ \ \ \ \ \ \ \ \ \left\{\left| \partial \ln u^*_A u_B(\bm{p})  +\partial \ln u^*_A u_B(\bm{q})+ \delta \ln f(\bm{p})- \delta \ln f(\bm{q}) \right|^2\right\} 
\end{align}
where we have defined short-handed notations $\partial \ln * = \frac{\partial * }{ *}$ and $\delta \ln f \equiv \frac{\delta f}{f_0}$, and it should be remembered that
\begin{align}
    \nu_0(\bm{k}) =& e^{-2 \Re z} |f_0(\bm{k})|^2 / [1+ e^{-2 \Re z}|f_0(\bm{k})|^2] \\
    f_0(\bm{k}) = & u^*_A(\bm{k}) /u_B(-\bm{k})
\end{align}

Before proceeding, let's recap the parity of the terms in the above expression: $u_A^*u_B(\bm{k})$ and $\partial \ln u_A^*u_B(\bm{k})$ are odd in $\bm{k}$, whereas $\delta \ln f(\bm{k})$ and all the other squared terms are even in $\bm{k}$. This implies that the Hartree energy cannot by optimized by varying $f$:
\begin{align}
     \delta^2 E_{\text{Hartree}}(\bm{p},\bm{q}) = & V \nu_0(\bm{p}) \left[\partial  u^*_Au_B(\bm{p})\right]^*  \nu_0(\bm{q})  \left[\partial u^*_Au_B(\bm{q})\right]
\end{align}
On the other hand, we can symmetrize over $(\bm{p},\bm{q})$ such that 
\begin{align}
      \delta^2 E_{\text{Fock}+\text{Cooper}}  (\bm{p},\bm{q})  = & \frac{V}{2} (1-\nu_0(\bm{q}))(1-\nu_0(\bm{p})) |u_A(\bm{p})|^2 |u_A(\bm{q})|^2 e^{-2 \Re z} \nonumber\\
        & \ \ \ \ \ \ \ \ \ \ \ \ \ \left\{\left| \partial \ln u^*_A u_B(\bm{p})  +\partial \ln u^*_A u_B(\bm{q}) \right|^2 +  \left|\delta \ln f(\bm{p})- \delta \ln f(\bm{q}) \right|^2\right\}
\end{align}
So that we see that $\delta f =0$ is already the optimal choice, this yields: 
\begin{align}
      \delta^2 E_{\text{Fock}+\text{Cooper}}    = & \frac{V}{2\mathsf{V}}\sum_{\bm{p},\bm{q}} (1-\nu_0(\bm{q}))(1-\nu_0(\bm{p})) |u_A(\bm{p})|^2 |u_A(\bm{q})|^2 e^{-2 \Re z} \left[\left| \partial \ln u^*_A u_B(\bm{p})\right|^2  + \left|\partial \ln u^*_A u_B(\bm{q}) \right|^2 \right] \\
       = &   \frac{V}{\mathsf{V}}\sum_{\bm{k}} (1-\nu_0(\bm{k})) |u_A(\bm{k})|^2 \left| \partial \ln u^*_A u_B(\bm{p})\right|^2  \cdot \sum_{\bm{k}} (1-\nu_0(\bm{k})) |u_A(\bm{k})|^2 e^{-2 \Re z}\\
       = & \frac{V}{\mathsf{V}}\sum_{\bm{k}} \nu_0(\bm{k}) \frac{\left| \partial u^*_A u_B(\bm{p})\right|^2}{|u_A(\bm{k})|^2}   \cdot \sum_{\bm{k}} [1-\nu_0(\bm{k})] |u_A(\bm{k})|^2 
\end{align}

Putting these together, we reach a simple result for stiffness upper bound:
\begin{align}
    \kappa_\text{var} = \frac{V}{2\mathsf{V}^2} \left\{\left|\sum_{\bm{k}} \nu_0(\bm{k}) \partial[u^*_A u_B(\bm{k})]\right|^2+
    \sum_{\bm{k} }  \nu_0(\bm{k})    \frac{\left|\partial \left[u^*_A u_B(\bm{k})\right]\right|^2 }{ |u_A(\bm{k})|^2 }   \sum_{\bm{k}} [1-\nu_0(\bm{k})] |u_A(\bm{k})|^2 \right\}
\end{align}
All the terms here only depend on the $P(\bm{k})$ projection matrix so is manifestly gauge invariant. We fixed the embedding to be the periodic embedding so it is also embedding-independent. It is simple to check that this expression is invariant under $z\leftrightarrow -z$, $A\leftrightarrow B$, which confirms the p-h duality between the two almost degenerate SC phases.

This bound is nothing but the familiar diamagnetic response:
\begin{align}
    \langle \partial^2 \hat{H}(\bm{A})\rangle_{\text{GS}}
\end{align}
The non-trivial thing here is that we have proven that this is already the optimal bound in the variational space with a BCS-type states. This should be contrasted with the case of quantum geometric nesting models where this does not produce the tightest bound within this ansatz space. 

In the dilute limit $\Re z\rightarrow \infty$, we can simplify the stiffness upper bound:
\begin{align}
    \kappa_\text{var} (\Re z\rightarrow +\infty)= \frac{V}{2\mathsf{V}^2} 
    \sum_{\bm{k} } \frac{e^{-2\Re z}\left|\partial \left[u^*_A u_B(\bm{k})\right]\right|^2}{|u_B(\bm{k})|^2}   \cdot  \sum_{\bm{k}} |u_A(\bm{k})|^2 
\end{align}
and the filling factor:
\begin{align}
    \nu(\Re z\rightarrow +\infty) = \frac{e^{-2\Re z}}{\mathsf{V}} \sum_{\bm{k}} \frac{|u_A(\bm{k})|^2}{|u_B(\bm{k})|^2}
\end{align}
Comparing with the exact result for two-particle bound state mass in Eq.~\ref{eq: two particle mass}, we find a simple relation:
\begin{align}
        \kappa_\text{var} (\Re z\rightarrow +\infty) = \frac{1}{2} \nu(\Re z\rightarrow +\infty)  \rho^\text{FB}_A [m_{2\text{p}}^{-1}]_{ij}
\end{align}
Applying the particle-hole duality, we find that the dilute-hole limit result can be related to the exact two-hole bound state mass in Eq.~\ref{eq: two hole mass}
\begin{align}
        \kappa_\text{var} (\Re z\rightarrow -\infty) = \frac{1}{2} [1-\nu(\Re z\rightarrow -\infty)]  \rho^\text{FB}_B [m_{2\text{h}}^{-1}]_{ij}
\end{align}

\subsection{Simplification for quantum geometric nesting models}

To compare the solvable model constructed in this work with the previous ones, here we discuss the results for the quantum geometric nesting (QGN) models~\cite{PhysRevX.14.041004}. These models have two key features which will lead to simplifications to the stiffness upper bound calculation: 1) the $\hat{S}_{\bm{R}}$ operators are hermitian, which implies that
\begin{align}
    B(\bm{p},\bm{q}) = B^\dagger(\bm{q},\bm{p})
\end{align}
and 2) the singular value spectrum of the ideal pairing form factor is degenerate and is the same for all momentum, which implies
\begin{align}
    F_0(\bm{k})F_0^\dagger(\bm{k}) = F_0^\dagger(\bm{k})F_0(\bm{k}) = |f_0|^2 \mathbbm{1} \implies \nu_{0,R}(\bm{k}) = \nu_{0,L}(\bm{k}) = |f_0|^2 e^{-2\Re z} /(1+e^{-2\Re z}|f_0|^2) \equiv \nu_{0}
\end{align}

The most important implication of these two features is that the perturbative Hartree energy now vanishes. To see this, we decompose
\begin{align}
    \delta \Xi(\bm{q}) \equiv & \delta \Xi_1(\bm{q})+ \delta \Xi_2(\bm{q})\\
   \delta \Xi_1(\bm{q}) \equiv & -\sum_{m=1}^\infty \partial B(\bm{q},\bm{q}) \nu_{0} \\
   \delta \Xi_2(\bm{q}) \equiv & - e^{-2\Re z}  B(\bm{q},\bm{q})  \delta F(\bm{q})F_0^{\dagger}\left[\mathbbm{1}-\nu_{0}\right]
\end{align}
Since $\delta \Xi_1(\bm{q})$ is a total derivative, it sums to zero in momentum space. On the other hand, 
\begin{align}
   \Tr[ \delta \Xi_2(\bm{q})] =& - e^{-2\Re z}  \Tr\left\{ B(\bm{q},\bm{q})  \delta F(\bm{q})F_0^{\dagger}\left[\mathbbm{1}-\nu_{0}\right]\right\} \\
    =&-e^{-2\Re z}  \Tr\left\{ B^\dagger(\bm{q},\bm{q})  \delta F(\bm{q})\left[\mathbbm{1}-\nu_{0}\right]F_0^{\dagger} \right\} \\
    =&e^{-2\Re z}  \Tr\left\{  B^*(-\bm{q},-\bm{q}) F_0^{*} \delta F^T(-\bm{q})\left[\mathbbm{1}-\nu_{0}\right] \right\} \\
    =&e^{-2\Re z}  \Tr\left\{    \left[\mathbbm{1}-\nu_{0}\right] \delta F(-\bm{q}) F_0^{\dagger}  B^\dagger(-\bm{q},-\bm{q})\right\}  \\
    =&e^{-2\Re z}  \Tr\left\{B(-\bm{q},-\bm{q})  \delta F(-\bm{q}) F_0^{\dagger}  \left[\mathbbm{1}-\nu_{0}\right]  \right\} = - \Tr[ \delta \Xi_2(-\bm{q})]
\end{align}
So it also sums to zero when evaluating $\delta^2 E_{\text{Hartree}}$. Thus we conclude
\begin{align}
    \delta^2 E_{\text{Hartree}} = 0
\end{align}
for QGN models, which is very different from the models constructed in this work.

Therefore, the energy only comes from the Fock + Cooper terms. Now that the $\nu$'s are now simply numbers, they can be pulled out of the trace in their expression
\begin{align}
\delta^2 E_{\text{Fock}+\text{Cooper}}  (\bm{p},\bm{q}) = &\frac{e^{-2\Re z} (1-\nu_0)^2}{2} V(\bm{q}-\bm{p})  \Tr \left[\delta \Upsilon'^\dagger(\bm{p},\bm{q}) \delta\Upsilon'(\bm{p},\bm{q}) \right] \\
\delta\Upsilon'(\bm{p},\bm{q}) \equiv &\left[\partial B(\bm{q},\bm{p}) F_0(\bm{p}) +\delta F(\bm{q})  B^T(-\bm{p}, -\bm{q})+B(\bm{q},\bm{p})\delta F(\bm{p}) +F_0(\bm{q})  \partial B^T(-\bm{p}, -\bm{q})\right] 
\end{align}
At an abstract level, if we regard the $\delta F$ on different momentum as a huge vector, the variational energy, before optimizing over all possible values of $\delta F$,  takes the form:
\begin{align}
    \delta^2 E_\text{var}[\vec{\delta F}]  \sim \left[\vec{\delta F}^\dagger \cdot M \cdot \vec{\delta F} + \vec{W}^\dagger \cdot \vec{\delta F}+  \vec{\delta F}^\dagger\cdot \vec{W} + D\right]
\end{align}
where $M$ is a matrix, $\vec{W}\propto f$ is a vector, and $D\propto |f|^2$ is a constant. This is sufficient for us to infer that the optimized value in the parenthesis must be proportional to $|f|^2$. To be concrete, we define
\begin{align}
    \varepsilon \equiv \frac{1}{\mathsf{V}|f_0|^2}\min_{\delta F} \frac{1}{2\mathsf{V}}\sum_{\bm{p}\bm{q}} V(\bm{q}-\bm{p})  \Tr \left[\delta \Upsilon'^\dagger(\bm{p},\bm{q}) \delta\Upsilon'(\bm{p},\bm{q}) \right] 
\end{align}

Therefore, the optimized variational energy should be equal to:
\begin{align}
    \delta^2 E_\text{var} = \mathsf{V}\varepsilon (1-\nu_0)^2 e^{-2\Re z}  |f_0|^2 = \mathsf{V}\varepsilon (1-\nu_0)\nu_0  
\end{align}

The interesting point is that $\varepsilon$ is actually proportional to the inverse of the two-particle mass, $m_2^{-1}$. To see this, we use the 2RDM result for projected BCS in Eq.~\ref{eq: two particle sector} to evaluate the energy as
\begin{align}
   \delta^2E_{N=1} = \frac{1}{\mathsf{N}_0|f_0|^2}\sum_{\bm{p}\bm{q}} \frac{1}{2\mathsf{V}}V(\bm{q}-\bm{p})  \Tr \left[\delta \Upsilon'^\dagger(\bm{p},\bm{q}) \delta\Upsilon'(\bm{p},\bm{q}) \right] 
\end{align}
We skip the details of the derivation of this expression, which is almost identical to the case of BCS states, but would like to point out some key steps: 1) the Hartree terms trivially vanish in this sector. 2) in principle, the normalization factor is $|f_\text{sq}|^2$; however, since each $|F(\bm{q})|^2$ dependence in this quantity is $\propto 1/\mathsf{V}$, all the corrections due to $\delta F(\bm{q}) \cdot \bm{A}$ could be dropped. Then the optimized energy in this sector is 
\begin{align}
    \delta^2E_{N=1}  = \frac{2}{\mathsf{N}_\text{flat}}\varepsilon
\end{align}
Since this variational ansatz already include all the states within this number and total momentum sector, this result is not only variational but actually {\it exact}: $E_\text{var}=E_\text{actual}$; this means that in the $N=1$ sector,
\begin{align}
    \frac{2}{\mathsf{N}_\text{flat}}\varepsilon = \delta^2 E_\text{var} = \delta^2 E_\text{actual} \equiv (m_2)^{-1}.
\end{align}
We thus reached a simple conclusion
\begin{align}
    \kappa_\text{var} \equiv \frac{1}{4\mathsf{V}}\delta^2 E_\text{var} = \frac{\mathsf{N}_\text{flat}}{8} (m_2)^{-1} (1-\nu_0)\nu_0  \succeq \kappa 
\end{align}

To see an explicit calculation within this class of models, we now compute the stiffness upper bound for the attractive Hubbard model with uniform pairing condition ($P_{\alpha\alpha} (\bm{k}) = \epsilon$ for any $\alpha$, $\bm{k}$). It is defined by $V(\bm{k})=|U|$ and a set of $\hat{S}^{(\alpha)}_{\bm{R}}$ operators with:
\begin{align}
B(\bm{p},\bm{q})_{nm;\sigma\sigma'} & = \sigma^z_{\sigma\sigma'} U^{\sigma, \dagger}_{n\mu} (\bm{p}) D^{(\alpha)}_{\mu\nu} U^{\sigma'}_{\nu m} (\bm{q}) = \sigma^z \otimes U^{\sigma,\dagger}(\bm{p}) D^{(\alpha)} U^{\sigma}(\bm{q}) \\
\partial B(\bm{p},\bm{q})_{nm;\sigma\sigma'} &= \sigma^z \otimes \left[\partial U^{\sigma,\dagger}(\bm{p}) D^{(\alpha)} U^{\sigma}(\bm{q})+U^{\sigma,\dagger}(\bm{p}) D^{(\alpha)} \partial U^{\sigma}(\bm{q})\right]
\end{align}
where $D_{\mu\nu}^{(\alpha)} = \delta_{\mu \alpha}\delta_{\nu \alpha}  $. The pairing form factor is:
\begin{align}
    F_{nm;\sigma\sigma'} = \sigma^y_{\sigma \sigma'} \otimes \mathbbm{1}_{nm}
\end{align}
such that $\nu_0(\bm{k}) = \frac{e^{-2\Re z} }{1+ e^{-2\Re z}}$ for any mode. 

For simplicity and clarity of the following derivation, we will consider case with a single spinful flat band with spin symmetry $u^\uparrow =u^\downarrow = u$ (the general case should be easily recoverable). Then the variation of $F(\bm{k})$ should be parametrized as $\sigma^y \delta f(\bm{k})$ where $ \delta f$ is a scalar function that is even in $\bm{k}$. This is a key difference from the spinless case. Then the energy is 
\begin{align}
\delta^2 E_\text{var} =& \sum_{\bm{p}\bm{q}} 
\delta^2 E_{\text{Fock}+\text{Cooper}}  (\bm{p},\bm{q})\\
\delta^2 E_{\text{Fock}+\text{Cooper}}  (\bm{p},\bm{q}) = &\frac{ (1-\nu_0)\nu_0}{2} |U| \sum_{\alpha} \Tr \left[\delta \Upsilon^{(\alpha),\dagger}(\bm{p},\bm{q}) \delta\Upsilon^{(\alpha)}(\bm{p},\bm{q}) \right] \\
\delta\Upsilon^{(\alpha)}(\bm{p},\bm{q}) \equiv & \sigma^x \left\{2 \left[\partial u^{*}_\alpha(\bm{q}) u_\alpha (\bm{p})+u^{*}_\alpha(\bm{q}) \partial u_\alpha(\bm{p})\right]  + \left[\delta f(\bm{p}) -\delta f(\bm{q})\right]u^{*}_\alpha(\bm{q})  u_\alpha(\bm{p}) \right\}
\end{align}

Varying $\delta f^*$, we find that the optimal energy can be obtained by taking
\begin{align}
    \delta f(\bm{k}) =  \left[  \partial u^{\dagger} \cdot u -   u^{\dagger} \cdot  \partial u  \right](\bm{k})
\end{align}
Note that this is a key difference from the case discussed in the last subsection: here the optimal $\delta f$ is not $0$!

Plugging this optimal choice back in yields the optimal energy:
\begin{align}
    \delta^2 E_\text{var} =&  (1-\nu_0)\nu_0 |U| \sum_{\alpha,\bm{p}\bm{q}} \left\{2 \left[\partial u^{*}_\alpha(\bm{q}) u_\alpha (\bm{p})+u^{*}_\alpha(\bm{q}) \partial u_\alpha(\bm{p})\right]  \right.\nonumber\\
&\ \ \ \ \ \ \ \ \ \ \ \ \ \ \ \ \ \ \ \ \ \ \ \ \ \ \ \ \ \ \ \ \ \ \ \ \ \ \ \ \ \ \ \ \left. + \left[ \partial u^{\dagger} \cdot u(\bm{p}) -   u^{\dagger} \cdot  \partial u (\bm{p}) - \partial u^{\dagger} \cdot u(\bm{q}) +   u^{\dagger} \cdot  \partial u (\bm{q})\right]u^{*}_\alpha(\bm{q})  u_\alpha(\bm{p}) \right\}^2 
\end{align}
This is very complicated, by it can be greatly simplified by using the minimal embedding in which the quantum metric is minimized and~\cite{PhysRevB.106.014518}
\begin{align}
    \sum_{\bm{k}} u^*_\alpha \partial u_\alpha (\bm{k}) =  \sum_{\bm{k}}  \partial u^*_\alpha u_\alpha (\bm{k}) = 0
\end{align}
Using this identity,  we obtain:
\begin{align}
     \delta^2 E_\text{var}   =&  
8\mathsf{V} (1-\nu_0)\nu_0 |U| g \epsilon \implies \kappa_\text{var} = 4 \mathsf{V} (1-\nu_0)\nu_0 |U| g \epsilon
\end{align}
where $g \equiv \frac{1}{\mathsf{V}}\sum_{\bm{k}} \partial u^\dagger \cdot  (\mathbbm{1}-uu^\dagger)\cdot \partial u $ is the integrated minimal quantum metric. This recovers the mean field result obtained previously~\cite{PhysRevB.106.014518}.

\section{Upper bound on excitation spectrum}

Here we derive the energy of BdG quasiparticles in a BCS state, which have definite momentum but not definite charge. Therefore, these results should be viewed as upper bound on the excitation energy in the corresponding sector.

The state with an excitation on momentum $\bm{k}_0$ in the mode $i_0$ is given by (the answer will be the same if we use $\hat{\xi}_{\bm{k}_0,i_0}$)
\begin{align}
    |z;(\bm{k}_0,i_0)\rangle 
    \equiv &  \hat{\xi}_{\bm{k}_0,i_0}^\dagger|z\rangle
\end{align}
Its energy can be computed as
\begin{align}
    E_{i_0}(\bm{k}_0) &\equiv  \frac{\langle z| \hat{\xi}_{\bm{k}_0,i_0} \hat{H}\hat{\xi}_{\bm{k}_0,i_0}^\dagger |z\rangle}{\langle z| \hat{\xi}_{\bm{k}_0,i_0} \hat{\xi}_{\bm{k}_0,i_0}^\dagger |z\rangle} \\
    &=\frac{1}{\mathsf{V}}  \sum_{\substack{\bm{p}\bm{q},\bm{p}'\bm{q}' \\ nmkl\leq \mathsf{N}_\text{flat}}} \delta_{\bm{p}'-\bm{q}',\bm{q}-\bm{p}} V(\bm{q}-\bm{p})  S^*_{mn}(\bm{q},\bm{p}) S_{kl}(\bm{p}',\bm{q}') \frac{\langle z| \left[\hat{\xi}_{\bm{k}_0,i_0} ,\hat{\gamma}^{\dagger}_{\bm{p},n} \hat{\gamma}_{\bm{q},m}  \right] \left[\hat{\gamma}^{\dagger}_{\bm{p}',k} \hat{\gamma}_{\bm{q}',l} ,\hat{\xi}_{\bm{k}_0,i_0}^\dagger\right] |z\rangle}{\langle z| \hat{\xi}_{\bm{k}_0,i_0} \hat{\xi}_{\bm{k}_0,i_0}^\dagger |z\rangle} \\
    &=\frac{1}{\mathsf{V}}  \sum_{\substack{\bm{q} \\ j;nmkl\leq \mathsf{N}_\text{flat}}}  V(\bm{q}-\bm{k}_0)  S^*_{mn}(\bm{q},\bm{k}_0) S_{kl}(\bm{q},\bm{k}_0) v_{n}^{(i_0),*}(\bm{k}_0) v^{(j)}_m (\bm{q}) v^{(j),*}_k (\bm{q}) v_{l}^{(i_0)}(\bm{k}_0)   \frac{\langle z|  \hat{\xi}_{\bm{q},j}  \hat{\xi}^{\dagger}_{\bm{q},j}  |z\rangle}{\langle z| \hat{\xi}_{\bm{k}_0,i_0} \hat{\xi}_{\bm{k}_0,i_0}^\dagger |z\rangle} \\
    &=\frac{1}{\mathsf{V}}  \sum_{\substack{\bm{q} \\ j;nmkl\leq \mathsf{N}_\text{flat}}}  V(\bm{q}-\bm{k}_0)  S^*_{mn}(\bm{q},\bm{k}_0) S_{kl}(\bm{q},\bm{k}_0) v_{n}^{(i_0),*}(\bm{k}_0) v^{(j)}_m (\bm{q}) v^{(j),*}_k (\bm{q}) v_{l}^{(i_0)}(\bm{k}_0)  \frac{1-\nu^{(j)}(\bm{q})}{1-\nu^{(i_0)}(\bm{k}_0)}
\end{align}
where the second equality is because that the elementary fermion bilinear operators $\hat{P}\hat{S}_{\bm{K}}\hat{P}$ annihilate the ground state.

This expression can be simplified for the case we constructed in Sec.~\ref{sec: construction: inversion symmetric systems}:
\begin{align}
    E(\bm{k}_0) 
    &=\frac{V}{\mathsf{V}}  \sum_{\bm{q} }  |u_A(\bm{q})|^2 |u_B(\bm{k}_0)|^2  \frac{1+e^{-2\Re z} |u_A(\bm{k}_0)|^2/|u_B(\bm{k}_0)|^2 }{1+e^{-2\Re z} |u_A(\bm{q})|^2/|u_B(\bm{q})|^2 } \\
    & = \frac{V}{\mathsf{V}}  \sum_{\bm{q} }     \frac{|u_B(\bm{k}_0)|^2+e^{-2\Re z} |u_A(\bm{k}_0)|^2 }{1/|u_A(\bm{q})|^2+e^{-2\Re z} /|u_B(\bm{q})|^2 }
\end{align}
Taking $e^{-\Re z}$ to $0$ or $\infty$ will reduce to the exact result in the single-electron or single-hole sector. It is simple to check that this expression is invariant under $z\leftrightarrow -z$, $A\leftrightarrow B$, which confirms the p-h duality between the two almost degenerate SC phases.

\section{Toy models}

\subsection{A two-band toy model example for chiral topological superconductor}

We showcase the power of this construction with a very simple two-orbital ($\alpha = \uparrow,\downarrow$) model defined by hopping matrix
\begin{align}\label{eq: two band}
    t(\bm{k})  &= \begin{bmatrix}
       -\frac{\Delta}{2}+  \frac{\bm{k}^{2} }{2m_\downarrow}& v_F \bar{k} \\
       v_F k &  \frac{\Delta}{2} +  \frac{\bm{k}^{2} }{2m_\uparrow}
    \end{bmatrix} = h_0(\bm{k}) \tau^0 +  \vec{h}(\bm{k}) \cdot \vec{\tau}\\
   h_0(\bm{k})& \equiv  \frac{(m_\uparrow+m_\downarrow)\bm{k}^{2} }{4 m_\uparrow m_\downarrow} \\
   \vec{h}(\bm{k}) &\equiv  \begin{pmatrix}
        v_F k_x \\ v_F k_y \\ \frac{\Delta}{2} + \frac{(m_\downarrow-m_\uparrow)\bm{k}^{2} }{4m_\uparrow m_\downarrow} 
   \end{pmatrix}
\end{align}
where $\tau^{0,x,y,z}$ are Pauli matrices, and $k \equiv k_x +\mathrm{i} k_y$ and $\bar{k}\equiv k_x -\mathrm{i} k_y $. When $\frac{\Delta(m_\uparrow-m_\downarrow) }{m_\uparrow m_\downarrow} > 0$, the bands are Chern bands; the upper/lower band has Chern number $\pm \text{sgn}(\Delta)$. Furthermore, we note that, when
\begin{align}
    v_F^2 = \frac{\Delta (m_\uparrow -m_\downarrow)}{2m_\uparrow m_\downarrow}
\end{align}
the bands admit quadratic dispersions 
\begin{align} 
    \epsilon_{1,2}(\bm{k}) =  \frac{(m_\uparrow+m_\downarrow)\bm{k}^{2} }{4 m_\uparrow m_\downarrow} \mp \left|\frac{\Delta}{2} + \frac{(m_\uparrow-m_\downarrow)\bm{k}^{2} }{4m_\uparrow m_\downarrow}\right|
\end{align}
and thus when $\Delta>0$ and $m_\uparrow \rightarrow \infty$, or $\Delta < 0$ and $m_\downarrow\rightarrow \infty$, we find the lower band is perfectly flat and is separated from the upper band by a gap $|\Delta|$. Note that in reaching this flat (lower) band limit, we tuned two parameters of $m_\uparrow,m_\downarrow$; the rest two parameters $v_F, \Delta$ then control the overall length and energy scales of the problem. For simplicity we will set $\Delta= v_F =1$ in the following discussion, without loss of generality. The model then simplifies to
\begin{align}\label{eq: two band}
    t(\bm{k})  &= \begin{bmatrix}
       - \frac{1}{2} +  \bm{k}^{2} & \bar{k} \\
       k &  \frac{1}{2}
    \end{bmatrix} = h_0(\bm{k}) \tau^0 +  \vec{h}(\bm{k}) \cdot \vec{\tau}\\
   h_0(\bm{k})& \equiv  \frac{\bm{k}^{2} }{2} \\
   \vec{h}(\bm{k}) &\equiv  \begin{pmatrix}
        k_x \\  k_y \\ \frac{1}{2} - \frac{\bm{k}^{2} }{2} 
   \end{pmatrix}
\end{align}
which has eigenvalues
\begin{align}
    \epsilon_1(\bm{k}) = & -\frac{1}{2}\\
    \epsilon_2(\bm{k}) = & \frac{1}{2} + \bm{k}^2
\end{align}
and eigenvectors:
\begin{align}
    U_{n=1}(\bm{k}) & = \begin{pmatrix}
        1 \\  -k
    \end{pmatrix}  /\sqrt{1+\bm{k}^2} \\
    U_{n=2}(\bm{k}) & = \begin{pmatrix}
        \bar{k} \\ 1
    \end{pmatrix}  /\sqrt{1+\bm{k}^2}
\end{align}
So the lower band is perfectly flat and has $C_\text{FB}=-1$.

Now applying the above general construction in the previous section with orbital identification $A=\uparrow, B=\downarrow$, we find the ideal interaction could be as simple as (cf. Eq.~\ref{eq: one flatband Hint2 lowest})
\begin{align}\label{eq: ideal int model 1}
    \hat{H}_\text{int} &=  V \sum_{\bm{R}} \left(  \hat{n}_{\bm{R}\downarrow} - \hat{n}_{\bm{R}\uparrow}\hat{n}_{\bm{R}\downarrow}\right)
\end{align}

The suggested pairing operator is 
\begin{align}
\hat{\eta}^{\dagger} = \frac{1}{\mathsf{V}}\sum_{\bm{k}} \bar{k} \ \hat{\gamma}^\dagger_{1,\bm{k}} \hat{\gamma}_{1,-\bm{k}}.
\end{align}
Mapping to a BdG wavefunction, we find 
\begin{align}
    w(\bm{k}) = \begin{pmatrix}
       - e^{-z} \bar{k}\\
       e^{-z} \bm{k}^2 \\
       1 \\
       \bar{k}
    \end{pmatrix} /\sqrt{1+\bm{k}^2} \sqrt{1+e^{-2\Re z}\bm{k}^2} 
\end{align}
which yields $C^\text{BdG} =0$ for any $\alpha$. This means that the superconductor has $C^\text{maj}=-1$ - a topological $p+\mathrm{i}p$ superconductor!

The particle-hole dual construction can be made with orbital identification $A=\downarrow, B=\uparrow$. The corresponding ideal interaction could be as simple as (cf. Eq.~\ref{eq: one flatband Hint2 lowest})
\begin{align}\label{eq: ideal int model 2}
    \hat{H}_\text{int} &=  - V \sum_{\bm{R}}   \hat{n}_{\bm{R}\uparrow}\hat{n}_{\bm{R}\downarrow}
\end{align}

The suggested pairing operator is 
\begin{align}
\hat{\eta}^{\dagger} = \frac{1}{\mathsf{V}}\sum_{\bm{k}} \frac{1}{k} \ \hat{\gamma}^\dagger_{1,\bm{k}} \hat{\gamma}_{1,-\bm{k}}.
\end{align}
Mapping to a BdG wavefunction, we find 
\begin{align}
    w(\bm{k}) = \begin{pmatrix}
        e^{-z} \\
        -e^{-z}k \\
       k \\
       \bm{k}^2
    \end{pmatrix} /\sqrt{1+\bm{k}^2} \sqrt{1+e^{-2\Re z}\bm{k}^2} 
\end{align}
which yields $C^\text{BdG} =0$ for any $\alpha$ and thus $C^\text{maj}=-1$. This means that the two p-h dual constructions, which only differ by some fermion bilinear terms, are in the same topological SC phase. 

It is straightforward to make a higher Chern number $C=-M$ generalization of this model:  
\begin{align}\label{eq: two band}
    t(\bm{k})  &= \begin{bmatrix}
       - \frac{1}{2} +  \bm{k}^{2M} & \bar{k}^M \\
       k^M &  \frac{1}{2}
    \end{bmatrix} 
\end{align}
which has eigenvalues
\begin{align}
    \epsilon_1(\bm{k}) = & -\frac{1}{2}\\
    \epsilon_2(\bm{k}) = & \frac{1}{2} + \bm{k}^{2M}
\end{align}
and eigenvectors:
\begin{align}
    U_{n=1}(\bm{k}) & = \begin{pmatrix}
        1 \\  -k^M
    \end{pmatrix}  /\sqrt{1+\bm{k}^{2M}} \\
    U_{n=2}(\bm{k}) & = \begin{pmatrix}
        \bar{k}^M \\ 1
    \end{pmatrix}  /\sqrt{1+\bm{k}^{2M}}
\end{align}
Now the lower band is still perfectly flat and has Chern number $C_\text{FB}=-M$. When $L$ is odd, the interactions in Eq.~\ref{eq: ideal int model 1}\&\ref{eq: ideal int model 2} are still exactly solvable and give rise to $C^\text{BdG} =0$. This amounts to $C=-M$ for the resulting superconductors.

\subsection{A Multi-band generalization: potential relevance to multi-layer rhombohedral graphene}

In this section, we construct a simple, multi-band generalization of the above model, which we propose to serve as a simple effective model for $L$-layer rhombohedral graphene.

We start with a discussion of the band structure of $L$-layer rhombohedral graphene for a single flavor (valley and spin). A minimal model of the band structure is given by the following $2L \times 2L$ hopping matrix:
\begin{align}
    t(\bm{k}) =\begin{bmatrix}
        u_d & v_F \bar{k} \\
        v_F k & u_d  & t_1 \\
         & t_1 & 2 u_d & v_F \bar{k} \\
         & & v_F k & 2 u_d  & t_1 \\
        & &  & \ddots & \ddots & \ddots  \\
         & & & & t_1 & L u_d & v_F \bar{k} \\
        & & & & & v_F k & L u_d 
    \end{bmatrix}
\end{align}
where $k,\bar{k} \equiv k_x \pm \mathrm{i}k_y$ and the basis is chosen such that
\begin{align}
    \psi \equiv & (\psi_{\mathcal{A}1}, \psi_{\mathcal{B}1},\psi_{\mathcal{A}2}, \psi_{\mathcal{B}2},\dots ,\psi_{\mathcal{A}L}, \psi_{\mathcal{B}L})^T.
\end{align}
where $\mathcal{A},\mathcal{B}$ label the graphene sub-lattice index.

The dispersion of the two lowest-energy bands are remarkably flat inside the regime $|\bm{k}| \lesssim k_0 \equiv t_1/v_F$, which also set the momentum scale of nontrivial berry curvature/Fubini-Study  metric concentration in momentum space.

We are mainly concerned with the parameter regime $ u_d \sim 15 \text{meV} \ll t_1 \approx 355 \text{meV}$ which is relevant to experiment. In this limit, and when $|\bm{k}| \lesssim k_0$, we find that the lowest-energy hole and electron bands (the $L$-th and $L+1$-th bands) are separated by a gap 
\begin{align}
    \Delta \approx (L-1) u_d
\end{align}
and their wavefunctions admit a simple approximated form. For example, the hole band has
\begin{align}
    \psi_{\mathcal{A}l} \approx & (-k/k_0)^{l-1} \\
    \psi_{\mathcal{B}l} \approx & -l \frac{u_d}{t_1}(-k/k_0)^{l}
\end{align}
whereas the electron band has
\begin{align}
    \psi_{\mathcal{B}l} \approx & (-\bar{k}/k_0)^{L-l} \\
    \psi_{\mathcal{A}l} \approx & - (L-l+1) \frac{u_d}{t_1}(-\bar{k}/k_0)^{L-l+1}
\end{align}
without loss of generality we will hereinafter consider the hole band.

Here we aim to study the SC order observed in this system, highlighting the potential role of quantum geometry. Since the experimentally relevant carrier density corresponds to partial filling the flat regime $|\bm{k}| \lesssim k_0$, we are motivated to assume that the band dispersion is unimportant and thus can be assumed to be completely flat; this is also consistent with the fact that most of the non-trivial quantum geometry is concentrated therein. Moreover, since in this regime the weight of the wavefunction is mostly carried by the lowest $\mathcal{A}$ orbitals, we can neglect the $\mathcal{B}$ sub-lattice and approximately consider an even simpler wavefunction
\begin{align}
   u_{l=1,\dots,L} \propto & (-k/k_0)^{l-1} 
\end{align}
as a caricature of the system. We note that there indeed exist a well defined hopping matrix that gives rise to this state as an exact eigenstate at a fixed energy. Most remarkably, the Schur complement of the $\mathcal{B}$ sublattice hopping matrix in the full hopping matrix:
\begin{align}
    t_\text{Schur}(\bm{k}) \equiv& t_{\mathcal{A}\mathcal{A}} - t_{\mathcal{A}\mathcal{B}} t_{\mathcal{B}\mathcal{B}}^{-1} t_{\mathcal{B}\mathcal{B}}\nonumber\\
    = &u_d \begin{bmatrix}
        1 & \\
        & 2 \\
        & & \ddots \\
        & & & L
    \end{bmatrix} - \frac{t^2_1}{u_d} \begin{bmatrix}
        \frac{(|\bm{k}|/k_0)^2}{1} &  \bar{k}/k_0 \\
        k/k_0 & \frac{(|\bm{k}|/k_0)^2}{2}  + 1 & \bar{k}/(2k_0) \\
         & k/(2k_0) & \frac{(|\bm{k}|/k_0)^2}{3}  + \frac{1}{2}  & \bar{k}/(3k_0) \\
        & & \ddots & \ddots & \ddots  \\
         &  & & k/[(L-2)k_0] & \frac{(|\bm{k}|/k_0)^2}{L-1}  + \frac{1}{L-2}  & \bar{k}/[(L-1)k_0] \\
        & & & & k/[(L-1)k_0] & \frac{(|\bm{k}|/k_0)^2}{L}  + \frac{1}{L-1}  
    \end{bmatrix}
\end{align}
almost exactly features this state as the exact eigenstate at energy $u_d$; dropping the terms $\sim u_d$ and the $\frac{t_1^2(|\bm{k}|/k_0)^2}{Lu_d}$ in $\tilde{t}_{LL}$ makes the eigenstate and flat band exact, the latter of which is equivalent to turning off the intra-layer hopping on the last layer and is controllable in the large $L$ limit. This leads to a simple $L$-band model generalization of the two band model in the previous section (set $k_0=1$, $t_1^2/u_d=1$ as the units):
\begin{align}\label{eq: multi band}
    \tilde{t}(\bm{k})  = -\begin{bmatrix}
        \frac{\bm{k}^2}{1} &  \bar{k} \\
        k & \frac{\bm{k}^2}{2}  + 1 & \frac{\bar{k}}{2} \\
         & \frac{k}{2} & \frac{\bm{k}^2}{3}  + \frac{1}{2}  & \frac{\bar{k}}{3} \\
        & & \ddots & \ddots & \ddots  \\
         &  & & \frac{k}{L-2} & \frac{\bm{k}^2}{L-1}  + \frac{1}{L-2}  & \frac{\bar{k}}{L-1} \\
        & & & & \frac{k}{L-1} &  \frac{1}{L-1}  
    \end{bmatrix}
\end{align}
or more concretely
\begin{align}
t_{\alpha\beta}(\bm{k})  &=\begin{cases}
    \frac{1}{\alpha-1} + \frac{\bm{k}^2}{\alpha}  & \alpha=\beta \neq 1, L \\
    \bm{k}^2 & \alpha=\beta =1\\
    \frac{1}{L-1} & \alpha=\beta =L\\
    \bar{k}/\alpha & \alpha +1 = \beta \\
    k/\beta & \alpha -1 =\beta    
\end{cases}    
\end{align}
Taking $L=2$ reduces to the previous model.

The highest band of this model is perfectly flat and has eigenstate $u_{l=1,\dots,L} \propto  (-k/k_0)^{l-1}$ and Chern number $C_\text{FB} = -L+1$. Since at small $\bm{k}$, most of the electron wavefunction weight is concentrated at small $\alpha$, we may assume that the interaction at the lowest two layers is the most important. Guided by this thought, we apply the general construction approach by choosing $A = 2$, $B=1$, then we find the suggested pairing operator is
\begin{align}
    \hat{\eta}^{\dagger} = \frac{1}{\mathsf{V}}\sum_{\bm{k}} k \hat{\gamma}^\dagger_{1,\bm{k}} \hat{\gamma}_{1,-\bm{k}}.
\end{align}
which has $C^\text{BdG}=L-2$ and thus the total Chern number of the superconductor $C^\text{maj} = 2L-3$. The corresponding solvable interaction is simply (for $\mathsf{N}_\text{layer}>2$)
\begin{align}
    \hat{H}_\text{int} = - V \sum_{\bm{R}} \hat{n}_{\bm{R},\alpha = 1} \hat{n}_{\bm{R},\alpha = 2}
\end{align}

Instead, if we choose $A = 1$, $B=2$, then we find the suggested pairing operator is
\begin{align}
    \hat{\eta}^{\dagger} = \frac{1}{\mathsf{V}}\sum_{\bm{k}} \frac{1}{\bar{k}} \hat{\gamma}^\dagger_{1,\bm{k}} \hat{\gamma}_{1,-\bm{k}}.
\end{align}
which has $C^\text{BdG}=-L+2$ and thus the total Chern number of the superconductor $C^\text{maj} = 1$. The corresponding solvable interaction is simply
\begin{align}
    \hat{H}_\text{int} = - V \sum_{\bm{R}} \hat{n}_{\bm{R},\alpha = 1} \hat{n}_{\bm{R},\alpha = 2}
\end{align}
which is the same as the previous one! This means that the two superconducting GSs are exactly degenerate for this model, but they have different Chern numbers.

\subsection{A lattice model with tunable quantum geometry}

The above models are defined in open momentum space so that they are difficult to simulate in numerics. Here we regularize the two band models in a hexagonal brillioun zone (BZ). To achieve this, we simply need to define the flat band wavefunction:
\begin{align}
    u(\bm{k}) &= \begin{pmatrix} 1\\
    [k_0\hat{\zeta}(\bm{k})]^{-M} \end{pmatrix}\sqrt{1+\left| k_0 \hat{\zeta}(\bm{k})\right|^{-2M} } 
\end{align}
where
\begin{align}
    \hat{\zeta}(\bm{k})  \equiv \zeta(k) -\pi \bar{k} /A_\text{BZ} 
\end{align}
is a non-holomorphic but periodic modification of the Weierstrass $\zeta$ function~\cite{10.1063/1.5042618}, $k,\bar{k} \equiv k_x \pm \mathrm{i}k_y$, and $A_\text{BZ} = \frac{\sqrt{3}}{2}|\bm{G}_{1,2}|^2$ is the area of the BZ determined by the primitive reciprocal lattice vectors. In this work, we take the convention that $\bm{G}_{1,2}=2\pi (1,\pm 1/\sqrt{3}))$ such that the half-periods of Weierstrass function is $g,\bar{g}=(\pi(1+\mathrm{i}/\sqrt{3}),\pi(1-\mathrm{i}/\sqrt{3}))$.

The reason this is a good compactification of the infinite-momentum-space models is that, when $|\bm{k}|\lesssim k_0 < |\bm{G}_1|$, 
\begin{align}
    u(\bm{k}) \approx \begin{pmatrix}
        1 \\ (-k/k_0)^M
    \end{pmatrix}
\end{align}
The parameter $k_0$ controls the radius within which the wavefunction is approximately holomorphic as well as the berry curvature concentration. This wavefunction describes a Chern band with Chern number $C_\text{FB}=-M$.

Assuming this band is the lowest, the proposed unprojected Hamiltonian is given by Eq.~\ref{eq: one flatband Hint2 lowest}. The projected, exactly solvable model is given by Eq.~\ref{eq: inversion solvable H}.

\bibliographystyle{apsrev4-1} 
\bibliography{ref}